\documentclass[11pt,english]{article}

\usepackage[utf8]{inputenc} 

\usepackage{dsfont}

\usepackage[left=1in,right=1in,top=1in,bottom=1in]{geometry}
\usepackage{authblk}
\usepackage{enumitem}
\usepackage{ulem} 
\usepackage{empheq}
\usepackage[none]{hyphenat}

\usepackage{graphicx}
\usepackage{subfigure}
\usepackage{float}
\usepackage{wrapfig} 

\usepackage{xcolor}
\usepackage{color}
\definecolor{darkgreen}{rgb}{0,0.5,0}
\definecolor{darkblue}{rgb}{0,0,0.6}
\definecolor{purple}{rgb}{0.4,.2,0.7}
\usepackage{colortbl}      

\usepackage{amsmath}
\usepackage{amssymb}

\usepackage{tensor} 
\usepackage{slashed}
\usepackage{mathtools}
\usepackage{leftidx}
\usepackage{esint}
\usepackage{amsfonts,amsthm,bm}

\usepackage{physics}
\usepackage{feynmf}
\usepackage{braket}
\usepackage{ytableau}

\usepackage{prettyref}
\usepackage[colorlinks=true,citecolor=darkgreen,linkcolor=purple,urlcolor=purple]{hyperref}

\usepackage{tikz}

\usepackage[numbers,sort&compress]{natbib}

\newcommand{\nn}{\nonumber}

\numberwithin{equation}{section}
\numberwithin{figure}{section}
\numberwithin{table}{section}

\begin{document}

\emergencystretch 3em

\title{\LARGE\textsc{ De Sitter Horizon Edge Partition Functions}}

\author[]{\normalsize Y.T.\ Albert Law}
\affil[]{ \it \normalsize	 Leinweber Institute for Theoretical Physics at Stanford, 382 Via Pueblo, Stanford, CA 94305, USA}

\date{}
\maketitle

\begin{center}
	\vskip-10mm
	{ \href{mailto:ytalaw@stanford.edu}{ytalaw@stanford.edu}}
\end{center}

\vskip10mm

\thispagestyle{empty}

\begin{abstract}
	
	One-loop $S^{d+1}$ path integrals were shown to factorize into two parts: a bulk thermal ideal gas partition function in a $dS_{d+1}$ static patch and an edge partition function associated with degrees of freedom living on $S^{d-1}$. Here, we analyze the $\mathfrak{so}(d)$ structure of the edge partition functions for massive and massless totally symmetric tensors of arbitrary rank in any $d\geq 3$. For linearized Einstein gravity on $S^{d+1}$, we find that the edge partition function receives contributions from  shift-symmetric vector and scalar fields on $S^{d-1}$, suggesting a possible interpretation in terms of an embedded $S^{d-1}$ brane.

\end{abstract}


\newpage

\tableofcontents


\newpage

\section{Introduction}

Envisioning precision tests on candidate microscopic models for the de Sitter (dS) horizon, considerable efforts have recently been made to study quantum corrections to its thermodynamics \cite{Anninos:2020hfj,Law:2020cpj,David:2021wrw,Anninos:2021ihe,Anninos:2021ene,Anninos:2021eit,Grewal:2021bsu,Muhlmann:2022duj,Bobev:2022lcc,Castro:2023dxp,Castro:2023bvo,Bourne:2024ded,bandaru2024quantumsitterentropysphere}, refining the original Gibbons-Hawking proposal \cite{Gibbons:1976ue}. In some cases, matching with a microscopic calculation up to the logarithmic correction has been reported  \cite{Shyam:2021ciy,Coleman:2021nor,Bobev:2022lcc}.

The work \cite{Anninos:2020hfj} considered the leading quantum corrections arising from free matter fields and graviton, in both Lorentzian and Euclidean signatures. In Lorentzian signature, \cite{Anninos:2020hfj} computed a ``quasicanonical" ideal gas thermal partition function $Z_\text{bulk} (\beta)$ for any free fields living on a $dS_{d+1}$ static patch,
\begin{align}
	ds^2=-\left( \ell_\text{dS}^2 - r^2\right)dt^2+\frac{dr^2}{1-\frac{r^2}{\ell_\text{dS}^2}} +r^2 d\Omega^2 \;, \qquad 0\leq r<\ell_\text{dS} \;, 
\end{align}
at any inverse temperature $\beta$.\footnote{$\beta$ is normalized such that $\beta=2\pi$ corresponds to the dS or Hawking temperature \cite{Rahman:2024vyg}.} For bosonic fields, $Z_\text{bulk} (\beta)$ takes the general form
\begin{align}\label{introeq:Zbulk}
	\log Z_\text{bulk} (\beta)\equiv\log \widetilde{\Tr} \, e^{-\beta \hat H}\equiv\int_0^\infty \frac{dt}{2t} \frac{1+e^{-2\pi t/\beta}}{1-e^{-2\pi t/\beta}}\,\chi(t) \; .
\end{align}
Here $\chi(t)$ is the so-called Harish-Chandra character of the dS boost generator for a given $SO(1,d+1)$ unitary irreducible representation (UIR) \cite{10.3792/pja/1195523378,10.3792/pja/1195523460}, rigorously defined as a distribution on $SO(1,d+1)$ \cite{10.2307/1992907,bams/1183525024,10.3792/pja/1195522333}. The explicit formula for the graviton is
\begin{align}
	\chi(t)	
	=  \left[ \frac{(d+2)(d-1)}{2}  \frac{e^{-dt} +1}{|1-e^{-t}|^d} - d \frac{e^{-(d+1)t} +e^{t}}{|1-e^{-t}|^d} \right]_+ \;. 
\end{align}
The notation $\left[ \cdots \right] _+$  is defined in \eqref{eq:flipnotation} with $q=e^{-t}$. Physically, $\chi(t)$ encodes the quasinormal mode (QNM) spectrum for the given free field,
\begin{align}\label{intro:qnmchar}
	\chi(t)= \sum_z N_z \, e^{-iz |t|}  \; ,
\end{align}
where $z$ and $N_z$ are the QNM frequencies and degeneracies. $\chi(t)$ can also be understood as an integrated Green function \cite{Grewal:2024emf,spin}. The main observation in \cite{Anninos:2020hfj} is that the Fourier transform $\tilde\rho(\omega)=\int_{-\infty}^\infty \frac{dt}{2\pi} e^{-i\omega t}\chi(t)$ can be assigned as a spectral density on the continuous normal mode spectrum, leading to the definition of the quasi-trace \eqref{introeq:Zbulk}.\footnote{$\tilde\rho(\omega)$ can be understood in terms of scattering phases associated with the reduced scattering problems descending from the free field equations \cite{Law:2022zdq}. Relatedly, $\widetilde{\Tr}$ in \eqref{introeq:Zbulk} can be understood as a relative or renormalized trace. A short review of these ideas in the dS context can be found in  \cite{Law:2023ohq}.}

\begin{figure}[H]
	\centering
	\includegraphics[width=0.8\textwidth]{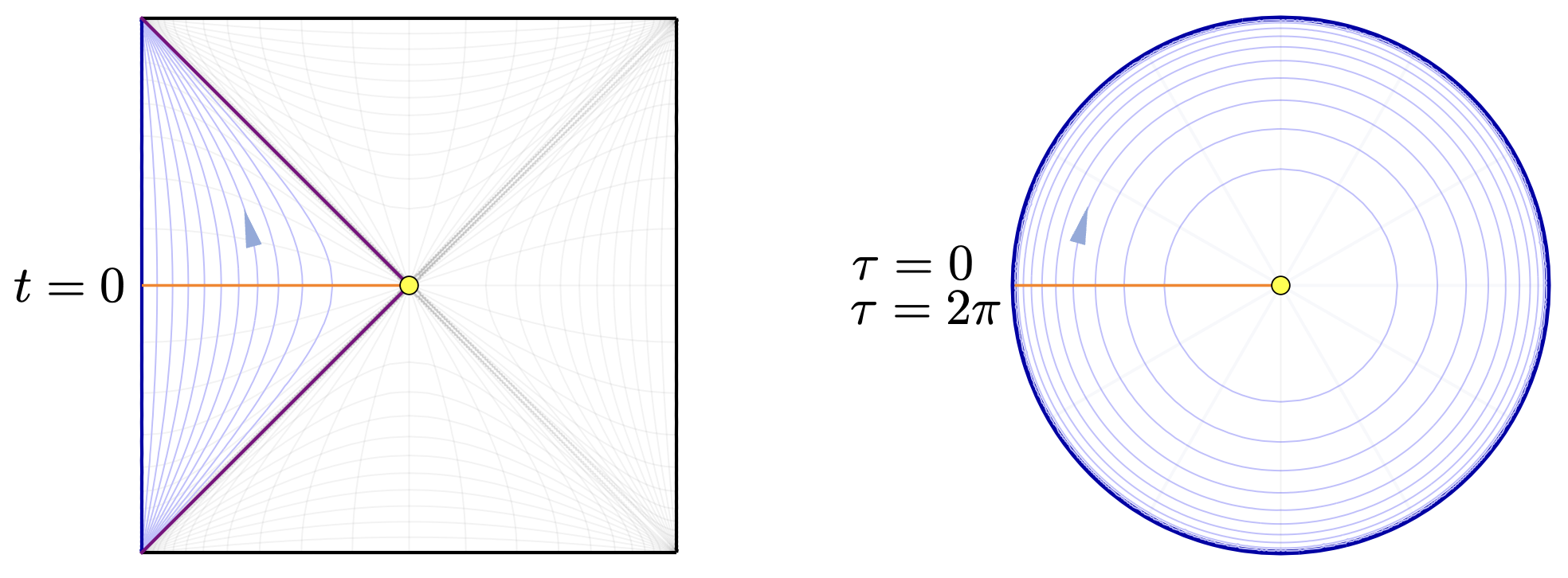}
	\caption{Left: Penrose diagram for $dS_{d+1}$ and a static patch (highlighted in blue). The yellow dot indicates the bifurcation surface at $r=\ell_\text{dS}$. Right: Upon the Wick rotation $t \to -i\tau$, $\tau\sim\tau+2\pi$, the static patch becomes the round sphere, with $\tau$ parametrizing the thermal circle.}
	\label{pic:staticpatch}
\end{figure}

Traditionally, quantum corrected dS entropy was proposed to be computed by a Euclidean gravitational path integral with a positive cosmological constant \cite{Gibbons:1976ue}. \cite{Anninos:2020hfj} studied the 1-loop partition functions $Z_{\rm PI}$ for matter fields and gravitons on the round $S^{d+1}$ saddle of radius $\ell_\text{dS}$. For a real scalar with mass $M^2$, $Z_{\rm PI}=\det \left(-\nabla_0^2+M^2 \right)^{-\frac12}$ where $-\nabla_0^2$ is the scalar Laplacian on $S^{d+1}$, and it is found that 
\begin{align}\label{introeq:PIscalar}
	Z_\text{PI} =Z_{\rm bulk} (\beta=2\pi) \qquad \qquad \text{(Scalar)}\;. 
\end{align}
This is consistent with the intuition that a $S^{d+1}$ partition function should compute a thermal partition function in a $dS_{d+1}$ static patch at inverse temperature $2\pi$, given the fact that the round sphere $S^{d+1}$ is the Wick-rotation of a $dS_{d+1}$ static patch (see figure \ref{pic:staticpatch}). 
Interestingly, for any fields with spin $s\geq 1$,  \eqref{introeq:PIscalar} is modified by some ``edge'' contribution\footnote{We note that our notation for $Z_{\rm edge}$ is different from \cite{Anninos:2020hfj}.}
\begin{align}\label{introeq:PIsplit}
	Z_\text{PI} =Z_{\rm bulk} (\beta=2\pi) Z_{\rm edge} \qquad \qquad \text{(spin $s\geq 1$)}\;. 
\end{align}
The reason the extra factor was dubbed ``edge" is because $Z_{\rm edge}$ takes the form of a path integral on a co-dimension-2 sphere $S^{d-1}$ of radius $\ell_\text{dS}$, naturally associated with the bifurcation surface in the Lorentzian siganture or the origin in the Euclidean signature (the yellow dots in figure \ref{pic:staticpatch}). For Maxwell theory, $Z_{\rm edge}$ is the $S^{d-1}$ path integral for a ghost compact scalar with a target $U(1)$ circle \cite{RiosFukelman:2023mgq,Ball:2024hqe}. More generally, $Z_{\rm edge}$ for a $p$-form gauge theory is a path integral for a ghost $(p-1)$-form gauge theory on $S^{d-1}$ \cite{David:2021wrw,Mukherjee:2023ihb,Ball:2024xhf}. On the other hand, the explicit formula for graviton is
\begin{align}\label{introeq:Zedgeoriginal}
	\log Z_{\rm edge}	= & \log   \frac{i^{d+3} }{ \text{Vol}(SO(d+2))_c } \left(\frac{32\pi^3 G_N}{\text{Vol}(S^{d-1})\ell_\text{dS}^{d-1}}\right)^{\frac{\text{dim}SO(d+2)}{2}}\nn\\
	& \qquad  -\int_0^\infty \frac{dt}{2t}\frac{1+e^{-t}}{1-e^{-t}}\left[(d+2)\frac{e^{-(d-1)t}+e^t}{(1-e^{-t})^{d-2}}-\frac{e^{-dt}+e^{2t}}{(1-e^{-t})^{d-2}} \right]_+ \;.
\end{align}
Explicit formulas for the unit round $S^n$ volume and the 'canonical' volume of $SO(d+2)$ are given in \eqref{eq:volumes}. The first factor accounts for  the logarithmic correction to the dS entropy: $\mathcal{S}_{\rm dS } =\frac{A}{4G_N}- \frac{\text{dim}SO(d+2)}{2}\log \frac{A}{4G_N} +\cdots $, but was not included as part of the edge partition function in \cite{Anninos:2020hfj}. However, as we will see in section \ref{sec:gravity}, incorporating this factor aligns with a natural path integral interpretation of \eqref{introeq:Zedgeoriginal}.

For Maxwell theory,  when interpreted in terms of entanglement entropy across the bifurcation surface, the bulk-edge split \eqref{introeq:PIsplit} aligns with the idea of ``edge modes" in studies of Maxwell entanglement \cite{Kabat:1995eq,Donnelly:2011hn, Donnelly:2012st, Eling:2013aqa, Radicevic:2014kqa, Donnelly:2014gva, Donnelly:2014fua, Huang:2014pfa, Ghosh:2015iwa, Hung:2015fla, Aoki:2015bsa, Donnelly:2015hxa, Radicevic:2015sza, Pretko:2015zva, Soni:2015yga,Zuo:2016knh, Soni:2016ogt, Delcamp:2016eya, Agarwal:2016cir, Blommaert:2018rsf, Blommaert:2018oue, Freidel:2018fsk}. More recently, this has been analyzed in Lorentzian signature \cite{Ball:2024hqe} in a framework where the horizon is excised by a 't Hooft brick wall \cite{tHooft:1984kcu}. Incorporated by a ``Dynamical Edge Mode" (DEM) boundary condition, edge modes can be understood as physical gauge transformations with support on the boundary, with the normal component of the electric field as its conjugate momentum. Applied to the case of a dS horizon, it is shown that $Z_{\rm edge}$ can indeed be understood as a thermal canonical partition function for these modes. The DEM approach has been generalized to $p$-form gauge theories \cite{Ball:2024xhf}.

In contrast to the Maxwell case, gravitational edge modes remain an active area of research.\footnote{See \cite{Donnelly:2016auv,Geiller:2017xad,Speranza:2017gxd,Geiller:2017whh,Freidel:2019ees,Takayanagi:2019tvn,Freidel:2020xyx,Freidel:2020svx,Freidel:2020ayo,Donnelly:2020xgu,Ciambelli:2021vnn,Carrozza:2021gju,Ciambelli:2021nmv,Carrozza:2022xut,Ciambelli:2022cfr,Mertens:2022ujr,Wong:2022eiu,Donnelly:2022kfs,Lee:2024etc,Fliss:2025kzi} for a partial list of this great body of work, and also \cite{Ciambelli:2022vot} for a review.} Despite natural suspicions of a connection between \eqref{introeq:Zedgeoriginal} and gravitational edge modes, it is not immediately clear. One challenge is that much of the existing work on gravitational edge modes focuses on classical phase space and symmetries, while explicit results demonstrating how gravitational edge modes contribute to thermal partition functions or entanglement entropy are scarce, even in the case of linearized gravity (see however \cite{Benedetti:2019uej,David:2022jfd,Balasubramanian:2023dpj,Blommaert:2024cal}). This is not too surprising given the technical complexity of explicit graviton calculations on curved spacetime backgrounds, together with the broader conceptual baggage associated with the full nonlinear theory.

As an initial step toward uncovering its underlying dynamical principles, it is crucial to obtain a more refined version of the formula \eqref{introeq:Zedgeoriginal}. Unlike the $p$-form case, the $\mathfrak{so}(d)$ structure (or the field content on $S^{d-1}$) underlying \eqref{introeq:Zedgeoriginal} remains obscure. In this work, we take on the challenge of analyzing and elucidating this structure.

\paragraph{The idea}

A static patch in dS preserves only the subgroup $SO(1,1)\times SO(d)$ of the full dS group $SO(1,d+1)$, which becomes $U(1)\times SO(d)$ upon Wick-rotation to the Euclidean signature. It is therefore natural to organize our calculation according to this subgroup. One approach is to recompute the spectrum, but this becomes increasingly cumbersome for higher-spin fields.

Instead, we aim to uncover the $\mathfrak{so}(d)$ contents of $Z_{\rm PI}$ through direct manipulation. To illustrate the idea, consider a massive scalar whose $Z_{\rm PI}$ takes the form 
\begin{align}\label{introeq:scalarPI}
	\log Z_{\rm PI}=-\frac12\log \det \left( -\nabla^2_0+M^2\right) = \int \frac{d\tau}{2\tau} \Tr \, e^{-\left( -\nabla^2_0+M^2\right)   \tau}
\end{align}
in the heat kernel representation. The trace Tr runs over eigenfunctions of $ -\nabla^2_0$, i.e. spherical harmonics, which furnish finite-dimensional $\mathfrak{so}(d+2)$ unitary irreducible representations (UIRs) $ \rho^{d+2} _{L}$ labeled by an integer $L\geq 0$. This motivates the introduction of an abstract generating function
\begin{align}\label{introeq:genfn}
	\bigoplus_{L= 0}^\infty\rho^{d+2} _{L}  q^L 
\end{align}
where the variable $q$ tracks the contributions from $\rho^{d+2} _{L}$. To proceed, we employ the branching rule for $\mathfrak{so}(d+2)$-modules into $\mathfrak{u}(1)\oplus \mathfrak{so}(d)$-modules \cite{branch}, enabling us to express \eqref{introeq:genfn} as an infinite sum of the latter.  As we will demonstrate in section \ref{sec:Lspin0PI}, we can manipulate \eqref{introeq:genfn} into
\begin{align}\label{introeq:genresult}
	\bigoplus_{L=0}^\infty \rho^{d+2}_L q^L =\left(\bigoplus_{k=-\infty}^\infty \rho_{k}^2 q^{|k|}\right)\otimes  \left(\bigoplus_{l=0}^\infty \rho_l^d \sum_{n=0}^\infty  q^{2n+l}\right) \; .
\end{align}
Here $\rho_{k}^2$ denotes the $\mathfrak{so}(2)\simeq \mathfrak{u}(1)$-module labeled by $k\in\mathbb{Z}$, while $\rho_l^d$ represents the $\mathfrak{so}(d)$-module labeled by $l\geq 0$. The point is that this analysis allows us to explicitly track all the $\mathfrak{so}(d)$ contents. For this scalar example, applying \eqref{introeq:genresult} to \eqref{introeq:scalarPI}, it becomes evident that the Bose-Einstein factor in \eqref{introeq:Zbulk} (at $\beta=2\pi$) originates from the sum over $\mathfrak{u}(1)$-modules, while the character $\chi(t)$ arises from the infinite sum over $\mathfrak{so}(d)$-modules.

For spinning fields, we do not expect their analogous generating functions to exhibit the simple factorized form \eqref{introeq:genresult}. However, since their bulk partition function \eqref{introeq:Zbulk} takes the same form as the scalar case, we can isolate a part that takes the factorized form \eqref{introeq:genresult}. Whatever remains then naturally leads to $Z_{\rm edge}$, with its $\mathfrak{so}(d)$ content kept explicit throughout the analysis.

We focus on symmetric tensor fields, but the methods naturally extend to general mixed-symmetry fields. Our primary analysis considers $d \geq 3$ as $d \leq 2$ deviates from the higher-dimensional pattern.\footnote{In fact, the topological nature of the $d\leq 2$ case makes it an ideal arena for understanding gravitational edge modes. For example, the Chern-Simons formulation of $dS_3$ Einstein gravity enables the derivation of an all-loop expression for the $dS_3$ gravity partition function on $S^3$ \cite{Anninos:2020hfj}. Building on this, \cite{Anninos:2021ihe} explored the hypothesis that the all-loop $dS_3$ entropy corresponds to a topological entanglement entropy \cite{Kitaev:2005dm,Levin:2006zz} arising from edge modes.} Since our focus is on the algebraic structures of the $S^{d+1}$ path integrals, we will proceed formally, omitting discussions of UV divergences. A detailed discussion of UV regularization and the extraction of universal parts of integrals of the form \eqref{introeq:Zbulk} and \eqref{introeq:Zedgeoriginal} can be found, for instance, in Appendix C of \cite{Anninos:2020hfj}. We will also set the dS length to unity from now on: $\ell_\text{dS}=1$. 

A concise summary of the main results of this work, together with broader perspectives, appears in~\cite{Law:2026tuk}.

\subsection*{Overview  and summary of results}

	As motivated above, we will study QNMs in $dS_{d+1}$ and 1-loop $S^{d+1}$ partition functions, organizing both by their $SO(d)$ representation content. To help the reader navigate the various results, we briefly summarize what is already known in the literature. The $\mathfrak{so}(d)$ content of QNM spectra has been worked out for massive scalars, Maxwell fields and gravitons in \cite{Lopez-Ortega:2006aal}, and massive vectors \cite{Rahman:2024mok}. In addition, the bulk-edge split \eqref{introeq:PIsplit} is known to hold for all free bosonic theories, while explicit edge-mode spectra have so far been fully determined only for massive and massless $p$-forms \cite{Anninos:2020hfj,RiosFukelman:2023mgq,Ball:2024hqe,David:2021wrw,Mukherjee:2023ihb,Ball:2024xhf}. Our aim is to extend these results to massive and (partially) massless totally symmetric tensor fields. Mixed-symmetry fields and fermions remain outside the scope of this work.

In section \ref{ref:qnmsod}, we work out the $\mathfrak{so}(d)$ contents of QNMs for massive spin-$s$ fields in $dS_{d+1}$ building upon the algebraic method in \cite{Sun:2020sgn}. A key result is the non-trivial prediction \eqref{eq:massspinqnm} for their QNM spectra one would obtain by solving the massive spin-$s$ equations of motion in a $dS_{d+1}$ patch.

In section \ref{ref:branching}, we carry out our analysis of 1-loop $S^{d+1}$ partition functions for massive spin-$s$ fields using the branching rule $\mathfrak{so}(d+2)\to \mathfrak{u}(1)\oplus \mathfrak{so}(d)$. The main result is the formula \eqref{eq:spinsZedge} for $Z_{\rm edge}$, which reveals that it receives contributions from ghost massive fields with spin $0,1,\dots,s-1$ on $S^{d-1}$. 

In section \ref{sec:Maxwell}, we shift our focus to massless vectors. While the Maxwell case is well-studied, we extend the analyses from previous sections to highlight unique subtleties of the massless case. In section \ref{sec:YM}, we discuss Yang-Mills (YM) theory. We investigate the possibility that $Z^\text{YM,1-loop}_{\rm edge}$  arises from an underlying interacting theory. We find that $Z^\text{YM,1-loop}_{\rm edge}$ is equal to the 1-loop partition function of a ghost sigma model (SM) on $S^{d-1}$:
\begin{align}
	Z^\text{YM,1-loop}_\text{edge} \left( S^{d+1}\right)  = \frac{1}{Z^\text{1-loop}_{\rm SM} \left( S^{d-1}\right)  } \; . 
\end{align}

In section \ref{sec:gravity}, we apply our methods to linearized Einstein gravity. Our main result is the following significantly more refined version of \eqref{introeq:Zedgeoriginal}:
\setlength\fboxsep{12pt} 
\begin{empheq}[box=\fbox]{align}\label{introeq:Zedgegravresult}
	&\qquad \qquad \qquad \qquad Z_{\rm edge }= Z^\text{det}_\text{edge} Z^\text{non-det}_\text{edge} \nn\\
	Z^\text{det}_\text{edge} & = \det\nolimits'_{-1} \left| -\nabla_1^2-\left( d-2\right)  \right|^{\frac12}  \det\nolimits' \left| -\nabla_0^2-\left( d-1\right)  \right| \det\nolimits'  \left(-\nabla_0^2 \right)^{\frac12} \nn\\
	Z^\text{non-det}_\text{edge} &= \frac{i^{d+3}}{\text{Vol}(SO(d+2))_c}\left( \frac{16\pi^2 G_N } {\text{Vol}(S^{d-1})} \right)^{\frac{\text{dim}SO(d+2)}{2}} d^{\frac{\text{dim}SO(d)+2d}{2}}\left( d-2 \right)^{\frac{1}{2}}  \; . 
\end{empheq}
The determinants correspond to ghost fields on $S^{d-1}$ with special tachyonic masses, which exhibit shift symmetries \cite{Bonifacio:2018zex,Bonifacio:2019hrj}. The appearance of shift-symmetric theories suggests that $Z_{\rm edge}$ is linked to a spontaneous symmetry-breaking phenomenon.  In section \ref{sec:gravZedgecan}, we explore a possible interpretation in terms of an embedded $S^{d-1}$ brane, where all the tachyonic fields admit natural geometric descriptions.

In section \ref{ref:HS}, we extend our considerations to partially massless (PM) gauge fields. For PM fields with maximal depth, we find that their $Z_{\rm edge}$ are captured by lower-spin fields that nonlinearly realize the global higher-spin symmetries, generalizing the YM and gravity case. Curiously, for PM fields with non-maximal depth, in addition to shift-symmetric fields, other fields must be included. In section \ref{sec:conical} we discuss how our branching rule method can be used to uplift existing $S^{d+1}$ results to obtain 1-loop partition functions on a sphere $S^{d+1}_\beta$ with a thermal periodicity $\beta \neq 2\pi$. Section \ref{sec:outlook} concludes with open problems and directions for future research.

We include several appendices. Appendix \ref{sec:dSUIR} summarizes essential facts about the de Sitter group $SO(1,d+1)$ and its unitary irreducible representations (UIRs). Appendix \ref{sec:sod} provides useful details about $\mathfrak{so}(d)$. In appendix \ref{sec:charintegral}, we outline the conventions for our path integral measure and review the derivation of the 1-loop sphere partition function $Z_{\rm PI}$.


\section{Quasinormal modes on $dS_{d+1}$ and their $\mathfrak{so}(d)$ contents}\label{ref:qnmsod}

In this section, we analyze the QNMs of massive spin-$s$ symmetric tensor fields in $dS_{d+1}$, building on the methods in \cite{Sun:2020sgn}. As detailed in \cite{Sun:2020sgn}, the QNMs fall into lowest-weight representations of the $\mathfrak{so}(1,d+1)$ algebra, which underlie the algebraic constructions in \cite{Ng:2012xp,Jafferis:2013qia,Tanhayi:2014kba}. Appendix \ref{sec:dSUIR} summarizes relevant facts about $SO(1,d+1)$ and its UIRs. Our primary goal here is to determine the $\mathfrak{so}(d)$ contents of the QNMs, and therefore, the explicit realizations of the QNMs as static patch mode functions will be omitted. Useful mathematical facts about $\mathfrak{so}(d)$  and symmetric transverse traceless (STT) spherical harmonics are collected in appendix \ref{sec:sod}.

\subsection{Massive scalar}

The QNMs fall into two lowest-weight representations of the $\mathfrak{so}(1,d+1)$ algebra  \eqref{eq:conformal algebra}. To construct these, we start with the lowest-weight states
\begin{align}\label{eq:scalarprimaries}
	\ket{\Delta} \qquad \text{and} \qquad \ket{\bar\Delta}\;  .
\end{align}
Here, the $\mathfrak{so}(1,1)$ weight $\Delta$ is related to the scalar mass through $M^2=\Delta \bar\Delta =\Delta \left( d-\Delta\right) $. The states \eqref{eq:scalarprimaries} are primary, meaning\footnote{To align with the conventional understanding of QNMs as mode functions, we note that \eqref{eq:scalarprimaries} can be realized as solutions to the Klein-Gordon equation, subject to \eqref{eq:primary conditions} as translated into additional differential equations.}
\begin{align}\label{eq:primary conditions}
	D \ket{\Delta} = \Delta \ket{\Delta}\; , \qquad K_i \ket{\Delta} =0 \;, \qquad M_{ij} \ket{\Delta} =0 \; ,
\end{align}
and similarly for $\ket{\bar\Delta}$. The descendants of \eqref{eq:scalarprimaries}, obtained by successive actions of the creation operators $P_i$ ($i=1,\dots, d$), 
\begin{align}\label{eq:scalarqnmmodule}
	\alpha{\rm -tower:} \quad  P_{i_1}\cdots P_{i_N}\ket{\Delta} \qquad \text{and} \qquad \beta{\rm -tower:} \quad P_{i_1}\cdots P_{i_N}\ket{\bar\Delta}\; , \quad N = 0 , 1, \dots \; ,
\end{align}
give the rest of the QNMs. Following the terminology of \cite{Sun:2020sgn}, we refer to these two towers of QNMs as the $\alpha$- and $\beta$-towers.

\paragraph{The $\mathfrak{so}(d)$ content}

The two towers of symmetric tensors \eqref{eq:scalarqnmmodule} can be decomposed into a direct sum of $\mathfrak{so}(d)$ irreducible representations (irreps); it suffices to focus on the traceless parts of the products $P_{i_1}\cdots P_{i_N}$, since the traces can be captured by acting with powers of $P^2$.\footnote{This is essentially the problem of branching rank-$N$ symmetric tensor representations of $\mathfrak{gl}(d)$ into traceless symmetric tensor representations of $\mathfrak{so}(d)$.} Explicitly, recall that spherical harmonics on $S^{d-1}$ can be represented by degree-$l$ symmetric homogeneous polynomials in an ambient $\mathbb{R}^d$:
\begin{align}\label{eq:scalarY}
	Y_0^l (X)=E_{i_1 \cdots i_l}X^{i_1}\cdots X^{i_l} \;, \qquad l \geq 0\; ,  
\end{align}
restricted to the hypersurface $X^2=1$. The tensor $E_{i_1 \cdots i_l}$ is totally symmetric and traceless. With \eqref{eq:scalarY}, we can then reorganize the two modules \eqref{eq:scalarqnmmodule} as 
\begin{align}\label{scalar harm}
	P^{2n}Y_0^l (P)\ket{\Delta} \qquad \text{and} \qquad 	P^{2n}Y_0^l (P)\ket{\bar\Delta} \; ,
\end{align}
where $P^{2n}$ means we act with $P^2 = P^i P_i$ by $n$ times. The QNM frequencies are the eigenvalues of the operator $D$ divided by $i$, and can be read off as the total number of powers of $P_i$ acting on $\ket{\Delta}$: 
\begin{align}
	i \omega_{nl} = \Delta + 2n + l  \qquad \text{and} \qquad  \bar\Delta + 2n + l \; ,
\end{align}
which are the well-known spectra of QNMs obtained by directly solving the Klein-Gordon equation the static patch \cite{Lopez-Ortega:2006aal}. 

\paragraph{Harish-Chandra character}

As mentioned in the introduction, the Harish-Chandra character for a given $SO(1,d+1)$ UIR encodes QNMs as in \eqref{intro:qnmchar}. In this scalar case,
\begin{align}\label{eq:scalarHC}
	\chi(q)=\sum_{N=0}^\infty \binom{N+d-1}{N}  q^{N} \left( q^\Delta +q^{\bar{\Delta}}\right) =\sum_{n=0}^\infty \sum_{l=0}^\infty D_l^d  q^{2n+l} \left( q^\Delta +q^{\bar{\Delta}}\right) \; , \qquad q\equiv e^{-|t|} \;.
\end{align}
One can check that the two sums are equal
\begin{align}
	\chi(q)=\frac{q^\Delta +q^{\bar\Delta}}{\left( 1-q\right) ^d} \;,
\end{align}
consistent with the equivalence between the \eqref{eq:scalarqnmmodule} and \eqref{scalar harm} as bases for the two lowest-weight $\mathfrak{so}(1,d+1)$ representations.

\subsection{Massive vector}\label{sec:spin1char}

Similar to the scalar case, the starting point for constructing massive spin-1 QNMs involves two primary states labeled by the $\mathfrak{so}(1,1)$ weights $\Delta$ and $\bar\Delta$, related to the mass through $M^2=\left( \Delta-1\right) \left(  \bar\Delta -1\right) =\left( \Delta -1\right) \left( d-\Delta-1\right) $, and $\bar\Delta$ and an additional $\mathfrak{so}(d)$  vector index $i$:
\begin{align}\label{eq:vecprimary}
	 \ket{\Delta}_i   \qquad \text{and} \qquad \ket{\bar\Delta}_i \; .
\end{align}
These satisfy
\begin{align}
	D \ket{\Delta}_i = \Delta \ket{\Delta}_i\; , \qquad K_j \ket{\Delta}_i =0 \;, \qquad M_{ij} \ket{\Delta}_k =\delta_{ik}\ket{\Delta}_j - \delta_{jk}\ket{\Delta}_i \; ,
\end{align}
and similarly for $\ket{\bar\Delta}_i$. Two towers of QNMs are constructed by acting on \eqref{eq:vecprimary} with $P_i$:
\begin{align}\label{eq:vecdesc}
	\alpha{\rm -tower:} \quad P_{i_1}\cdots P_{i_N}\ket{\Delta}_j  \qquad \text{and} \qquad  \beta{\rm -tower:} \quad P_{i_1}\cdots P_{i_N}\ket{\bar\Delta}_j \; , \qquad N = 0 , 1, \dots \;.
\end{align}
It is useful to introduce an auxiliary vector $U^i \in \mathbb{R}^d$ to encode the vector structure of \eqref{eq:vecprimary}: 
\begin{align}
	\ket{\Delta}_i \to \ket{\Delta,U}_1 \equiv\ket{\Delta}_i U^i \; , \qquad \ket{\bar\Delta}_i \to \ket{\bar\Delta,U}_1 \equiv\ket{\bar\Delta}_i U^i 
\end{align}
and thus the descendants \eqref{eq:vecdesc}.

\paragraph{The $\mathfrak{so}(d)$ content}

As before we decompose the tensors \eqref{eq:vecdesc} into direct sums of $\mathfrak{so}(d)$ irreps.  The problem then reduces to decomposing the tensor product:\footnote{See, e.g., \cite{king_modification_1971} for the general rule for tensor product decomposition for $\mathfrak{so}(N)$-modules, nicely reviewed in \cite{Bekaert:2006py}.}
\begin{align}
	\rho^d_N \otimes \rho^d_1 
	=
	\begin{cases}
		 \rho^d_{N,1} \oplus \rho^d_{N+1} \oplus   \rho^d_{N-1} \; &  , \qquad N \geq 1 \\
		\qquad \quad \; \rho^d_1 & , \qquad N = 0
	\end{cases}
	\; . 
\end{align}
This holds uniformly for $d=3$ as well, as long as we take into account the modification \eqref{eq:so3reps}. This leads to three sectors of $\mathfrak{so}(d)$ irreps, explicitly constructed as follows. 

First, we have the vector sector carrying the representations $\rho^d_{N,1}$ 
\begin{align}\label{eq:vecvectortype}
	\textbf{V:} \qquad P^{2n}Y_1^l (P,\partial_U)\ket{\Delta,U}_1\; , \quad \text{and} \quad P^{2n}Y_1^l (P,\partial_U)\ket{\bar\Delta,U}_1\; , \quad n\geq 0 \, , \quad  l\geq 1 \; ,
\end{align}
the vector spherical harmonics $Y_1^l$ are represented as homogeneous polynomials in $\mathbb{R}^d$:
\begin{align}
	Y_1^l (X,U)=E_{i_1 \cdots i_l, j}X^{i_1}\cdots X^{i_l} U^{j} \; .
\end{align}
The tensor $E_{i_1 \cdots i_l, j}$ has a structure of a traceless Young diagram with $l$ boxes in the first row and 1 box in the second. It is straightforward to read off the frequencies for \eqref{eq:vecvectortype}: 
\begin{align}\label{eq:procaVqnm}
	i \omega_{nl}^{\textbf{V} } =  \Delta  + 2n + l  \qquad \text{and} \qquad  \bar\Delta   + 2n + l \; , \qquad n\geq 0 \; , \quad l\geq 1 \; . 
\end{align}

Next, we have two scalar sectors. The first one is spanned by 
\begin{align}\label{eq:vecStype-}
	\textbf{S1:} \quad P^{2n}\partial_U\cdot \partial_{P}Y_0^l (P)\ket{\Delta,U}_1 \quad \text{and} \quad P^{2n}\partial_U\cdot \partial_{P}Y_0^l (P)\ket{\bar\Delta,U}_1 \; , \quad n\geq 0 \, , \quad  l\geq 1 \; ,
\end{align}
which carries the representations $ \rho^d_{N+1}$, with frequencies 
\begin{align}\label{eq:procaS1qnm}
	i \omega_{nl}^{\textbf{S1}} =  \Delta-1  + 2n + l  \qquad \text{and} \qquad  \bar\Delta -1  + 2n + l \; , \qquad n\geq 0 \; , \quad l\geq 1 \; . 
\end{align}
In \eqref{eq:vecStype-}, the dot denotes contraction using the standard metric of $\mathbb{R}^d$. Note that since $\partial_{P^i}Y_0^{l=0} (P)=0$, the spectrum starts from $l=1$ instead of $l=0$. 

The other scalar sector is spanned by
\begin{align}\label{eq:vecStype+}
	\textbf{S0:} \qquad P^{2n} \Pi \,  Y_0^l (P) P \cdot \partial_U \ket{\Delta,U}_1 \quad \text{and} \quad P^{2n} \Pi \, Y_0^l (P) P \cdot \partial_U \ket{\bar\Delta,U}_1\; , \quad n\geq 0\, , \quad  l\geq 0 \; ,
\end{align}
carrying the representations $ \rho^d_{N-1}$, with frequencies
\begin{align}\label{eq:procaS0qnm}
	i \omega_{nl}^{\textbf{S0} } =  \Delta+1  + 2n + l  \qquad \text{and} \qquad  \bar\Delta +1  + 2n + l \; , \qquad n\geq 0 \; , \quad l\geq 0 \; . 
\end{align}
In \eqref{eq:vecStype+}, we have introduced the projector 
\begin{align}\label{eq:projector}
	\Pi \left(  c_{i_1, \cdots , i_N} P^{i_1} \cdots P^{i_N} \right) =c_{i_1, \cdots , i_N} \left(  P^{i_1} \cdots P^{i_N} - {\rm traces}\right) 
\end{align}
such that the polynomials in $P$ are traceless. For a simple example, 
\begin{align}
	 \Pi \,  Y^1 (P) P \cdot \partial_U \ket{\Delta,U}_1  = \Pi \,  E_i P^i P^j \ket{\Delta}_j 
	  =   \left( P^i P^j - \frac{1}{d} \delta^{ij} P^2\right) E_i \ket{\Delta}_j \; . 
\end{align}
We note that the QNM spectra \eqref{eq:procaVqnm}, \eqref{eq:procaS1qnm} and \eqref{eq:procaS0qnm} were obtained in \cite{Rahman:2024mok} by explicitly solving the Proca wave equation the static patch.

\paragraph{Harish-Chandra Character}

We now turn to the Harish-Chandra Character for a massive vector. From \eqref{eq:vecdesc}, it is clear that the Harish-Chandra character is equal to the scalar one times $d$, the number of polarizations of the vector, i.e. 
\begin{align}\label{eq:massspin1char}
	\chi_{[\Delta, 1]}(t) =d \frac{q^\Delta+q^{\bar{\Delta}} }{(1-q)^d}\; . 
\end{align}
In terms of the $\mathfrak{so}(d)$ irreps, it is straightforward to work out the contributions from each sector as well. The contribution from the $\alpha$-towers is given by
\begin{align}\label{vec char}
	\chi_\Delta^\mathbf{V}(t) &\equiv \sum_{n=0}^\infty \sum_{l=1}^\infty D_{l,1}^d q^{\Delta+2n+l}= q^\Delta \left[ d \frac{q }{(1-q)^d}-\frac{q^{-1}+q}{(1-q)^d}+\frac{q^{-1}}{1-q^2} \right] \nn\\
	\chi_\Delta^\mathbf{S1}(t)&\equiv \sum_{n=0}^\infty \sum_{l=1}^\infty D_l^d q^{ \Delta-1  + 2n + l } =q^\Delta \left[ \frac{q^{-1}}{(1-q)^d}-\frac{q^{-1}}{1-q^2}\right] \nn\\
	\chi_\Delta^\mathbf{S0}(t) &\equiv   \sum_{n=0}^\infty \sum_{l=0}^\infty D_l^d q^{ \Delta+1  + 2n + l }= \frac{q^{\Delta+1}}{(1-q)^d} \; ,
\end{align}
while that from the $\beta$-towers is given by replacing $\Delta\to\bar\Delta$. It is clear that the sum of these ``sub-characters" reproduces the massive spin 1 character \eqref{eq:massspin1char}:
\begin{align}\label{eq:vecchar}
	\chi_{[\Delta, 1]}(t) = \chi_\Delta^\mathbf{V}(t)+\chi_\Delta^\mathbf{S1}(t)+\chi_\Delta^\mathbf{S0}(t)+\chi_{\bar\Delta}^\mathbf{V}(t)+\chi_{\bar\Delta}^\mathbf{S1}(t)+\chi_{\bar\Delta}^\mathbf{S0}(t)\; .
\end{align}
Before moving on, we note that for $d\geq 4$, if the $l$-sums in \eqref{vec char} are extended to $l=-1$, we have
\begin{align}
	\bar\chi_\Delta^\mathbf{V}(t) \equiv &\sum_{n=0}^\infty \sum_{l=-1}^\infty D_{l,1}^d q^{\Delta+2n+l}= q^\Delta \left[ d \frac{q }{(1-q)^d}-\frac{q^{-1}+q}{(1-q)^d} \right] \nn\\
	\bar\chi_\Delta^\mathbf{S1}(t)\equiv &\sum_{n=0}^\infty \sum_{l=-1}^\infty D_l^d q^{ \Delta-1  + 2n + l } =\frac{q^{\Delta-1}}{(1-q)^d}\nn\\
	\bar\chi_\Delta^\mathbf{S0}(t) \equiv &  \sum_{n=0}^\infty \sum_{l=-1}^\infty D_l^d q^{ \Delta+1  + 2n + l }= \frac{q^{\Delta+1}}{(1-q)^d} \; .
\end{align}
For sectors $\mathbf{V}$ and $\mathbf{S1}$ this re-distributes the last term in \eqref{vec char}; specifically, we have
\begin{align}
	\chi_\Delta^\mathbf{V}(t) + \chi_\Delta^\mathbf{S1}(t) =  \bar\chi_\Delta^\mathbf{V}(t) + \bar\chi_\Delta^\mathbf{S1}(t) \; . 
\end{align}
For the $\mathbf{S0}$ sector, the extension has no effect since $D^d_{-1}=0$. For $d=3$, this observation still holds, provided the definitions of $D^3_{l,1}$ and $D^3_l$ are extended for $l<1$ and $l<0$, respectively:
\begin{align}\label{eq:d=3modify}
	D^3_{l,1}=
	\begin{cases}
		2l+1 & ,\quad l\geq 1 \\
		0 & ,\quad l=0 \\
		-1 & ,\quad  l=-1
	\end{cases} \;, \qquad 
	D^3_{l} = 
	\begin{cases}
		2l+1 &, \quad l\geq 0 \\
		0 & ,\quad l=-1 
	\end{cases} \;. 
\end{align}

\subsection{Massive spin-2}\label{sec:spin2char}

We start with the massive spin-2 primary states, labeled by the $\mathfrak{so}(1,1)$ weights $\Delta$ and $\bar\Delta$, related to the mass through $M^2=\left( \Delta-2\right) \left(  \bar\Delta -2\right) =\left( \Delta -2\right) \left( d-\Delta-2\right) $, and $\bar\Delta$ and two additional $\mathfrak{so}(d)$  vector indices:
\begin{align}\label{eq:massivespin2primary}
 \ket{\Delta,U}_2\equiv\ket{\Delta}_{ij} U^i U^j\qquad \text{and} \qquad  \ket{\bar\Delta,U}_2\equiv\ket{\bar\Delta}_{ij} U^i U^j
\end{align}
where, similar to the vector case, we encode the tensor structures using the vector $U^i$. To furnish an irreducible representation of $\mathfrak{so}(d)$, \eqref{eq:massivespin2primary} must be traceless:
\begin{align}
	\partial_U \cdot \partial_U \ket{\Delta,U}_2 = \ket{\Delta}\indices{_i^i}= 0 =  \ket{\bar\Delta}\indices{_i^i} = 	\partial_U \cdot \partial_U \ket{\bar\Delta,U}_2 \;. 
\end{align}

\paragraph{The $\mathfrak{so}(d)$ content}

We again have an $\alpha$- and a $\beta$-tower of QNMs, obtained by acting with $P_i$ on the two primary states, respectively. The tensor product decomposition,
\begin{align}\label{eq:spin2tensor}
	\rho^d_N \otimes \rho^d_2
	=
	\begin{cases}
		 \rho^d_{N,2}  \oplus  \rho^d_{N+1,1}  \oplus \rho^d_{N-1,1} \oplus \rho^d_{N+2}\oplus \rho^d_{N} \oplus \rho^d_{N-2}  \; &  , \qquad N \geq 2 \\
		 \qquad \qquad \rho^d_{2,1} \; \;  \oplus \qquad \qquad  \;\,\rho^d_3 \, \; \;  \oplus \, \rho^d_{1}   \; & , \qquad N = 1\\
		\qquad \qquad \qquad \qquad \qquad \quad \;\;\,  \rho^d_2 & , \qquad N = 0
	\end{cases}
\end{align}
implies that we have 6 sectors of  $\mathfrak{so}(d)$ irreps. Note that with the modification \eqref{eq:so3reps}, \eqref{eq:spin2tensor} holds uniformly for $d=3$ as well. We focus on the $\alpha$-tower; the $\beta$-tower is simply given by replacing $\Delta$ with $\bar \Delta$. We first have the tensor sector ($\rho^d_{N,2}  $)
\begin{align}\label{eq:spin2T}
	\mathbf{T:} \qquad P^{2n}Y_2^l (P,\partial_U)\ket{\Delta,U}_2 \; , \qquad n\geq 0\; , \quad l\geq 2\; .
\end{align}
Here, $Y_2^l$ represents a spin-2 symmetric transverse traceless (STT) spherical harmonic with the homogeneous polynomial representation:
\begin{align}
	Y_2^l (X,U)=E_{i_1 \cdots i_l, j_1 j_2}X^{i_1}\cdots X^{i_l} U^{j_1}U^{j_2} \; ,
\end{align}
where the tensor $E_{i_1 \cdots i_l, j_1 j_2}$ has a structure of a traceless Young diagram with $l$ boxes in the first row and 2 boxes in the second. The frequency spectrum associated with \eqref{eq:spin2T} is
\begin{align}
	i \omega_{nl}^{\mathbf{T}} =  \Delta  + 2n + l  \; , \qquad n\geq 0 \; , \quad l\geq 2 \; . 
\end{align}
Next, we have two vector sectors ($\rho^d_{N+1,1}$ and $\rho^d_{N-1,1}$)
\begin{alignat}{2}\label{eq:spin2V}
	&\mathbf{V2:} \qquad P^{2n} \partial_U\cdot \partial_{P}Y_1^l(P,\partial_U)\ket{\Delta,U}_2 
	\; , \qquad && n\geq 0 \; , \quad l\geq 2 \nn\\
	&\mathbf{V1:}	\qquad P^{2n}  \Pi \, Y_1^l (P,\partial_U)P \cdot \partial_U \ket{\Delta,U}_2 
	\; , \qquad && n\geq 0 \; , \quad l\geq 1 \;.
\end{alignat}
The $\mathbf{V2}$ spectrum starts from $l=2$ instead of $l=1$ because $\partial_{P^i}Y^1_1 (P,U)=0$. We recall the projector $\Pi$ defined in \eqref{eq:projector}. The frequency spectra are
\begin{alignat}{2}
	i \omega_{nl}^{\mathbf{V2}} & =  \Delta -1 + 2n + l  
	\; , \qquad  && n\geq 0 \; , \quad l\geq 2 \nn\\
	i \omega_{nl}^{\mathbf{V1}} & =  \Delta  +1 + 2n + l  
	\; , \qquad && n\geq 0 \; , \quad l\geq 1    \; . 
\end{alignat}
Finally, we have three scalar sectors ($\rho^d_{N+2}$, $\rho^d_{N}$ and $\rho^d_{N-2}$)
\begin{alignat}{2}\label{eq:spin2S}
	&\mathbf{S2:} \qquad P^{2n}(\partial_U\cdot \partial_{P})^2Y_0^l (P)\ket{\Delta,U}_2 \; , \qquad && n\geq 0\; , \quad l\geq 2 \nn\\ 
	&\mathbf{S1:}  \qquad P^{2n}  \Pi \,\partial_U\cdot \partial_{P}Y_0^l (P) P \cdot \partial_U \ket{\Delta,U}_2  \; , \qquad && n\geq 0\; , \quad l\geq 1 \nn\\
	&\mathbf{S0:}  \qquad P^{2n}  \Pi \, Y_0^l (P) (P \cdot \partial_U)^2 \ket{\Delta,U}_2 	\; , \qquad && n\geq 0 \; , \quad l\geq 0 \; .
\end{alignat}
Since $\partial_{P^i}\partial_{P^j}Y^1 (P)=\partial_{P^i}Y^0 (P)=0$, the $\mathbf{S2}$ spectrum starts from $l=2$ instead of $l=0$, while the $\mathbf{S1}$ spectrum starts from $l=1$ instead of $l=0$. The frequency spectra are
\begin{alignat}{2}
	 i \omega_{nl}^{\mathbf{S2}} & =  \Delta -2 + 2n + l   \; , \qquad  && n\geq 0 \; , \quad l\geq 2 \nn\\
	 i \omega_{nl}^{\mathbf{S1}} & =  \Delta  + 2n + l 	 \; , \qquad && n\geq 0 \; , \quad l\geq 1  \nn\\
	  i \omega_{nl}^{\mathbf{S0}} & =  \Delta  +2+ 2n + l    \; , \qquad && n\geq 0 \; , \quad l\geq 0  \; . 
\end{alignat}

\paragraph{Harish-Chandra Character}

As before, it is easy to check that the sum of contributions from all 6 sectors reproduces the massive spin-2 character
\begin{align}\label{eq:spin2masschar}
	\chi_{\left[ \Delta, 2\right] }(t) =D_2^d \frac{q^\Delta+q^{\bar{\Delta}} }{(1-q)^d}= \chi_\Delta^{\mathbf{T}}(t)+\chi_\Delta^{\mathbf{V2}}(t)+\chi_\Delta^{\mathbf{V1}}(t)+\chi_\Delta^\mathbf{S2}(t)+\chi_\Delta^\mathbf{S1}(t) +\chi_\Delta^\mathbf{S0}(t) +\left( \Delta \leftrightarrow \bar\Delta\right) 
\end{align}
with each sub-character defined in an analogous way as \eqref{vec char}. For $d=3$, since spin-2 STT harmonics do not exist on $S^2$ (as reflected in their degeneracies \eqref{eq:so3degen}), the entire tensor sector \eqref{eq:spin2T} is absent. The sub-characters $\chi_\Delta^{\mathbf{T}}(t)$ and $\chi_{\bar\Delta}^{\mathbf{T}}(t)$ becomes trivial, but \eqref{eq:spin2masschar} still holds.

We conclude this section with the following observation for $d\geq 4$. As we did at the end of the last section, we extend the $l$-sums in the characters of all sectors so that they start from $l=-1$. We denote the extended sub-characters with the notation $\bar\chi$. One can check that 
\begin{align}
	\chi_\Delta^{\mathbf{T}}(t)+\chi_\Delta^{\mathbf{V2}}(t)+\chi_\Delta^\mathbf{S2}(t)& = \bar\chi_\Delta^{\mathbf{T}}(t)+\bar\chi_\Delta^{\mathbf{V2}}(t)+\bar\chi_\Delta^\mathbf{S2}(t) \nn\\
	\chi_\Delta^{\mathbf{V1}}(t)+\chi_\Delta^\mathbf{S1}(t) &=\bar\chi_\Delta^{\mathbf{V1}}(t)+\bar\chi_\Delta^\mathbf{S1}(t)\nn\\
	 \chi_\Delta^\mathbf{S0}(t) &= \bar\chi_\Delta^\mathbf{S0}(t) \;. 
\end{align}


\subsection{Massive spin-$s$}\label{sec:HSQNM}

For general spin $s$, we start with two massive spin-$s$ primary states
\begin{align}
	\ket{\Delta,U}_s\equiv \ket{\Delta}_{i_1 \cdots i_s} U^{i_1}\cdots U^{i_s} \qquad \text{and} \qquad \ket{\bar\Delta,U}_s\equiv \ket{\bar\Delta}_{i_1 \cdots i_s} U^{i_1}\cdots U^{i_s} 
\end{align}
where the $\mathfrak{so}(1,1)$ weights $\Delta$ and $\bar\Delta$ are related to the mass through \eqref{eq:massdim}, and the rank-$s$ tensors are totally symmetric and traceless:
\begin{align}
	\partial_U \cdot \partial_U \ket{\Delta,U}_s = 0=\partial_U \cdot \partial_U \ket{\bar\Delta,U}_s \;. 
\end{align}

\paragraph{The $\mathfrak{so}(d)$ content}

Using the tensor product decomposition
\begin{align}
	\rho^d_N \otimes \rho^d_s &= \bigoplus_{p=0}^{\min(N,s)} \bigoplus_{m=0}^{p} \rho^d_{N+s-2p+m,m} \; ,
\end{align}
we identify sectors of $\mathfrak{so}(d)$ irreps. As before, with the modification \eqref{eq:so3reps}, this tensor decomposition holds uniformly for $d=3$ as well.  Each sector is constructed using spin-$m$ STT spherical harmonics, with $0\leq m\leq s$. For a given spin-$m$, there is an additional parameter $p$, satisfying $m \leq p \leq s$, which distinguishes the $s-m+1$ distinct towers. Explicitly, the $\alpha$-tower takes the form
\begin{align}\label{s p sector}
	(m,p)\text{-type}:\quad P^{2n}\Pi\, (\partial_U\cdot \partial_{P})^{p-m} Y_m^l (P,\partial_U) (P \cdot \partial_U)^{s-p} \ket{\Delta,U}_s \; , \quad n\geq 0\; , \quad  l\geq p \; .
\end{align}
Here, $Y_m^l (P,\partial_U)$ are the spin-$m$ STT spherical harmonics represented by the homogeneous polynomial \eqref{appeq:STSHpoly} in $\mathbb{R}^d$. We also recall the projector $\Pi$ defined in \eqref{eq:projector}. Since 
\begin{align}
	\partial_{P^{i_1}}\cdots \partial_{P^{i_{p-m}}} Y_m^l (P,U)=0 \; , \qquad m\leq l\leq p-1\; ,
\end{align}
the spectrum \eqref{s p sector} starts from $l=p$ instead of $l=m$. Together with the $\beta$-tower, we write down the QNM frequency spectra for the $(m,p)$-sector:
\begin{align}\label{eq:massspinqnm}
	i \omega_{nl}^{(m,p)} =  \Delta+s+m-2p + 2n + l  \qquad \text{and} \quad  \bar\Delta +s+m-2p  + 2n + l \; , \quad n\geq 0 \; , \quad l\geq p \; . 
\end{align}
This can be viewed as a non-trivial prediction for the QNM spectra, derived by solving the massive spin-$s$
s equations of motion in a $dS_{d+1}$ static patch. $m$ and $l$ label the spherical harmonics, while $n$ is an ``overtone" number.

\paragraph{Harish-Chandra Character}

The sector \eqref{s p sector} contributes the sub-character
\begin{align}\label{eq:mpchar}
	\chi_\Delta^{(m,p)}(t)\equiv \sum_{n=0}^\infty \sum_{l=p}^\infty D_{l,m}^d q^{\Delta+s+m-2p + 2n + l}\;. 
\end{align}
Motivated by the observations at the end of sections \ref{sec:spin1char} and \ref{sec:spin2char}, one naturally suspects that for $d\geq 4$, at each fixed $p$, the sum of \eqref{eq:mpchar} over $p\leq m \leq s$ should equal the sum of the extended sub-characters
\begin{align}\label{eq:mpcharexgtend}
	\bar\chi_\Delta^{(m,p)}(t)\equiv  \sum_{n=0}^\infty \sum_{l=-1}^\infty D_{l,m}^d q^{\Delta+s+m-2p + 2n + l}\;, 
\end{align}
that is
\begin{align}\label{eq:charrelation}
	\sum_{m=0}^p \chi_\Delta^{(m,p)}(t) = \sum_{m=0}^p \bar\chi_\Delta^{(m,p)}(t) \; . 
\end{align}
It turns out to be true. To show this, note that \eqref{eq:charrelation} holds as long as
\begin{align}
	\sum_{m=0}^p \sum_{l=-1}^{p-1}D^d_{l,m}q^{m+l}=0 \; , 
\end{align}
which can be proved by induction on $p$. One will need to use the relation 
\begin{align}
	D^d_{s,n} = - D^d_{n-1,s+1} \; . 
\end{align}
For completeness, one can derive an elegant formula for \eqref{eq:mpcharexgtend} and \eqref{eq:charrelation}:
\begin{gather}
	\bar\chi_\Delta^{(m,p)}(t)=q^{\Delta+ s+m-2p} \frac{D_m^d -D_{m-1}^d \left( q^{-1}+q\right) +D_{m-2}^d }{(1-q)^d} \; , \nn\\
	\sum_{m=0}^p \chi_\Delta^{(m,p)}(t) = \sum_{m=0}^p \bar\chi_\Delta^{(m,p)}(t)  =q^{\Delta+ s-p}   \frac{D^d_p - D^d_{p-1} q }{(1-q)^d}\; .
\end{gather}
Using these, one can verify that the sum of contributions from all sectors reproduces the full character
\begin{align}\label{local char sum}
	\chi_{[s,\Delta]}(t)= D^d_s \frac{q^\Delta +q^{\bar{\Delta}}}{(1-q)^d}=\sum_{p=0}^s \sum_{m=0}^{p} \chi_\Delta^{(m,p)}(t)+\left( \Delta \leftrightarrow \bar\Delta \right)  =\sum_{p=0}^s \sum_{m=0}^{p}\bar\chi_\Delta^{(m,p)}(t) +\left( \Delta \leftrightarrow \bar\Delta \right)  \;  .
\end{align}
For $d=3$, all sectors \eqref{s p sector} with $m\geq 2$ become trivial due to the non-existence of STT harmonics with spin $m\geq 2$. One can check that the contributions from the remaining non-trivial sectors sum up to the total character \eqref{local char sum} with $d=3$.


\section{Branching rule analysis for 1-loop $S^{d+1}$ path integrals}\label{ref:branching}

We now turn to our main problem: analyzing the $\mathfrak{so}(d)$ content of 1-loop $S^{d+1}$ path integrals. For notations, conventions, and relevant mathematical background on $\mathfrak{so}(d)$ and symmetric transverse traceless (STT) spherical harmonics, we refer the reader to Appendix \ref{sec:sod}.

\subsection{Massive scalar}\label{sec:Lspin0PI}

We warm up with a scalar with mass $M^2>0$ on $S^{d+1}$, whose path integral is
\begin{align}\label{eq:scalarPI}
	\log Z_\text{PI} = \int_0^\infty \frac{d\tau}{2\tau} e^{-\frac{\epsilon^2}{4\tau}} \Tr \, e^{-\left(-\nabla_0^2+M^2 \right)\tau } = \int_0^\infty \frac{d\tau}{2\tau} e^{-\frac{\epsilon^2}{4\tau}} \sum_{L=0}^\infty D^{d+2}_L \, e^{-\left(\lambda_L+M^2 \right)\tau } \; .
\end{align}
Here $e^{-\frac{\epsilon^2}{4\tau}}$ is a UV regulator, and we will set $\epsilon=0$ from now on. The trace Tr sums over the spectrum of the Laplacian $-\nabla_0^2$, which are nothing but spherical harmonics $f_L$, with eigenvalue $\lambda_L$, furnishing finite-dimensional $\mathfrak{so}(d+2)$ representations $\rho^{d+2}_L$.

\subsubsection{Branching rule analysis}

We would like to understand how the infinite sum \eqref{eq:scalarPI} organizes according to the $\mathfrak{u}(1)\oplus \mathfrak{so}(d)$ subalgebra. This motivates the introduction of a formal generating function
\begin{align}\label{scalar irrep gen fn}
	\bigoplus_{L=0}^\infty \rho^{d+2}_L q^L \; , 
\end{align}
with an auxiliary variable $q$. Next, we invoke the branching law of $\mathfrak{so}(d+2)$ irreps into $\mathfrak{u}(1)\oplus \mathfrak{so}(d)$ irreps, which  says \cite{branch} (see appendix \ref{branch})
\begin{align}\label{scalar irrep}
	\rho^{d+2}_L = \bigoplus_{l=0}^L \left( \bigoplus_{p=0}^{L-l}\rho_{L-l-2p}^2 \otimes  \rho_l^d\right).
\end{align}
Here $\rho_k^2$ denotes the  irreps of $\mathfrak{u}(1)\simeq \mathfrak{so}(2)$ labeled by $k\in \mathbb{Z}$, and $\rho_l^d$ the $\mathfrak{so}(d)$ irreps ($d\geq 3$) labeled by $l\geq 0$. We can think of $k$ as labeling the Matsubara frequencies associated with the dS temperature $\beta = 2\pi $ while $l$ the scalar spherical harmonics on $S^{d-1}$. A useful trick is to represent $\rho_k^2$ with an auxiliary variable $x$ raised to the $k$-th power:
\begin{align}\label{x rule}
	x^k \leftrightarrow \rho_{k}^2 \; .
\end{align}
With this notation, a direct sum of $\rho_k^2$ is equivalent to an ordinary sum in $x^k$, the tensor product $\otimes$ becomes a simple scalar multiplication, and \eqref{scalar irrep} reads
\begin{align}\label{scalar irrep x}
	\rho^{d+2}_L \to \bigoplus_{l=0}^L   \frac{x^{L-l+1}-x^{-L+l-1}}{x-x^{-1}} \rho_l^d \; .
\end{align}
Now, we manipulate the generating function \eqref{scalar irrep gen fn} as follows:\footnote{In performing these formal manipulations, one should keep in mind that we have a Taylor expansion in small $q$. On the other hand, understanding $x$ as $e^{i\theta}$ for some variable $\theta$, the expansion in $x$ is in fact a Fourier expansion in $\theta$.
	}
\begin{align}\label{eq:scalarmanipulate}
	\bigoplus_{L=0}^\infty \rho^{d+2}_L q^L\to\bigoplus_{L=0}^\infty \bigoplus_{l=0}^L   \frac{x^{L-l+1}-x^{-L+l-1}}{x-x^{-1}} \rho_l^d q^L
	=\bigoplus_{l=0}^\infty \frac{q^l}{(1-qx^{-1})(1-qx)} \rho_l^d \; .
\end{align}
In the first equality we substituted the decomposition \eqref{scalar irrep x}; in the second we exchanged the summations $\bigoplus_{L=0}^\infty \bigoplus_{l=0}^L = \bigoplus_{l=0}^\infty \bigoplus_{L=l}^\infty$ and performed the sum over $L$. To proceed, we expand
\begin{align}\label{expansion}
	\frac{1}{(1-qx^{-1})(1-qx)}= \sum_{n=0}^\infty\sum_{k=0}^\infty q^{k+n}x^{k-n}=\sum_{k=-\infty}^\infty x^k q^{|k|}\sum_{n=0}^\infty q^{2n}\; . 
\end{align}
Putting \eqref{expansion} back into \eqref{eq:scalarmanipulate} and using \eqref{x rule}, we finally have
\begin{align}\label{scalar decom}
	\bigoplus_{L=0}^\infty \rho^{d+2}_L q^L =\left(\bigoplus_{k=-\infty}^\infty \rho_{k}^2 q^{|k|}\right)\otimes  \left(\bigoplus_{l=0}^\infty \rho_l^d \sum_{n=0}^\infty  q^{2n+l}\right) \; .
\end{align}
In summary, the entire generating function is factorized into a $\mathfrak{u}(1)$ part and a $\mathfrak{so}(d)$ part.

\subsubsection{Character formula}

Now let us get back to the path integral \eqref{eq:scalarPI}. It has been shown in \cite{Anninos:2020hfj} that after a series of manipulations (reviewed in appendix \ref{sec:charintegral}), \eqref{eq:scalarPI} can be recast into 
\begin{align}\label{eq:scalarPIrewrite}
	\log Z_\text{PI}=\int_0^\infty \frac{dt}{2t} \left( q^\Delta +q^{\bar{\Delta}}\right)\sum_{L=0}^\infty D^{d+2}_{L} q^L \; ,\qquad q=e^{-t} \; . 
\end{align}
Since the infinite $L$-sum is same as the generating function \eqref{scalar irrep gen fn}, with $\rho^{d+2}_L $ replaced by their dimensions $D^{d+2}_{L} $, we can immediately use \eqref{scalar decom} to write
\begin{align}
	\log Z_\text{PI}=\int_0^\infty \frac{dt}{2t} \left(\sum_{k=-\infty}^\infty  q^{|k|}\right) \left[\sum_{n=0}^\infty \sum_{l=0}^\infty D_l^d  q^{2n+l}\left( q^\Delta +q^{\bar{\Delta}}\right)\right] \;.
\end{align}
The sum in the first bracket yields the factor capturing the bosonic statistics
\begin{align}
	\sum_{k=-\infty}^\infty  q^{|k|} =\frac{1+q}{1-q}
\end{align}
and the quantity in the second bracket is exactly the QNM character \eqref{eq:scalarHC}. In other words,
\begin{align}\label{eq:massscalarPI}
	\log Z_\text{PI}=\int_0^\infty \frac{dt}{2t}\frac{1+q}{1-q}\chi(t)\;.
\end{align}


\subsection{Massive vector}\label{sec:Lspin1PI}

We now study the case of a Proca field with mass $M^2=\left( \Delta-1\right) \left(  \bar\Delta -1\right) =\left( \Delta -1\right) \left( d-\Delta-1\right) >0$ on $S^{d+1}$, whose path integral is \cite{Anninos:2020hfj,Law:2020cpj}
\begin{align}\label{eq:procaPI}
	\log Z_\text{PI} = \int_0^\infty \frac{d\tau}{2\tau}\Tr_{-1}  e^{-\left(-\nabla_1^2+M^2 +d\right)\tau } = \int_0^\infty \frac{d\tau}{2\tau} \sum_{L=-1}^\infty D^{d+2}_{L,1}\, e^{-\left( \lambda_{L,1}+M^2+d\right) \tau}\; .
\end{align}
We have suppressed the UV regulator. The $L\geq 1$ part of the sum corresponds to transverse vector spherical harmonics furnishing the representations $\rho^{d+2}_{L,1}$. The subscript $-1$ in $\Tr_{-1}$ means that we extend the sum from $L\geq 1$ to $L\geq -1$, which originates from the path-integration over off-shell longitudinal modes of the local vector field \cite{Law:2020cpj}.\footnote{This is one of the reasons why we use the term `1-loop path integral' instead of `1-loop determinant', because the former generically contains more than the determinant of a Laplacian, due to off-shell or zero mode issues \cite{Law:2020cpj}.} Since $D^{d+2}_{0,1}=0$, the $L=0$ term is in fact trivial; the $L=-1$ term on the other hand contributes as
\begin{align}
	D^{d+2}_{-1,1}\, e^{-\left( \lambda_{-1,1}+M^2+d\right) \tau} = - e^{-M^2 \tau}\; .
\end{align}

\subsubsection{Branching rule analysis}\label{sec:procabranch}

As in the scalar case, we package all the irreps $\rho^{d+2}_{L,1}$ with $L\geq 1$ into a formal generating function 
\begin{align}\label{eq:procagenfn}
	\bigoplus_{L=1}^\infty \rho^{d+2}_{L,1} q^L \; . 
\end{align}
To eventually connect with \eqref{eq:procaPI}, we will need to incorporate the extension to $L=-1$, on which we will comment later. The relevant $\mathfrak{so}(d+2)\to \mathfrak{u}(1)\oplus \mathfrak{so}(d)$  branching law for this case says \cite{branch}
\begin{align}\label{v irrep x}
	\rho^{d+2}_{L,1} \to  &\left( \frac{x^{L}-x^{-L}}{x-x^{-1}} \rho_0^d \right) \oplus \left( \bigoplus_{l=1}^L\bigoplus_{m=0}^1  \frac{x^{L-l+1}-x^{-L+l-1}}{x-x^{-1}}\frac{x^{2-m}-x^{m-2}}{x-x^{-1}} \rho_{l,m}^d \right) \; , \quad L\geq 1 \; .
\end{align}
Now, we manipulate \eqref{eq:procagenfn} as follows: 
\begin{align}\label{eq:procagenman}
	\bigoplus_{L=1}^\infty \rho^{d+2}_{L,1} q^L 
	\to &\, \bigoplus_{L=1}^\infty q^L \left[\left( \frac{x^{L}-x^{-L}}{x-x^{-1}} \rho_0^d \right) \oplus \left( \bigoplus_{l=1}^L\bigoplus_{m=0}^1  \frac{x^{L-l+1}-x^{-L+l-1}}{x-x^{-1}}\frac{x^{2-m}-x^{m-2}}{x-x^{-1}} \rho_{l,m}^d \right)\right]\nn\\
	=&\, \frac{1}{(1-qx^{-1})(1-qx)} \left[ q\rho_0^d\oplus \bigoplus_{l=1}^\infty (x+x^{-1})q^l  \rho_{l}^d\oplus \bigoplus_{l=1}^\infty q^l \rho_{l,1}^d \right] \; .
\end{align}
In the second equality we have switched the sums over $L$ and $l$ and performed the sum over $L$. To proceed, we use the following relation
\begin{align}\label{relation}
	\frac{x^p+x^{-p}}{\left( 1-qx^{-1}\right) \left( 1-qx\right) }=\frac{q^p+q^{-p}}{\left( 1-qx^{-1}\right) \left( 1-qx\right)}-q^{-p}\left(\frac{1-(qx)^p}{1-qx}\right)\left(\frac{1-(qx^{-1})^p}{1-qx^{-1}}\right)
\end{align}
where $p$ is a non-zero integer. The second term on the right-hand side expands into a finite number of terms in powers of $qx^{\pm 1}$. Using this with $p=1$, we can simplify
\begin{align}\label{eq:procagensim}
	\bigoplus_{L=1}^\infty \rho^{d+2}_{L,1} q^L
	= \frac{1}{(1-q x^{-1})(1-qx)} \left(\bigoplus_{l=1}^\infty q^l \rho_{l,1}^d \oplus \bigoplus_{l=0}^\infty q^{l+1} \rho_l^d \oplus \bigoplus_{l=1}^\infty q^{l-1} \rho_l^d \right) \ominus \bigoplus_{l=1}^\infty q^{l-1} \rho_l^d
\end{align}
where the formal minus $\ominus$ means $Y \ominus Y =\emptyset$. Note that the $l=0$ term in the second sum in the bracket corresponds to the first term in \eqref{eq:procagenman}.

\paragraph{Longitudinal mode extension}

Let us get back to the question of extending to $L=-1$. It is accomplished by formally defining
\begin{align}\label{neg rep}
	\rho^{d+2}_{n,s} \equiv \ominus \rho^{d+2}_{s-1,n+1}\; , \qquad n<s \; . 
\end{align}
As we will see, this works for spin $s\geq 2$ as well. The definition \eqref{neg rep} implies
\begin{align}
	\rho^{d+2}_{s-1,s}=\emptyset \; .
\end{align}
In particular, if we extend \eqref{eq:procagenfn} to $L=-1$, i.e. 
\begin{align}\label{eq:procagenfnextend}
	\bigoplus_{L=-1}^\infty \rho^{d+2}_{L,1} q^L \; ,
\end{align}
the $L=0$ term is trivial, while the $L=-1$ term contributes as
\begin{align}
	\rho^{d+2}_{-1,1} q^{-1}= \ominus \rho^{d+2}_{0,0}q^{-1} =\ominus \rho_0^2 \otimes \rho_0^d   q^{-1} \; . 
\end{align}
This term naturally combines with the last term of \eqref{eq:procagensim} as an extension to $l=0$. Finally, using the expansion \eqref{expansion}, we arrive at the bulk-edge split
\begin{align}\label{full vec dec}
	\bigoplus_{L=-1}^\infty \rho^{d+2}_{L,1} q^L = V_{\rm bulk} \ominus V_{\rm edge}\;.
\end{align}
Here the bulk sector
\begin{align}\label{eq:bulksector}
	V_{\rm bulk} = \left( \bigoplus_{k=-\infty}^\infty \rho_{k}^2 q^{|k|}\right) \otimes\left[\left(   \bigoplus_{l=1}^\infty \rho_{l,1}^d \sum_{n=0}^\infty  q^{2n+l}\right)\oplus\left(   \bigoplus_{l=1}^\infty \rho_{l}^d \sum_{n=0}^\infty  q^{2n+l-1}\right)\oplus\left(   \bigoplus_{l=0}^\infty \rho_{l}^d \sum_{n=0}^\infty  q^{2n+l+1}\right)\right] 
\end{align}
factorizes into a $\mathfrak{u}(1)$ and a $\mathfrak{so}(d)$ part as in the scalar case. The new ``edge" sector
\begin{align}\label{eq:vecedgesector}
	V_{\rm edge} =  \rho_0^2 \otimes \bigoplus_{l=0}^\infty \rho_l^d   q^{l-1} \; ,
\end{align}
has $\mathfrak{u}(1)$-weight $k=0$.

\subsubsection{Character formula}

Similar to the scalar case, the Proca path integral \eqref{eq:procaPI} can be recast into \cite{Anninos:2020hfj}
\begin{align}\label{eq:spin1SH}
	\log Z_\text{PI}=\int_0^\infty \frac{dt}{2t} \left( q^\Delta +q^{\bar{\Delta}}\right)\sum_{L=-1}^\infty D^{d+2}_{L,1} q^L \; ,\qquad q=e^{-t} \; . 
\end{align}
Using \eqref{full vec dec}, we can rewrite the $L$-sum, leading to a split of the path integral:
\begin{align}\label{vec PI}
	\log Z_\text{PI}=\log Z_\text{bulk} +\log Z_\text{edge} \; . 
\end{align}
Here the quasicanonical bulk partition function 
\begin{align}\label{eq:vecZbulk}
	\log Z_\text{bulk}=\int_0^\infty \frac{dt}{2t}\frac{1+q}{1-q} \chi(t) 
\end{align}
descends from \eqref{eq:bulksector}, with the quantity 
\begin{align}\label{eq:PIbulkcharproca}
	\chi(t)= \left( q^\Delta +q^{\bar{\Delta}}\right) \left(\sum_{n=0}^\infty \sum_{l=1}^\infty D_{l,1}^d q^{2n+l}+\sum_{n=0}^\infty \sum_{l=1}^\infty D_{l}^d q^{2n+l-1}+\sum_{n=0}^\infty\sum_{l=0}^\infty D_{l}^d q^{2n+l+1}\right)
\end{align}
capturing the sum over QNMs analyzed in section \ref{sec:spin1char}. Specifically, the sums in the bracket correspond to the $\mathbf{V}$, $\mathbf{S1}$ and $\mathbf{S0}$ sectors respectively. Thus, \eqref{eq:PIbulkcharproca} is precisely the Harish-Chandra character \eqref{eq:massspin1char} for a massive vector. On the other hand, \eqref{eq:vecedgesector} leads to 
\begin{align}\label{eq:vecZedge}
	\log Z_\text{edge}= -\int_0^\infty \frac{dt}{2t}\left( q^\Delta +q^{\bar{\Delta}}\right)\sum_{l=0}^\infty D^{d}_{l} q^{l-1} \;. 
\end{align}
We observe that \eqref{eq:vecZedge} is same as \eqref{eq:scalarPIrewrite} with $d\to d-2$, which implies that $Z_\text{edge}$ is the partition function for a ghost scalar with mass $M^2$ on $S^{d-1}$.


\subsection{Massive spin-2}

Our next case is a spin-2 field with mass $M^2=\left( \Delta-2\right) \left(  \bar\Delta -2\right) =\left( \Delta -2\right) \left( d-\Delta-2\right) >0$ on $S^{d+1}$, whose path integral is \cite{Law:2020cpj}
\begin{align}\label{eq:spin2PI}
	\log Z_\text{PI} = \int_0^\infty \frac{d\tau}{2\tau}  \Tr_{-1}  e^{-\left(-\nabla_2^2+M^2 +2\right)\tau } = \int_0^\infty \frac{d\tau}{2\tau} \sum_{L=-1}^\infty D^{d+2}_{L,2}\, e^{-\left( \lambda_{L,2}+M^2+2\right) \tau}\; .
\end{align}
The UV regulator is again suppressed. The $L\geq 2$ part of the sum captures the spin-2 STT spherical harmonics on $S^{d+1}$, which carry the representations $\rho^{d+2}_{L,2}$. Similar to the vector case, integrating over the off-shell longitudinal modes of the local spin-2 field introduces additional contributions, effectively extending the sum from $L\geq 2$ to $L\geq -1$ \cite{Law:2020cpj}.

\subsubsection{Branching rule analysis}\label{sec:spin2branch}

This time the relevant $\mathfrak{so}(d+2)\to \mathfrak{u}(1)\oplus \mathfrak{so}(d)$ decompositions are
\begin{align}
	\rho^{d+2}_{L,2} \to  &\, \left(\bigoplus_{l=0}^1 \bigoplus_{m=0}^l \left( \frac{x^{L-1}-x^{1-L}}{x-x^{-1}}\frac{x^{l-m+1}-x^{-l+m-1}}{x-x^{-1}} \rho_{l,m}^d \right)\right) \nn\\
	&\oplus \left( \bigoplus_{l=2}^L\bigoplus_{m=0}^2  \frac{x^{L-l+1}-x^{-L+l-1}}{x-x^{-1}}\frac{x^{3-m}-x^{m-3}}{x-x^{-1}} \rho_{l,m}^d \right).
\end{align}
Packaging these into a generating function and manipulating as before, we obtain
\begin{align}
	\bigoplus_{L=2}^\infty \rho^{d+2}_{L,2} q^L =&\,  \frac{q^2}{(1-q x^{-1})(1-qx)}\left( \rho_{0}^d\oplus \left(x+x^{-1} \right)\rho_{1}^d \oplus \rho_{1,1}^d\right)\nn\\
	&\oplus \bigoplus_{l=2}^\infty\frac{q^l}{(1-q x^{-1})(1-qx)}\left( \left(x^2+1+x^{-2} \right)\rho_{l}^d\oplus \left(x+x^{-1} \right)\rho_{l,1}^d \oplus \rho_{l,2}^d\right) \; . 
\end{align}
Next, we rewrite this using the relation \eqref{relation}. After some calculations, one finds
\begin{align}\label{eq:spin2branchinter}
	&\bigoplus_{L=2}^\infty \rho^{d+2}_{L,2} q^L \nn\\
	=&\,  \frac{1}{(1-q x^{-1})(1-qx)}\left[\bigoplus_{l=2}^\infty q^l\left(\rho_{l,2}^d\oplus q^{-1}\rho_{l,1}^d\oplus q^{-2}\rho_{l}^d \right)\oplus \bigoplus_{l=1}^\infty q^{l+1}\left(\rho_{l,1}^d\oplus q^{-1}\rho_{l}^d \right)\oplus \bigoplus_{l=0}^\infty q^{l+2}\rho_{l}^d \right]\nn\\
	&\ominus \left(1+(x+x^{-1})q \right)\bigoplus_{l=2}^\infty q^{l-2}\rho_{l}^d \ominus \bigoplus_{l=1}^\infty q^l \rho_l^d \ominus \bigoplus_{l=2}^\infty q^{l-1} \rho_{l,1}^d \; .
\end{align}
Including the extension to $L=-1$ with the formal definition \eqref{neg rep}, we have two extra terms
\begin{align}
	\rho^{d+2}_{0,2} q^0 =&\,  \ominus \rho^{d+2}_{1,1}=\ominus \left(\rho_0^2 \otimes \rho_0^d\right) \ominus \left(\left(\rho_1^2\oplus\rho_{-1}^2\right) \otimes \rho_{1}^d \right)\ominus  \left(\rho_0^2 \otimes \rho_{1,1}^d \right)\nn\\
	\rho^{d+2}_{-1,2} q^{-1}  =&\,  \ominus \rho^{d+2}_{1} q^{-1}=\ominus \left(\left(\rho_1^2\oplus\rho_{-1}^2\right)  \otimes \rho_0^d q^{-1}\right) \ominus \left(\rho_0^2\otimes \rho_{1}^d q^{-1}\right)\; ,
\end{align}
which are naturally combined into the second line of \eqref{eq:spin2branchinter}. 

Finally, using the expansion \eqref{expansion}, we have 
\begin{align}\label{full spin2 dec}
	\bigoplus_{L=-1}^\infty \rho^{d+2}_{L,2} q^L=&\,  V_\text{bulk} \ominus V_\text{edge} \;, 
\end{align}
where the bulk sector 
\begin{align}\label{eq:2Vbulk}
	V_\text{bulk} =&\, V_\text{bulk}^{T} \oplus V_\text{bulk}^{V} \oplus V_\text{bulk}^S\nn\\
	V_\text{bulk}^{T}=&\, \bigoplus_{k=-\infty}^\infty \rho_{k}^2q^{|k|}\otimes \left(\bigoplus_{l=2}^\infty \rho_{l,2}^d \sum_{n=0}^\infty q^{2n+l}\right)\nn\\
	V_\text{bulk}^V =&\, \bigoplus_{k=-\infty}^\infty \rho_{k}^2q^{|k|}\otimes \left[\left(\bigoplus_{l=1}^\infty\rho_{l,1}^d\sum_{n=0}^\infty q^{2n+l+1}\right)\oplus\left( \bigoplus_{l=2}^\infty\rho_{l,1}^d\sum_{n=0}^\infty q^{2n+l-1}\right) \right]\nn\\
	V_\text{bulk}^S =&\, \bigoplus_{k=-\infty}^\infty \rho_{k}^2q^{|k|}\otimes \left[\left(\bigoplus_{l=0}^\infty\rho_{l}^d\sum_{n=0}^\infty q^{2n+l+2}\right)\oplus\left( \bigoplus_{l=1}^\infty\rho_{l}^d\sum_{n=0}^\infty q^{2n+l}\right)\oplus\left( \bigoplus_{l=2}^\infty\rho_{l}^d\sum_{n=0}^\infty q^{2n+l-2}\right) \right]
\end{align}
factorizes into a $\mathfrak{u}(1)$ and a $\mathfrak{so}(d)$ part as before. The edge sector 
\begin{align}\label{eq:2Vedge}
	V_\text{edge} =&\,  V_\text{edge}^{V} \oplus V_\text{edge}^{S}\nn\\
	V_\text{edge}^{V}=&\, \rho_{0}^2q^0\otimes\left( \bigoplus_{l=1}^\infty\rho_{l,1}^d q^{l-1}\right)\nn\\
	V_\text{edge}^{S}=&\, \rho_{0}^2q^0\otimes \left[\left(\bigoplus_{l=0}^\infty\rho_{l}^d q^{l}\right)\oplus\left( \bigoplus_{l=1}^\infty\rho_{l}^d q^{l-2}\right)\right]\oplus (\rho_{1}^2+\rho_{-1}^2)q^1\otimes\left( \bigoplus_{l=0}^\infty\rho_{l}^d q^{l-2}\right) 
\end{align}
consists of modules with $\mathfrak{u}(1)$-weights $k=0,1$.

\subsubsection{Character formula}

For a massive graviton we will have
\begin{align}\label{eq:spin2SH}
	\log Z_\text{PI}=\int_0^\infty \frac{dt}{2t} \left( q^\Delta +q^{\bar{\Delta}}\right)\sum_{L=-1}^\infty D^{d+2}_{L,2} q^L\; ,\qquad q=e^{-t}\; .
\end{align}
Using \eqref{full spin2 dec}, we again split the path integral into a bulk and edge part:
\begin{align}\label{spin2 PI}
	\log Z_\text{PI}=\log Z_\text{bulk}+\log Z_\text{edge} \; .
\end{align}
The bulk part descends from \eqref{eq:2Vbulk}, leading to the bulk partition function
\begin{align}\label{eq:spin2Zbulk}
	\log Z_\text{bulk}	=\int_0^\infty \frac{dt}{2t}\frac{1+q}{1-q}  \chi(t)  
\end{align}
with the QNM character
\begin{align}
		\chi(t)& =\left( q^\Delta +q^{\bar{\Delta}}\right)\bigg(\sum_{n=0}^\infty\sum_{l=2}^\infty D_{l,2}^d q^{2n+l}+\sum_{n=0}^\infty\sum_{l=2}^\infty D_{l,1}^d q^{2n+l-1}+\sum_{n=0}^\infty\sum_{l=1}^\infty D_{l,1}^d q^{2n+l+1}\nn\\
	&\qquad +\sum_{n=0}^\infty\sum_{l=2}^\infty D_{l}^d q^{2n+l-2}+\sum_{n=0}^\infty\sum_{l=1}^\infty D_{l}^d q^{2n+l} +\sum_{n=0}^\infty\sum_{l=0}^\infty D_{l}^d q^{2n+l+2}\bigg)
\end{align}
expressed as a sum of sub-characters as in \eqref{eq:spin2masschar}. The edge sector \eqref{eq:2Vedge} on the other hand leads to the edge partition function 
\begin{align}\label{eq:spin2Zedge}
	\log Z_\text{edge}&=-\int_0^\infty \frac{dt}{2t}  \left( q^\Delta +q^{\bar{\Delta}}\right) \left( \sum_{l=1}^\infty D_{l,1}^d q^{l-1} +\sum_{l=0}^\infty D_{l}^d q^{l}+\sum_{l=1}^\infty D_{l}^d q^{l-2}+2q\sum_{l=0}^\infty D_{l}^d q^{l-2}\right)\nn\\
	&=-\int_0^\infty \frac{dt}{2t}  \left( q^\Delta +q^{\bar{\Delta}}\right) \left( \sum_{l=-1}^\infty D_{l,1}^d q^{l-1} +\sum_{l=-1}^\infty D_{l}^d q^{l}+\sum_{l=-1}^\infty D_{l}^d q^{l-2}+2q\sum_{l=-1}^\infty D_{l}^d q^{l-2}\right)
\end{align}
where the factor of 2 in the last term comes from the fact that $D_1^2=D_{-1}^2=1$. In the second line we have extended all the $l$-sums to $l=-1$, which introduces non-trivial terms only in the first and third sum.\footnote{This works uniformly for $d=3$ with the extended definition \eqref{eq:d=3modify} of the degeneracies.} Evaluating all these sums, we recover the edge partition function found in \cite{Anninos:2020hfj}:
\begin{align}\label{eq:ADLSZedge}
	\log Z_\text{edge}
	=-\int_0^\infty \frac{dt}{2t}   \frac{1+q}{1-q} \left(D_1^{d+2} \frac{q^{\Delta-1} +q^{\bar{\Delta}-1}}{(1-q)^{d-2}}\right) \; , 
\end{align}
We stress that, without our analysis, there is no clear guiding principle to distinguish the $\mathfrak{so}(d)$  contributions from \eqref{eq:ADLSZedge}. For instance, \eqref{eq:ADLSZedge} could arise from $D_1^{d+2}$ scalars on $S^{d-1}$. In contrast, \eqref{eq:spin2Zedge} allows us to conclude that $Z_\text{edge}$ receives contributions from one vector and four scalars on $S^{d-1}$. To underscore the non-triviality of this point, we reverse the steps in appendix \ref{sec:heatkernel} that lead to \eqref{eq:charPIstart}, rewriting \eqref{eq:spin2Zedge} explicitly in terms of determinants on $S^{d-1}$:
\begin{align}
	Z_\text{edge} & = \det\nolimits_{-1}\left(-\nabla_1^2+(\Delta-1)(\bar \Delta-1)+1  \right)^{\frac12}  \det \left(-\nabla_0^2+(\Delta-1)(\bar \Delta-1) \right)\nn\\
	& \qquad \qquad \times \det \left(-\nabla_0^2+\Delta (\bar \Delta-2) \right)^{\frac12}\det \left(-\nabla_0^2+\bar\Delta (\Delta-2) \right)^{\frac12}\;. 
\end{align}
Interestingly, the mass terms in these determinants are not necessarily real and positive. For instance, in the last two determinants, the mass terms become complex for the principal series \eqref{eq:principaldelta}. We leave the interpretation of these complex masses for future work.


\subsection{Massive spin-$s$}\label{sec:HSPI}

The $S^{d+1}$  path integral for a massive field with general spin $s\geq 1$ is \cite{Anninos:2020hfj,Law:2020cpj}
\begin{align}\label{eq:spinsPI}
	\log Z_\text{PI} = \int_0^\infty \frac{d\tau}{2\tau} \Tr_{-1}  e^{-\left(-\nabla_s^2+M^2 + M_s^2\right)\tau } = \int_0^\infty \frac{d\tau}{2\tau} \sum_{L=-1}^\infty D^{d+2}_{L,s}\, e^{-\left( \lambda_{L,2}+M^2+M_s^2\right) \tau}\; ,
\end{align}
where $M_s^2 = s-(s-2)(s+d-2)$. We recall the relation \eqref{eq:principaldelta} between the mass and the corresponding $\mathfrak{so}(1,1)$ weight. The trace runs over the spectrum of the STT spin-$s$ Laplacian $-\nabla_s^2$ on $S^{d+1}$, whose eigenfunctions furnish the representations $\rho^{d+2}_{L,s}$. To account for longitudinal mode contributions, we extend the sum from $L\geq s$ to $L\geq -1$ in \eqref{eq:spinsPI}. In the following, we focus on $d\geq 4$, though the $d=3$ case can be treated analogously with minor modifications at specific steps.

\subsubsection{Branching rule analysis}

As in the lower spin case, we introduce the generating function
\begin{align}\label{eq:spinsgenfn}
	\bigoplus_{L=-1}^\infty \rho^{d+2}_{L,s} q^L \; ,
\end{align}
where we have also included the extension $-1\leq L\leq s-1$ defined through \eqref{neg rep}. 

The relevant $\mathfrak{so}(d+2)\to \mathfrak{u}(1)\times \mathfrak{so}(d)$  branching law for this case says \cite{branch}
\begin{align}
	\rho^{d+2}_{L,s} \to  &\left(\bigoplus_{l=0}^{s-1} \bigoplus_{m=0}^l \left( \frac{x^{L-s+1}-x^{-L+s-1}}{x-x^{-1}}\frac{x^{l-m+1}-x^{-l+m-1}}{x-x^{-1}} \rho_{l,m}^d \right)\right) \nn\\
	&\oplus \left( \bigoplus_{l=s}^L\bigoplus_{m=0}^s  \frac{x^{L-l+1}-x^{-L+l-1}}{x-x^{-1}}\frac{x^{s-m+1}-x^{-s+m-1}}{x-x^{-1}} \rho_{l,m}^d \right)
\end{align}
for $ L\geq s$, and 
\begin{align}
	\rho^{d+2}_{L,s} \equiv \ominus\rho^{d+2}_{s-1,L+1}\to  &\ominus\left(\bigoplus_{l=0}^{L} \bigoplus_{m=0}^l \left( \frac{x^{s-1-L}-x^{-s+1+L}}{x-x^{-1}}\frac{x^{l-m+1}-x^{-l+m-1}}{x-x^{-1}} \rho_{l,m}^d \right)\right) \nn\\
	&\ominus \left( \bigoplus_{l=L+1}^{s-1}\bigoplus_{m=0}^{L+1}  \frac{x^{s-l}-x^{-s+l}}{x-x^{-1}}\frac{x^{L-m+2}-x^{-L+m-2}}{x-x^{-1}} \rho_{l,m}^d \right)
\end{align}
for $-1\leq L\leq s-2$. Substituting these into \eqref{eq:spinsgenfn}, we have
\begin{align}\label{eq:spinstrans}
	\bigoplus_{L=s}^\infty \rho^{d+2}_{L,s} q^L =&\, \bigoplus_{l=0}^{s-1} \bigoplus_{m=0}^l \frac{x^{l-m+1}-x^{-l+m-1}}{x-x^{-1}}   \frac{q^s }{(1-q x^{-1})(1-qx)}\rho_{l,m}^d \nn\\
	&\oplus \bigoplus_{l=s}^\infty \bigoplus_{m=0}^s \frac{x^{s-m+1}-x^{-s+m-1}}{x-x^{-1}}  \frac{q^l}{(1-q x^{-1})(1-qx)} \rho_{l,m}^d \; , 
\end{align}
and
\begin{align}\label{eq:spinslong}
	\bigoplus_{L=-1}^{s-2} \rho^{d+2}_{L,s} q^L =&\, \ominus \bigoplus_{L=0}^{s-2}q^{L}\left(\bigoplus_{l=0}^{L} \bigoplus_{m=0}^l  \left( \frac{x^{s-1-L}-x^{-s+1+L}}{x-x^{-1}}\frac{x^{l-m+1}-x^{-l+m-1}}{x-x^{-1}} \rho_{l,m}^d \right)\right) \nn\\
	&\ominus \bigoplus_{L=0}^{s-1}q^{L-1} \left( \bigoplus_{l=L}^{s-1}\bigoplus_{m=0}^{L}  \frac{x^{s-l}-x^{-s+l}}{x-x^{-1}}\frac{x^{L-m+1}-x^{-L+m-1}}{x-x^{-1}} \rho_{l,m}^d \right)\; .
\end{align}
In \eqref{eq:spinslong}, the sum over $L$ on the first line can be done after exchanging the sums over $L$ and $l$. To do the $L$-sum on the second line, we exchange twice
\begin{align}
	\bigoplus_{L=0}^{s-1}\bigoplus_{l=L}^{s-1}\bigoplus_{m=0}^{L}=\bigoplus_{l=0}^{s-1}\bigoplus_{L=0}^{l}\bigoplus_{m=0}^{L}=\bigoplus_{l=0}^{s-1}\bigoplus_{m=0}^{l}\bigoplus_{L=m}^{l} \; .
\end{align}
Performing these sums, one finds terms that are naturally combined together, yielding 
\begin{align}\label{eq:spinlonginter}
	\bigoplus_{L=-1}^{s-2} \rho^{d+2}_{L,s} q^L =&\, \ominus\bigoplus_{l=0}^{s-1} \bigoplus_{m=0}^l \frac{x^{l-m+1}-x^{-l+m-1}}{x-x^{-1}}  \frac{q^s }{(1-q x^{-1})(1-qx)}\rho_{l,m}^d\nn\\
	&\ominus \bigoplus_{l=0}^{s-1} \bigoplus_{m=0}^l \frac{x^{s-l}-x^{-s+l}}{x-x^{-1}}  \frac{q^{m-1} }{(1-q x^{-1})(1-qx)}\rho_{l,m}^d\nn\\
	&\oplus\bigoplus_{l=-1}^{s-1} \bigoplus_{m=0}^l \frac{x^{s-m+1}-x^{-s+m-1}}{x-x^{-1}}  \frac{q^l}{(1-q x^{-1})(1-qx)} \rho_{l,m}^d \; .
\end{align}
To further simplify this, we note that upon exchanging the labels $m$ and $l$ and using \eqref{neg rep}, the second line is naturally combined with the third line, so that
\begin{align}\label{eq:spinslongcon}
	\bigoplus_{L=-1}^{s-2} \rho^{d+2}_{L,s} q^L =&\, \ominus \bigoplus_{l=0}^{s-1} \bigoplus_{m=0}^l \frac{x^{l-m+1}-x^{-l+m-1}}{x-x^{-1}}  \frac{q^s }{(1-q x^{-1})(1-qx)}\rho_{l,m}^d\nn\\
	&\oplus\bigoplus_{l=-1}^{s-1} \bigoplus_{m=0}^s \frac{x^{s-m+1}-x^{-s+m-1}}{x-x^{-1}}  \frac{q^l}{(1-q x^{-1})(1-qx)} \rho_{l,m}^d \; .
\end{align}
Combining this with \eqref{eq:spinstrans}, we arrive at a very simple expression
\begin{align}\label{HS decom}
	\boxed{
		\bigoplus_{L=-1}^\infty \rho^{d+2}_{L,s} q^L= \bigoplus_{m=0}^s\bigoplus_{l=-1}^\infty  \frac{x^{s-m+1}-x^{-s+m-1}}{x-x^{-1}}  \frac{q^l}{(1-q x^{-1})(1-qx)} \rho_{l,m}^d \; .}
\end{align}
It is then straightforward to identify the bulk and edge sectors:
\begin{align}\label{eq:HSVsplit}
	\bigoplus_{L=-1}^{\infty} \rho^{d+2}_{L,s} q^L =&\,  V_\text{bulk} \ominus V_\text{edge}\; .
\end{align}

\paragraph{The bulk sector}

The bulk sector consists of the following
\begin{align}
	V_\text{bulk} = \bigoplus_{m=0}^s\bigoplus_{l=-1}^\infty  \frac{q^{s-m+1}-q^{-s+m-1}}{q-q^{-1}}  \frac{q^l}{(1-q x^{-1})(1-qx)} \rho_{l,m}^d\; .
\end{align}
Using \eqref{expansion} and expanding
\begin{align}
	\frac{q^{s-m+1}-q^{-s+m-1}}{q-q^{-1}} =\sum_{p=m}^{s} q^{s+m-2p} \; ,
\end{align}
we have the nice factorization
\begin{align}\label{eq:HSVbulk}
	V_\text{bulk} = \left(\bigoplus_{k=-\infty}^\infty \rho_{k}^2 q^{|k|}\right)\otimes \left(\bigoplus_{p=0}^s\bigoplus_{m=0}^p\bigoplus_{l=-1}^\infty  \rho_{l,m}^d  \sum_{n=0}^\infty q^{s+m-2p + 2n + l}\right) \; . 
\end{align}
Here we have exchanged the $m$ and $p$ sums.

\paragraph{The edge sector}

The edge sector consists of the following
\begin{align}
	V_\text{edge}=\bigoplus_{m=0}^{s-1}\bigoplus_{l=-1}^\infty  \left(\frac{q^{s-m+1}-q^{-s+m-1}}{q-q^{-1}}  - \frac{x^{s-m+1}-x^{-s+m-1}}{x-x^{-1}} \right)\frac{q^l}{(1-q x^{-1})(1-qx)} \rho_{l,m}^d.
\end{align}
Note that there is no term with $m=s$. Using the formula
\begin{align}\label{edge exp}
	&\left(\frac{q^{N+1}-q^{-N-1}}{q-q^{-1}}  - \frac{x^{N+1}-x^{-N-1}}{x-x^{-1}} \right)\frac{1}{(1-qx^{-1})(1-qx)} \nn\\ 
	=&\,  q^{-1}\sum_{k=-(N-1)}^{N-1} x^{k} \sum_{p=0}^{\lfloor \frac{N-1-|k|}{2}\rfloor}  \frac{q^{N-|k|-2p}-q^{-N+|k|+2p}}{q-q^{-1}} \; , 
\end{align}
we can write
\begin{align}\label{eq:HSVedge}
	V_\text{edge}=&\, \bigoplus_{m=0}^{s-1}\bigoplus_{l=-1}^\infty q^{l-1} \rho_{l,m}^d \sum_{k=-(s-m-1)}^{s-m-1}x^{k}\sum_{p=0}^{\lfloor \frac{s-m-1-|k|}{2}\rfloor}  \frac{q^{s-m-|k|-2p}-q^{-s+m+|k|+2p}}{q-q^{-1}}\nn\\
	=&\, \bigoplus_{k=-(s-1)}^{s-1} x^k \bigoplus_{m=0}^{s-1-|k|}\bigoplus_{l=-1}^\infty  \rho_{l,m}^d \sum_{p=0}^{\lfloor \frac{s-m-1-|k|}{2}\rfloor} \sum_{n=0}^{s-m-1-|k|-2p}q^{l-s+m+|k|+2n+2p}\nn\\
	=&\, \bigoplus_{k=-(s-1)}^{s-1}\rho^2_{k} q^{|k|}\otimes   \bigoplus_{n=0}^{s-1-|k|} \bigoplus_{m=0}^{s-1-|k|-n}\bigoplus_{l=-1}^\infty\rho_{l,m}^d \sum_{p=s-\lfloor \frac{s-m-1-|k|-n}{2}\rfloor}^{s}q^{l+s+m+2n-2p} \; .
\end{align}
In the second line, we have expanded the ratio in the first line, and exchanged the $m$- and $k$-sums; in the third we have used 
\begin{align}
	\sum_{m=0}^r\sum_{p=0}^{\lfloor \frac{r-m}{2}\rfloor} \sum_{n=0}^{r-m-2p}=\sum_{m=0}^r\sum_{n=0}^{r-m} \sum_{p=0}^{\lfloor \frac{r-m-n}{2}\rfloor} = \sum_{n=0}^r\sum_{m=0}^{r-n} \sum_{p=0}^{\lfloor \frac{r-m-n}{2}\rfloor} \; , 
\end{align}
changed the variables $p\to s-p$, and used \eqref{x rule}. Note that for $s\leq 2$ the $p$-sum in the last line can only have one term, namely $p=s$.

\subsubsection{Character formula}

The 1-loop $S^{d+1}$ partition function for a massive spin-$s$ field takes the form \eqref{eq:charPIstart}. Using \eqref{eq:HSVsplit}, we can again split the path integral into a bulk and edge part: $\log Z_\text{PI}=\log Z_\text{bulk}+\log Z_\text{edge}$. The bulk part descends from \eqref{eq:HSVbulk}, leading to the bulk partition function
\begin{align}\label{eq:spinsZbulk}
	\log Z_\text{bulk}
	=\int_0^\infty \frac{dt}{2t}\frac{1+q}{1-q}  \chi(t)  
\end{align}
with the QNM character
\begin{align}
	\chi(t)& =\, \left( q^\Delta +q^{\bar{\Delta}}\right)\sum_{p=0}^s\sum_{m=0}^p\sum_{l=-1}^\infty   \sum_{n=0}^\infty D_{l,m}^d  q^{s+m-2p + 2n + l}
\end{align}
expressed as a sum of sub-characters as in \eqref{local char sum}. On the other hand, \eqref{eq:HSVedge}  leads to 
\begin{align}\label{eq:spinsZedge}
	&\log Z_\text{edge}\nn\\
	=& -\int_0^\infty \frac{dt}{2t}  \left( q^\Delta +q^{\bar{\Delta}}\right) \sum_{k=-(s-1)}^{s-1}  \sum_{n=0}^{s-1-|k|} \sum_{m=0}^{s-1-|k|-n}\sum_{l=-1}^\infty  \sum_{p=s-\lfloor \frac{s-m-1-|k|-n}{2}\rfloor}^{s} D_{l,m}^d\, q^{|k|+l+s+m+2n-2p} \;. 
\end{align}
One can check that this reproduces the edge partition function found in \cite{Anninos:2020hfj}
\begin{align}
	\log Z_\text{edge}
	=-\int_0^\infty \frac{dt}{2t}   \frac{1+q}{1-q} \left(D_{s-1}^{d+2} \frac{q^{\Delta-1} +q^{\bar{\Delta}-1}}{(1-q)^{d-2}}\right) \; .
\end{align}
We note that the refined $Z_{\rm edge}$ \eqref{eq:spinsZedge} has the general structure as anticipated in \cite{Grewal:2022hlo} for general static black holes. Similar to the spin-2 case, \eqref{eq:spinsZedge} can in principle be written in terms of determinants on $S^{d-1}$, but we leave this to future work.


\section{Maxwell theory on $dS_{d+1}$ and $S^{d+1}$}\label{sec:Maxwell}

After analyzing the massive case, we move on to massless vectors, as a final warm-up before the gravity case. Applying our methods to the Maxwell case, we will highlight new features present for massless gauge fields. After that, we slightly generalize the discussion to Yang-Mills theory. 

\subsection{Quasinormal modes and characters}\label{sec:MaxQNM}

In $SO(1,d+1)$ representation theory, massless vectors fall into the exceptional series II representations \cite{Sun:2021thf}, whose construction is more intricate than that of their massive counterparts. This intricacy also manifests itself in the algebraic construction of QNMs \cite{Sun:2020sgn}, where, in the massless limit ($\Delta \to d-1$ and $\bar{\Delta} \to 1$), certain sectors of QNMs become unphysical, while new physical QNMs emerge.

\paragraph{$\alpha$-tower}

First, one finds that the \textbf{S0}-sector \eqref{eq:vecStype+} becomes pure gauge in the $\Delta \to d-1$ limit \cite{Sun:2020sgn}. Meanwhile, sectors \eqref{eq:vecvectortype} and \eqref{eq:vecStype-} remain physical:
\begin{alignat}{2}\label{eq:Maxphyalpha}
	\textbf{V:} &\qquad P^{2n+l}Y_1^l (\hat{P},\partial_U)\ket{d-1,U}_1\; , \qquad & n\geq 0 \, , \quad  l\geq 1 \; , \nn\\
		\textbf{S:} &\quad P^{2n+l-1}\partial_U\cdot \partial_{\hat{P}}Y_0^l (\hat{P})\ket{d-1,U}_1  \; , \qquad & n\geq 0 \, , \quad  l\geq 1 \; .
\end{alignat}
with frequencies
\begin{align}\label{eq:alphaMaxfreq}
	i \omega_{nl}^{\alpha,\textbf{V} } =  d-1 + 2n + l  \; , \qquad i \omega_{nl}^{\alpha,\textbf{S}}  =  d-2  + 2n + l   \; , \qquad n\geq 0 \; , \quad l\geq 1 \; . 
\end{align}

\paragraph{$\gamma$-tower} 

The shadow tower undergoes a more dramatic change in the limit $\bar\Delta \to 1$. In particular, the entire $\beta$-tower becomes pure gauge in this limit. However, a new $\gamma$-tower emerges and can be thought of as generated by the descendants of the primary modes\footnote{While we use the notation $\ket{\bar\Delta \to 1}_{i}$, we stress that when realized as mode functions, obtaining the $\gamma$-primary from the $\beta$-primary involves an intricate procedure of stripping off a divergent factor when taking the massless limit \cite{Sun:2020sgn}; once again, however, such explicit construction is irrelevant to our purpose here.}
\begin{align}\label{eq:Maxgammaprimary}
	\ket{\gamma}_{ij} \equiv P_{i} \ket{\bar\Delta \to 1}_{j} - P_{j} \ket{\bar\Delta \to 1}_{i} \; ,
\end{align}
which carries the $\mathfrak{so}(d)$ irrep $\rho^d_{1,1}$. Since $\ket{\gamma}_{ij}$ is anti-symmetric in $i$ and $j$, we need to introduce a new auxiliary vector $Z^i \in \mathbb{R}^d$ to encode the tensor structure 
\begin{align}
	\ket{\gamma}_{ij}  \to \ket{\gamma,U ,Z}_2 \equiv \ket{\gamma}_{ij} U^i Z^j \; .
\end{align}
The rest of the $\gamma$-tower is generated by acting $P_i$ on \eqref{eq:Maxgammaprimary}. Following the same logic in section \ref{sec:spin1char}, the $\mathfrak{so}(d)$ contents can be worked out by decomposing the tensor product $\rho^d_N \otimes \rho^d_{1,1}$. However, as pointed out in \cite{Sun:2020sgn}, a lot of descendants of \eqref{eq:Maxgammaprimary} are actually trivial for the following reason. From a future boundary point of view, the primary mode \eqref{eq:Maxgammaprimary} can be thought of as a field strength-like object; as such, this satisfies a Bianchi identity: $P_i \ket{\gamma}_{jk}+P_j \ket{\gamma}_{ki}+P_k \ket{\gamma}_{ij}=0$. The non-trivial descendants correspond to the 2-row representations in the decomposition, i.e.  
\begin{align}
	\rho^d_N \otimes \rho^d_{1,1} \supset	 \bigoplus_{p=0}^{\min(N,1)} \rho^d_{1-p+N,1-p}=
	\begin{cases}
		\rho^d_{N+1,1}  \oplus \rho^d_{N} & , \qquad N\geq 1 \\
		\quad \rho^d_{1,1}   & , \qquad N=0
	\end{cases}\; .
\end{align}
This leads to the two sectors, explicitly given by 
\begin{alignat}{2}\label{eq:Maxgamma}
	\textbf{V:} &  \qquad  P^{2n} \partial_Z\cdot \partial_{P} Y_1^l (P,\partial_U)\ket{\gamma,U ,Z}_2 \; , \quad & n\geq 0 \, , \quad  l\geq 1 \; , \nn\\
	\textbf{S:} &  \qquad  P^{2n} \Pi \, \partial_Z\cdot \partial_{P}Y_0^l (P)P \cdot \partial_U \ket{\gamma,U ,Z}_2 \; , \quad & n\geq 0 \, , \quad  l\geq 1 \; ,
\end{alignat}
with frequency spectra
\begin{align}\label{eq:gamMaxfreq}
	i \omega_{nl}^{\gamma,\textbf{V} } =  1 + 2n + l  \; , \qquad 
	i \omega_{nl}^{\gamma,\textbf{S}}  =  2  + 2n + l   \; , \qquad n\geq 0 \; , \quad l\geq 1 \; . 
\end{align}
Comparing  \eqref{eq:alphaMaxfreq} and \eqref{eq:gamMaxfreq} with \cite{Lopez-Ortega:2006aal}, we note that the scalar spectra coincide with those of ``Physical modes I", while the vector ones coincide with those of ``Physical modes II".

\paragraph{Harish-Chandra character}

For the $\alpha$-towers \eqref{eq:Maxphyalpha}, we can simply take the sum of $\chi_\Delta^\mathbf{V}(t)$ and $\chi_\Delta^\mathbf{S1}(t)$ in \eqref{vec char} with $\Delta\to d-1$, so that 
\begin{align}\label{eq:Maxalphachar}
	\chi_\alpha(t) =  \chi_\alpha^\mathbf{V}(t)+\chi_\alpha^\mathbf{S}(t) = d \frac{q^{d-1} }{(1-q)^d}-\frac{q^{d}}{(1-q)^d} \; .
\end{align}
On the other hand, observe that the $\gamma$-towers \eqref{eq:Maxgamma} can be thought of as \eqref{eq:vecvectortype} and \eqref{eq:vecStype+} but with an extra $\partial_Z\cdot \partial_{P}$ (which projects out the $l=0$ modes from the $\textbf{S}$ sector). Therefore, we can simply take the sum of $\chi_{\bar \Delta}^\mathbf{V}(t)$ and $\chi_{\bar \Delta}^\mathbf{S0}(t)$ in \eqref{vec char} with $\bar\Delta\to 1$, with the $l$-sum in the latter starting from $1$ instead of $0$, so that 
\begin{align}\label{eq:Maxgammachar}
	\chi_\gamma(t) =  \chi_\gamma^\mathbf{V}(t)+\chi_\gamma^\mathbf{S}(t) = d \frac{q }{(1-q)^d}-\frac{1}{(1-q)^d}+1 \; . 
\end{align}
The last term $+1$ results from subtracting the $l=0$ term from $\chi_{\bar\Delta = 1}^\mathbf{S0}(t)$.

Summing \eqref{eq:Maxalphachar} and \eqref{eq:Maxgammachar} gives the character of a massless spin-1 field 
\begin{align}\label{eq:maxwellchar}
	\chi(t) = d \frac{q^{d-1}+q }{(1-q)^d}-\frac{q^{d}+1}{(1-q)^d} +1 \;. 
\end{align}

\subsection{Bulk-edge split of the $S^{d+1}$ path integral}\label{sec:MaxPI}

The 1-loop path integral for Maxwell theory on $S^{d+1}$ is given by \cite{Giombi:2015haa,Law:2020cpj}
\begin{gather}\label{eq:MaxPIheat}
	\log Z_\text{PI}= \log \frac{1}{{\rm Vol}\left( U(1)\right)_{\rm PI}}+\log  Z_\text{det}\nn\\
	\log Z_\text{det}\equiv  \int_0^\infty \frac{d\tau}{2\tau} \left(\Tr \, e^{-\left( -\nabla_1^2 +d\right) \tau} - \Tr'  e^{-\left( -\nabla_0^2 \right) \tau} \right)  \; . 
\end{gather}
Here the spin-1 Laplacian $-\nabla_1^2 +d$ is the $M^2\to 0$ limit of the massive one \eqref{eq:procaPI}; the subtraction of a scalar path integral accounts for the gauge invariance that arises in this limit. The zero mode is omitted from the determinant, which we denote by a prime on the trace $\Tr'$. The factor 
\begin{align}\label{eq:u1groupvolume}
	\frac{1}{{\rm Vol}\left( U(1)\right)_{\rm PI}} \equiv \frac{ \mathrm{g}}{ \sqrt{2\pi\text{Vol}\left( S^{d+1}\right) }} \; , \qquad  \text{Vol}\left( S^n\right) =\frac{2\pi^\frac{n+1}{2}}{\Gamma\left( \frac{n+1}{2}\right) }\;, 
\end{align}
stems from the integration over the global $U(1)$ transformation, with fundamental charge $\mathrm{g}$.

Since the determinant part $Z_{\rm det}$ is composed of the spin-0 and spin-1 determinants analyzed in sections \ref{sec:Lspin0PI} and \ref{sec:Lspin1PI}, we can readily apply the results thereof. To begin, we rewrite \eqref{eq:MaxPIheat} 
\begin{align}
	&\log Z_\text{det} \nn\\
	=& \, \int_0^\infty \frac{dt}{2t}\left[  \left( q^{d-1} +q\right)\sum_{L=-1}^\infty D^{d+2}_{L,1} q^L - \left( q^{d} +1\right)\sum_{L=-1}^\infty D^{d+2}_{L} q^L +\left(q^{d-2}+1 \right) +\left( q^d+1\right) \right]  \; ,
\end{align}
with $q=e^{-t}$. The two infinite sums in the square bracket are the $\Delta \to d-1$ and $\Delta \to d$ limits of \eqref{eq:spin1SH} and \eqref{eq:scalarPIrewrite} respectively; the last two terms serve to cancel the $L=-1$ and $L=0$ terms in the two sums. We can write
\begin{align}\label{eq:MaxPIdetsplit}
	\log Z_\text{det} = \log Z^{\rm naive}_\text{bulk} +\log Z^{\rm naive}_\text{edge} +\int_0^\infty \frac{dt}{2t}\left(q^{d-2}+1+ q^d+1\right)  \; . 
\end{align}
Here the naive bulk partition function
\begin{align}
	\log Z^{\rm naive}_{\rm bulk} = \int_0^\infty \frac{dt}{2t} \frac{1+q}{1-q} \hat\chi(t) \; , 
\end{align}
with
\begin{align}\label{eq:spin1naivechi}
	\hat \chi(t) &=  \left( q^{d-1} +q\right)\sum_{n=0}^\infty  \left( \sum_{l=1}^\infty D_{l,1}^d q^{2n+l}+\sum_{l=0}^\infty D_{l}^d q^{2n+l+1}+ \sum_{l=1}^\infty D_{l}^d q^{2n+l-1}\right) -  \left( q^{d} +1\right)\sum_{n=0}^\infty \sum_{l=0}^\infty D_l^d  q^{2n+l} \nn\\
	& = \left( q^{d-1} +q\right) \sum_{n=0}^\infty \sum_{l=1}^\infty D_{l,1}^d q^{2n+l}+\left( q^2+q^{d-2}\right)  \sum_{n=0}^\infty\sum_{l=1}^\infty D_{l}^d q^{2n+l}-1 \; , 
\end{align}
comes from taking the $\Delta \to d-1$ and $\Delta \to d$ limits of \eqref{eq:vecZbulk} and \eqref{eq:massscalarPI} respectively. Notice how the would-be $\alpha$-tower $\mathbf{S0}$ and $\beta$-tower $\mathbf{S1}$ contributions are canceled by the ghost subtraction. Apart from the $-1$ term, \eqref{eq:spin1naivechi} gives exactly the massless spin-1 character \eqref{eq:maxwellchar}. We are then led to write
\begin{gather}\label{eq:MaxZbulkspli}
	\log Z^{\rm naive}_{\rm bulk} =\log Z_{\rm bulk}+\log Z^0_{\rm bulk} \nn\\
	\log Z_{\rm bulk} =\int_0^\infty \frac{dt}{2t} \frac{1+q}{1-q} \chi(t)\; , \qquad \log Z^0_{\rm bulk} =  -\int_0^\infty \frac{dt}{2t} \frac{1+q}{1-q}\;,
\end{gather}
with $\chi(t)$ the character \eqref{eq:maxwellchar}. 

On the other hand, the naive edge partition function
\begin{align}\label{eq:Maxnaiveedge}
	\log Z^{\rm naive}_\text{edge}= -\int_0^\infty \frac{dt}{2t}\left( q^{d-1} +q\right) \sum_{l=0}^\infty D^{d}_{l} q^{l-1} = -\int_0^\infty \frac{dt}{2t} \frac{1+q}{1-q} \frac{1+q^{d-2}}{\left( 1-q\right)^{d-2} }\;
\end{align}
arises from taking the $\Delta \to d-1$ limit of \eqref{eq:vecZedge}. In other words, \eqref{eq:Maxnaiveedge} corresponds to a massless scalar on $S^{d-1}$. Notice that in the first equality, the term at $l = 0$ includes a $q^0$ contribution, which leads to an IR divergence in the integral; this is the well-known zero-mode subtlety for the sphere path integral of a free massless scalar, which potentially leads to pathologies.\footnote{Such a subtlety is closely related to the non-existence of dS-invariant vacuum in the Lorentzian signature \cite{Allen:1985ux,Allen:1987tz}.} Meanwhile, the integrals in the last term of \eqref{eq:MaxPIdetsplit} and $\log Z^0_{\rm bulk}$ are also IR-divergent. Naturally combining all these factors together with the group volume factor \eqref{eq:u1groupvolume}, we find
\begin{align}
	\log Z_{\rm PI} = \log Z_{\rm bulk} +\log Z_{\rm edge}\;,  
\end{align}
where the bulk part is \eqref{eq:MaxZbulkspli} and 
\begin{align}\label{eq:MaxedgePI}
	\log Z_\text{edge}=&\,  \log \frac{ \mathrm{g}}{ \sqrt{2\pi\text{Vol}\left( S^{d+1}\right) }}  + \log  Z^{\rm naive}_\text{edge} +\log Z^0_{\rm bulk} + \int_0^\infty \frac{dt}{2t}\left(q^{d-2}+1+ q^d+1\right) \; ,\nn\\
	=&\,  \log \frac{ \mathrm{g}}{ \sqrt{2\pi\text{Vol}\left( S^{d-1}\right) }}  -\int_0^\infty \frac{dt}{2t} \left[ \frac{1+q}{1-q} \frac{1+q^{d-2}}{\left( 1-q\right)^{d-2} }-\left(q^{d-2}+1\right)\right]  \; .
\end{align}
In obtaining this, we have separated out and regularized the terms 
\begin{align}
	 \int_0^\infty \frac{dt}{2t}\left(- \frac{1+q}{1-q}+q^d+1\right) \to f (z) = \frac{1}{\Gamma (z)}\int_0^\infty \frac{dt}{2t}t^z \left(- \frac{1+q}{1-q}+q^d+1\right) \; ,
\end{align}
to compute their UV-finite contribution 
\begin{align}\label{eq:Maxedgefactor}
	\int_0^\infty \frac{dt}{2t}\left(- \frac{1+q}{1-q}+q^d+1\right) \Bigg|_{\rm finite} = f' (0) = \sqrt{\frac{2\pi}{d}} \; .
\end{align}
Combining this with the group volume factor using $\text{Vol}\left( S^{d+1}\right) =\frac{2\pi}{d}\text{Vol}\left( S^{d-1}\right) $ results in \eqref{eq:MaxedgePI}. As noted in \cite{Ball:2024hqe}, \eqref{eq:MaxedgePI} has the interpretation in terms of a {\it compact} scalar. In fact, the integral on the second line can be rewritten in terms of a functional determinant of the Laplacian on $S^{d-1}$ (with the zero mode removed), so that $Z_{\rm edge}$ reduces to the inverse of the $S^{d-1}$ partition function for a compact scalar with target circle of radius $\frac{2\pi}{\mathrm{g}}$ \cite{Law:2020cpj}
\begin{align}\label{eq:Maxedgecomp}
	Z^{\rm U(1)}_\text{edge}\left( S^{d+1}\right) = \frac{ \mathrm{g}}{ \sqrt{2\pi\text{Vol}\left( S^{d-1}\right) }}  \det\nolimits' \left( -\nabla_0^2\right)^{\frac{1}{2} } = \frac{1}{Z_{\rm compact}(S^{d-1})}\;. 
\end{align}

\subsection{Yang-Mills on $S^{d+1}$ and its $Z_{\rm edge}$}\label{sec:YM}

The considerations above extend to Yang-Mills (YM) theory with a (compact) gauge group $G$:\footnote{Throughout this paper, we use the shorthand notation $\int_{S^{d+1}} \equiv \int_{S^{d+1}} d^{d+1}x \sqrt{g}$.}
\begin{align}\label{eq:YM ex} 
	Z^{\rm YM}_\text{PI}=\int  \frac{\mathcal{D} A}{\text{Vol} (\mathcal{G})}  \, e^{-S_E[A]} \; , \qquad 
	S_E[A]= \frac{1}{4\mathrm{g}^2}\int_{S^{d+1}} \Tr F^2= \frac{1}{4\mathrm{g}^2}\int_{S^{d+1}} F_{\mu\nu}^a F^{a, \mu\nu} \; ,
\end{align}
where $F_{\mu\nu}\equiv\partial_\mu A_\nu-\partial_\nu A_\mu+[A_\mu,A_\nu]$ with $A_\mu=A_\mu^a T_a$. Here $T_a$ are some standard basis anti-Hermitian matrices that satisfy the algebra 
\begin{align}\label{eq:Lie al}
	\left[ T_a,T_b\right] =f\indices{_{ab}^c}T_c \; , 
\end{align}
and Tr in \eqref{eq:YM ex} is the trace in the adjoint representation with the overall normalization fixed by requiring $T_a$ to be unit normalized: 
\begin{align}\label{eq:trace def}
	\text{Tr}\left( T_a T_b\right) \equiv\delta^{ab}\; .
\end{align} 
The YM action \eqref{eq:YM ex} is invariant under the non-Abelian gauge transformations $g=e^{\alpha}=e^{\alpha^a T_a}$
\begin{align}\label{eq:YMgaugetrans}
	A_\mu \to g^{-1} A_\mu g+g^{-1} \partial_\mu g = A_\mu+\underbrace{\partial_\mu \alpha+ \left[ A_\mu,\alpha \right]+ \cdots}_{\delta_\alpha A_\mu}  \; .
\end{align}
The 1-loop $S^{d+1}$ path integral for \eqref{eq:YM ex} around the background configuration $\bar A_\mu=0$ is \cite{Law:2020cpj}
\begin{gather}\label{eq:YMPIheat}
	\log Z^\text{YM,1-loop}_\text{PI}= \log \frac{1}{{\rm Vol} (G)_{\rm PI}} +\dim G \log  Z_\text{det} \; ,
\end{gather}
where the determinant part $Z_\text{det}$ is the same as \eqref{eq:MaxPIheat}. 

\subsubsection{The group volume factor}\label{sec:YMgroupvol}


Similar to Maxwell, we have a residual group volume factor ${\rm Vol} (G)_{\rm PI}$ associated with global $G$ transformations. This factor arises from the integration over $\text{dim }G$ constant modes in the gauge group volume division \eqref{eq:YM ex}. These constant modes can naturally be viewed as elements of the Lie algebra $\mathfrak{g}$, or equivalently, the tangent space of $G$ at the identity, on which our path integration measure induces the following bilinear form:
\begin{align}\label{vec PI bi form}
	\braket{{\bar\alpha}|{\bar\alpha}'}_\text{PI}=\frac{1}{2\pi \mathrm{g}^2} \int_{S^{d+1}} {\bar\alpha}^a {\bar\alpha}'^{ a}=\frac{\text{Vol}(S^{d+1})}{2\pi \mathrm{g}^2} {\bar\alpha}^a {\bar\alpha}'^{ a}\; ,
\end{align}
which determines the metric with respect to which ${\rm Vol} (G)_{\rm PI}$  is measured. As described in \cite{Anninos:2020hfj,Law:2020cpj}, this group volume factor can be related to a theory-independent ``canonical volume" ${\rm Vol} (G)_c$.

%
%
%
%

\paragraph{Global symmetry algebra}

To begin, we define a field transformation bracket $[[\cdot,\cdot ]]$ on the space of gauge transformations \eqref{eq:YMgaugetrans} by 
\begin{align}\label{eq:algebrabracket}
	\delta_\alpha \delta_{\alpha'}A_\mu- \delta_{\alpha'}\delta_\alpha A_\mu=\delta_{[[\alpha,\alpha']]}A_\mu \;. 
\end{align}
We stress that $[[\cdot,\cdot ]]$  depends on the explicit form of the field transformations \eqref{eq:YMgaugetrans}. In our convention, the bracket is equal to the negative of the matrix commutator\footnote{Had we rescaled $A_\mu \to \mathrm{g}A_\mu$ and $\alpha \to \mathrm{g}\alpha$ such that the action \eqref{eq:YM ex} is canonically normalized, the coupling constant $\mathrm{g}$ will show up in the gauge transformation: $	A_\mu \to A_\mu+\partial_\mu \alpha+ \mathrm{g} \left[ A_\mu,\alpha \right] + \cdots $, and thus the algebra bracket takes the form $	[[\alpha,\alpha']]=-\mathrm{g}[\alpha,\alpha']$ instead.}
\begin{align}\label{YM bracket}
	[[\alpha,\alpha']]=-[\alpha,\alpha']\; .
\end{align}
The algebra is defined on the space of all gauge transformations (parametrized by local fields $\alpha(x)$); however, we are mostly concerned with the {\it global} part of the gauge transformations, i.e. those $\bar\alpha$ that acts trivially on the background gauge field configuration, e.g. 
\begin{align}
	\delta_{\bar\alpha} \bar A_\mu= 0 \; .
\end{align} 
These form a closed subalgebra $\tilde{\mathfrak{g}}$ of the gauge algebra with a bracket $[[\cdot,\cdot ]]$ inherited from the latter, which is isomorphic to $\mathfrak{g}$.

\paragraph{Defining the canonical volume}

We define a theory-independent ``canonical'' invariant bilinear form $\braket{\cdot|\cdot}_\text{c}$ 
on $\tilde{\mathfrak{g}}$ as follows:
\begin{enumerate}
	\item First, pick a basis $M_a$ of $\tilde{\mathfrak{g}}$ such that they satisfy the same commutation relation as $T_a$: $[[M_a,M_b]]=f\indices{_{ab}^c}M_c$. This fixes the relative normalizations of $M_a$. 
	
	\item We then fix the overall normalization of $\braket{\cdot |\cdot}_\text{c}$ by requiring $M_a$ to be unit-normalized:
	\begin{align}
		\braket{M_a|M_b}_\text{c}=\delta_{ab} \; . 
	\end{align}
\end{enumerate}
We then define the canonical volume $\text{Vol}(G)_c$  as the volume measured with the metric $ds_c^2$. In our convention, we simply take $M_a =- T_a$.\footnote{If we use the convention $\delta A_\mu =\partial_\mu \alpha+ \mathrm{g} \left[ A_\mu,\alpha \right] + \cdots$ for the gauge transformation, we take $M_a =- \frac{1}{\mathrm{g}}T_a$  instead.} We then have $\braket{T_a|T_b}_\text{c}=\delta_{ab}$, and thus
\begin{align}
	\braket{{\bar\alpha}|{\bar\alpha}'}_\text{PI}=\frac{\text{Vol}(S^{d+1})}{2\pi \mathrm{g}^2}\braket{{\bar\alpha}|{\bar\alpha}'}_{\rm c}  \qquad \implies  ds_\text{PI}^2  =\frac{\text{Vol}(S^{d+1})}{2\pi \mathrm{g}^2} ds_c^2  \;. 
\end{align}
Hence, we can related the group volume $\text{Vol}(G)_\text{PI}$ to the canonical volume $\text{Vol}(G)_c$ through
\begin{align}\label{PI volume can}
	\text{Vol}(G)_\text{PI} = \left( \frac{\text{Vol}(S^{d+1})}{2\pi \mathrm{g}^2}\right)^\frac{\text{dim}(G)}{2}\text{Vol}(G)_c \;.
\end{align}

\subsubsection{$Z_{\rm edge}$ and the nonlinear realization of $G$}

Let us return to \eqref{eq:YMPIheat}. Proceeding as before, we can obtain a bulk-edge split: 
\begin{align}
	\log Z^\text{YM,1-loop}_{\rm PI} = \log Z^\text{YM,1-loop}_{\rm bulk} +\log Z^\text{YM,1-loop}_{\rm edge}\;.
\end{align}
The bulk part $\log Z^\text{YM,1-loop}_{\rm bulk}$ is simply $\dim G$ times the Maxwell bulk partition function \eqref{eq:MaxZbulkspli},
\begin{gather}
	\log Z^\text{YM,1-loop}_{\rm bulk}=\dim G \int_0^\infty \frac{dt}{2t} \frac{1+q}{1-q} \chi(t)\; , 
\end{gather}
with $\chi(t)$ the massless spin-1 character \eqref{eq:maxwellchar}. The edge part is 
\begin{align}\label{eq:YMedgePI}
	Z^\text{YM,1-loop}_\text{edge}=\frac{1}{{\rm Vol} (G)_c} \left( \frac{2\pi \mathrm{g}^2}{ \text{Vol}\left( S^{d-1}\right) }\right)^{\frac{\dim G}{2}}   \det\nolimits' \left( -\nabla_0^2\right)^{\frac{\dim G}{2} } \;. 
\end{align}
In obtaining this we have combined the $\dim G$ factors \eqref{eq:Maxedgefactor} with the group volume factor \eqref{PI volume can} using $\text{Vol}\left( S^{d+1}\right) =\frac{2\pi}{d}\text{Vol}\left( S^{d-1}\right) $.

While we do not expect a clean bulk-edge split for the full interacting YM path integral \eqref{eq:YM ex}, it is conceivable that  the 1-loop edge partition function \eqref{eq:YMedgePI} descends from the 1-loop approximation of a sector of the full YM theory. In the Maxwell case, $Z_{\rm edge}$ is the inverse of the partition function of a compact scalar on $S^{d-1}$ with target $U(1)$ circle, which nonlinearly realizes the global $U(1)$ symmetry. As we will see momentarily, \eqref{eq:YMedgePI} can be understood in the same way, namely that it comes from (the 1-loop approximation of) a theory that nonlinearly realizes the global $G$ symmetry.

We first write down a quadratic action $S^{d-1}$ that can give rise to the determinants \eqref{eq:YMedgePI}, namely that of $\dim G$ massless scalars 
\begin{align}\label{eq:preSMaction}
	S \left[ \pi\right]  =\frac{1}{\mathrm{g}^2} \int_{S^{d-1}} \frac{1}{2}\delta_{ab} \Big(\partial_\mu \pi^a \Big) \left(\partial^\mu \pi^b \right) \;, \qquad a,b =1 , 2, \dots , \dim G \; . 
\end{align}
Clearly, this is invariant under the shift symmetries
\begin{align}\label{eq:SMfreeinfintrans}
	\delta_C \pi^a(x) = C^a  
\end{align}
parametrized by a constant rank-$\dim G$ vector $C^a$, which form an abelian algebra. 

We seek to deform the shift symmetries \eqref{eq:SMfreeinfintrans} by adding positive powers of $\pi$, such that the resulting deformed symmetries form the group $G$. We start by pairing $\pi^a$ with the standard basis anti-Hermitian matrices $T_a$ that satisfy the algebra \eqref{eq:Lie al} to write $\pi=\pi^aT_a$. Then, we consider a group element of the form
\begin{align}\label{eq:Gelement}
	U(\pi) \equiv e^{\pi} = e^{\pi^a T_a}  \; .
\end{align}
We define the transformation laws for the scalars $\pi^a$ under $G$ by the left action:
\begin{align}\label{eq:SMtrans}
	e^{\pi_g^a(x) T_a} \equiv g \, e^{\pi^a(x) T_a} \; ,  \qquad g\in G \; . 
\end{align}
Parametrizing the group element by $g=e^{C}=e^{C^a T_a}$, one can work out the full nonlinear transformations of $\pi^a(x) $ order by order using the Baker–Campbell–Hausdorff formula: 
\begin{align}\label{eq:SMinfintrans}
	\delta_C\pi^a = C^a  +\frac{1}{2} f\indices{_{bc}^a} C^b \pi^c+ \cdots \;. 
\end{align}
This can be viewed as a deformation of the abelian shift symmetries \eqref{eq:SMfreeinfintrans}. On the space of symmetry parameters, we can define a commutator $\left[ \left[ \cdot , \cdot \right] \right] $ by
\begin{align}\label{eq:NLSMbracket}
	\delta_C  \delta_{C'} \pi- \delta_{C'} \delta_C   \pi  = \delta_{\left[ \left[ C, C' \right] \right] } \pi \; . 
\end{align}
It is clear that the algebra of deformed symmetries equipped with $\left[ \left[ \cdot , \cdot \right] \right]$ is isomorphic to the Lie algebra of $G$. With the building block \eqref{eq:Gelement}, the simplest two-derivative action invariant under the nonlinearly realized $G$ symmetry \eqref{eq:SMtrans} is given by 
\begin{align}\label{eq:SMaction}
	S^{\rm SM} \left[ \pi^a\right] = \frac{1}{\mathrm{g}^2} \int_{S^{d-1}} \tr \left( U^{-1} \partial_\mu U U^{-1} \partial^\mu U(\pi) \right) =  \frac{1}{\mathrm{g}^2} \int_{S^{d-1}} g_{ab}\left( \pi \right) \left(\partial_\mu \pi^a \right) \left(\partial^\mu \pi^b \right)
\end{align}
where the trace tr is over the adjoint representation and $g_{ab}\left( \pi \right)$ is the metric induced by the Killing form. \eqref{eq:SMaction} is nothing but the two-derivative action of a sigma model (SM) with a target group manifold $G$, whose overall size is controled by the coupling constant $\mathrm{g}$. By our bottom-up construction, expanding  \eqref{eq:SMaction} to the lowest order in $\pi^a$ recovers the quadratic action \eqref{eq:preSMaction}.

\paragraph{Sigma model on $S^{d-1}$}

Let us consider the $S^{d-1}$ partition function of the sigma model \eqref{eq:SMaction}:
\begin{align}\label{eq:SMPI}
	Z_{\rm SM}  \left( S^{d-1}\right)  = \int \mathcal{D}\pi \, e^{-	S^{\rm SM} [\pi] } = {\rm Vol}(G)^{\rm SM}_{\rm PI} \int \mathcal{D}'\pi \, e^{-	S^{\rm SM} [\pi] }  \; . 
\end{align}
Because of the invariance of the action under \eqref{eq:SMtrans}, the integrations over the constant modes are not weighted by the action. This finite-dimensional integral ${\rm Vol}(G)^{\rm SM}_{\rm PI} $ is the volume of $G$ measured with respect to a metric induced by the path integral measure. We can relate this to a theory-independent canonical volume $ {\rm Vol}(G)_c $ using the same idea discussed in section \ref{sec:YMgroupvol}. Because of our convention, the calculation goes exactly the same way except for shifting $d\to d-2$ since we are on $S^{d-1}$. In other words, we have
\begin{align}\label{eq:SMvolG}
	{\rm Vol}(G)^{\rm SM}_{\rm PI} = \left( \frac{{\rm Vol}(S^{d-1})}{2\pi \mathrm{g}^2}\right)^{\frac{\dim G}{2}} {\rm Vol}(G)_c \; . 
\end{align}
In 1-loop approximation, the non-zero mode integrations\eqref{eq:SMPI} receive contributions from the quadratic action \eqref{eq:preSMaction}, while the factor \eqref{eq:SMvolG} remains intact, containing the ``memory" of the parent theory \eqref{eq:SMaction}. In other words, we have
\begin{align}\label{eq:NLSMPI}
	Z^\text{1-loop}_{\rm SM}  \left( S^{d-1}\right)  =  \left( \frac{{\rm Vol}(S^{d-1})}{2\pi \mathrm{g}^2}\right)^{\frac{\dim G}{2}} {\rm Vol}(G)_c  \det\nolimits' \left( -\nabla_0^2\right)^{-\frac{\dim G}{2} } \; .
\end{align}
Comparing this with \eqref{eq:YMedgePI}, we arrive at the conclusion that the 1-loop part of \eqref{eq:SMPI} is equal to the inverse of \eqref{eq:YMedgePI}, i.e.
\begin{align}\label{eq:YMedgeSM}
	Z^\text{YM,1-loop}_\text{edge} \left( S^{d+1}\right)  = \frac{1}{Z^\text{1-loop}_{\rm SM} \left( S^{d-1}\right)  } \; . 
\end{align}
We stress again that in obtaining \eqref{eq:YMedgeSM} we did not have a sigma model to begin with. Our point of view is a bottom-up one: we only had access to the 1-loop answer \eqref{eq:NLSMPI}, the factor \eqref{eq:SMvolG} serves as a hint at the interaction structure of the original parent theory \eqref{eq:SMaction}. 

To conclude this section, we note that the edge partition function \eqref{eq:Maxedgecomp} for Maxwell theory has been interpreted as the thermal partition function of the edge sector incorporated by the dynamical edge mode (DEM) boundary condition in Lorentzian signature \cite{Ball:2024hqe}. The DEM boundary condition was recently generalized to the YM case in \cite{Ball:2024gti}, where edge modes are characterized as $G$-valued scalars. Furthermore, it was shown that the bulk and edge sectors are intrinsically coupled through their dynamics. Building on this, it would be intriguing to explore the extent to which \eqref{eq:YMedgeSM} remains valid beyond the 1-loop level, as well as to study the dynamical interplay between the bulk and edge sectors when interactions are included.


\section{Linearized gravity on $dS_{d+1}$ and $S^{d+1}$}\label{sec:gravity}

We now turn to linearized gravity, which is the most important application of our methods. In terms of $SO(1,d+1)$ representation theory, a graviton furnishes an exceptional series II representation \cite{Sun:2021thf}. From the mass formula \eqref{eq:massdim}, we see that the massless limit corresponds to taking
\begin{align}\label{eq:gravlimit}
	\Delta\to d \; , \qquad \bar\Delta \to 0 \; . 
\end{align}

\subsection{Quasinormal modes and characters}

We first work out the QNMs and their $\mathfrak{so}(d)$ contents. Similar to the massless vector, as we take the limit \eqref{eq:gravlimit}, some sectors of the $\alpha$-tower become pure-gauge, while the entire $\beta$-tower will be replaced by a new $\gamma$-tower \cite{Sun:2020sgn}.

\paragraph{$\alpha$-tower}

Among the original 6 sectors \eqref{eq:spin2T}, \eqref{eq:spin2V} and \eqref{eq:spin2S}, the sectors $\mathbf{V1}$, $\mathbf{S1}$ and $\mathbf{S0}$ become pure gauge in the limit  \eqref{eq:gravlimit}, leaving the physical sectors 
\begin{alignat}{2}\label{eq:gravphyalpha}
	 \mathbf{T:}& \qquad P^{2n+l}Y_2^l (\hat{P},\partial_U)\ket{d,U}_2 \qquad && n\geq 0\; , \quad l\geq 2 \nn\\
	\mathbf{V:} &\quad P^{2n+l-1} \partial_U\cdot \partial_{\hat{P}}Y_1^l(\hat{P},\partial_U)\ket{d,U}_2\qquad && n\geq 0 \; , \quad l\geq 2 \nn\\
	\mathbf{S:} &\quad P^{2n+l-2}(\partial_U\cdot \partial_{\hat{P}})^2Y_0^l (\hat{P})\ket{d,U}_2 \qquad && n\geq 0\; , \quad  l\geq 2\;, 
\end{alignat}
with frequency spectra
\begin{alignat}{2}\label{eq:gravphyalphaspec}
	i \omega_{nl}^{\alpha,\mathbf{T}} & = \quad  \; \; \, d  + 2n + l \; , \qquad && n\geq 0 \; , \quad l\geq 2 \nn\\
		i \omega_{nl}^{\alpha,\mathbf{V}} & =  d -1 + 2n + l   \; , \qquad  && n\geq 0 \; , \quad l\geq 2 \nn\\
	i \omega_{nl}^{\alpha,\mathbf{S}} & =  d -2 + 2n + l  \; , \qquad  && n\geq 0 \; , \quad l\geq 2 \; . 
\end{alignat}

\paragraph{$\gamma$-tower}

The $\gamma$-tower is generated by the descendants of the primary modes 
\begin{align}\label{eq:gravgamprim}
	\ket{\gamma}_{i_1j_1, i_2j_2} \equiv P_{[i_1 }P_{[i_2} \ket{\bar\Delta \to 0}_{j_1],j_2]} - {\rm traces}\; .
\end{align}
This corresponds to a field-strength-like object carrying the $\mathfrak{so}(d)$ irrep $\rho^d_{2,2}$, whose tensor structure can be encoded by two distinct auxiliary variables 
\begin{align}
	\ket{\gamma}_{i_1j_1, i_2j_2} \to \ket{\gamma,U ,Z}_4 \equiv \ket{\gamma}_{i_1j_1, i_2j_2} U^{i_1} U^{i_2} Z^{j_1} Z^{j_2} \; . 
\end{align}
Similar to the Maxwell case, a lot descendants of \eqref{eq:gravgamprim} are trivial because of the Bianchi identity it satisfies. The non-trivial ones can be read off as the 2-row representations in the tensor product decomposition 
\begin{align}
	\rho^d_N \otimes \rho^d_{2,2} \supset	 \bigoplus_{p=0}^{\min(N,2)} \rho^d_{2-p+N,2-p}=
	\begin{cases}
		\rho^d_{N+2,2} \oplus \rho^d_{N+1,1} \oplus \rho^d_{N} & , \qquad N\geq 2 \\
		\; \; \rho^d_{3,2} \; \;\; \oplus \; \; \rho^d_{2,1} & , \qquad N = 1 \\
		\; \;\rho^d_{2,2}   & , \qquad N=0 \\		
	\end{cases}\; .
\end{align}
These can be explicitly constructed as
\begin{alignat}{2}\label{eq:gravphygamma}
	\mathbf{T:} & \qquad P^{2n+l-2}\left( \partial_Z\cdot \partial_{\hat{P}}\right)^2 Y_2^l (\hat{P},\partial_U)\ket{\gamma,U ,Z}_4  \; , \qquad && n\geq 0\; , \quad l\geq 2 \nn\\
	\mathbf{V:} &\quad  P^{2n+l-2}  \Pi \, \left( \partial_Z\cdot \partial_{\hat{P}}\right)^2 Y_1^l (\hat{P},\partial_U)P \cdot \partial_U \ket{\gamma,U ,Z}_4 \; , \quad && n\geq 0 \; , \quad l\geq 2 \nn\\
	\mathbf{S:} & \quad P^{2n+l-2}  \Pi \, \left( \partial_Z\cdot \partial_{\hat{P}}\right)^2 Y_0^l (\hat{P}) (P \cdot \partial_U)^2\ket{\gamma,U ,Z}_4   \; , \quad && n\geq 0\; , \quad  l\geq 2 \; ,
\end{alignat}
with frequency spectra
\begin{alignat}{2}\label{eq:gravphygammaspec}
	i \omega_{nl}^{\gamma,\mathbf{T}} & =  \quad  \; \; \, 2n + l \; , \qquad && n\geq 0 \; , \quad l\geq 2 \nn\\
	i \omega_{nl}^{\gamma,\mathbf{V}} & =  1 + 2n + l   \; , \qquad  && n\geq 0 \; , \quad l\geq 2 \nn\\
	i \omega_{nl}^{\gamma,\mathbf{S}} & =  2 + 2n + l  \; , \qquad  && n\geq 0 \; , \quad l\geq 2 \; . 
\end{alignat}
The spectra \eqref{eq:gravphyalphaspec} and \eqref{eq:gravphygammaspec} exactly match with those obtained in \cite{Lopez-Ortega:2006aal} by explicitly solving the linearized gravity equations of motion.

\paragraph{Harish-Chandra character}

For the $\alpha$-towers \eqref{eq:gravphyalpha}, we can simply take the sum of $\chi_\Delta^\mathbf{T}(t)$, $\chi_\Delta^\mathbf{V2}(t)$ and $\chi_\Delta^\mathbf{S2}(t)$ in \eqref{eq:spin2masschar} with $\Delta\to d$, so that 
\begin{align}\label{eq:gravalphachar}
	\chi_\alpha(t) =  \chi_\alpha^\mathbf{T}(t)+\chi_\alpha^\mathbf{V}(t)+\chi_\alpha^\mathbf{S}(t) = D_2^d \frac{q^{d} }{(1-q)^d}-D_1^d \frac{q^{d+1}}{(1-q)^d} \; .
\end{align}
On the other hand, observe that the $\gamma$-towers \eqref{eq:gravphygamma} can be thought of as the $\bar\Delta\to 0$ limits of \eqref{eq:spin2T},  the $\mathbf{V1}$-sector in \eqref{eq:spin2V} and  the $\mathbf{S0}$-sector \eqref{eq:spin2S}, but with an extra $\left( \partial_Z\cdot \partial_{P}\right)^2$ (which projects out the $l\leq 1$ modes from the $\mathbf{V1}$- and $\mathbf{S0}$-sectors). Therefore, we can take the sum of $\chi_{\bar\Delta}^\mathbf{T}(t)$, $\chi_{\bar\Delta}^\mathbf{V1}(t)$ and $\chi_{\bar\Delta}^\mathbf{S0}(t)$ (with all the $l$-sum starting from $2$)  in \eqref{eq:spin2masschar} with $\bar\Delta\to 0$, so that 
\begin{align}\label{eq:gravgammachar}
	\chi_\gamma(t) =  \chi_\gamma^\mathbf{T}(t)+ \chi_\gamma^\mathbf{V}(t)+\chi_\gamma^\mathbf{S}(t) = \left[ D_2^d \frac{1 }{(1-q)^d}-D_1^d \frac{q^{-1}}{(1-q)^d} \right]_+ \; . 
\end{align}
Here we have introduced the notation
\begin{align}\label{eq:flipnotation}
	\left[\sum_k c_k q^k \right]_+ \equiv \sum_{k<0} \left(-c_k \right)q^{-k}  +\sum_{k>0} c_k q^{k}  = \sum_k c_k q^k -c_0 -\sum_{k<0} c_k \left( q^k + q^{-k}\right) \;. 
\end{align}
The sum of \eqref{eq:gravalphachar} and \eqref{eq:gravgammachar} gives the character of a massless spin-2 field:
\begin{align}\label{eq:spin2char}
	\chi(t)	=&\,  \left[ D_2^d \frac{q^{d} +1}{(1-q)^d} - D_1^d \frac{q^{d+1} +q^{-1}}{(1-q)^d} \right]_+\nn\\
	=&\,  D_2^d \frac{q^{d} +1}{(1-q)^d} - D_1^d \frac{q^{d+1} +q^{-1}}{(1-q)^d}- \left[ D_2^d - D_1^d \left( q^{-1}+d+q\right) \right]   \; .
\end{align}

\subsection{Bulk-edge split of the $S^{d+1}$ path integral}

We now consider the path integral for gravitons on $S^{d+1}$, which was discussed  extensively in the literature  \cite{Gibbons:1978ji, Christensen:1979iy,Fradkin:1983mq, Allen:1983dg,Taylor:1989ua,GRIFFIN1989295, Mazur:1989ch, Vassilevich:1992rk, Volkov:2000ih,Polchinski:1988ua} due to its various subtleties. After a careful analysis, one finds \cite{Anninos:2020hfj,Law:2020cpj}
\begin{gather}\label{eq:gravPIheat}
	\log Z_\text{PI}= \log \frac{i^{d+3}}{\text{Vol}(G)_{\rm PI}}  +\log  Z_\text{det}\nn\\
	\log Z_\text{det}\equiv   \int_0^\infty \frac{d\tau}{2\tau} \left(\Tr'_{-1} e^{-\left| -\nabla_2^2+2\right| \tau} - \Tr'_{-1}  e^{-\left| -\nabla_1^2-d\right| \tau} \right)   \; . 
\end{gather}
In the determinant part $Z_\text{det}$, the $-1$ in $\Tr_{-1}$ again denotes the extension of the eigenvalue sums to $-1$ as in \eqref{eq:spin2PI} and  \eqref{eq:procaPI}; the absolute value signs around the Laplace operators convert terms with would-be negative exponents into positive ones. We exclude terms with zero exponents, denoted by prime. These are associated with Killing vectors of $S^{d+1}$, the integration over which leads to the volume of the isometry group $SO(d+2)$ 
\begin{align}\label{eq:ZPIvolgrav}
		\frac{1}{\text{Vol}(G)_{\rm PI}}  =\frac{ 1 }{ \text{Vol}(SO(d+2))_c }\left( \frac{8\pi G_N  d(d+2)}{\text{Vol}(S^{d-1})}\right)^{\frac{D^{d+2}_{1,1}}{2}} 
\end{align}
measured with respect to the metric induced by the path integral measure, which depends on the Newton's constant $G_N$. We have related this to the canonical volume of $SO(d+2)$, measured with respect to the invariant group metric normalized such that	minimal $SO(2)$ orbits have lengths $2\pi$,
\begin{align}\label{eq:volumes}
	\text{Vol}(SO(d+2))_c = \prod_{k=2}^{d+2} \text{Vol}\left( S^{k-1}\right) \;, \qquad  \text{Vol}\left( S^n\right) =\frac{2\pi^\frac{n+1}{2}}{\Gamma\left( \frac{n+1}{2}\right) }\;,
\end{align}
using the ideas reviewed in section \ref{sec:YMgroupvol}. We refer the reader to \cite{Anninos:2020hfj,Law:2020cpj} for more details. 

Finally, in the graviton path integral, infinitely many conformal trace modes possess the wrong sign for their kinetic term, which is cured by Wick-rotations in field space in the standard prescription  \cite{Gibbons:1978ac}. Implementing these rotations in a manner consistent with locality leads to the phase $i^{d+3}$ first found by Polchinski \cite{Polchinski:1988ua}. In principle, a sign ambiguity $i^{\pm (d+3)}$ exists; here we adopt the sign choice argued to be correct in \cite{Maldacena:2024spf}.

\subsubsection{The determinant part}

We first focus on the determinant part, which can be written as
\begin{align}\label{eq:gravdet}
	\log Z_\text{det}=\log Z^{\rm naive}_\text{det}+ \log Z^0_{\rm det}  \; . 
\end{align}
Here 
\begin{align}\label{eq:gravnaivedet}
	\log Z^{\rm naive}_\text{det}= \int_0^\infty \frac{dt}{2t} \left[  \left( q^{d} +1\right)\sum_{L=-1}^\infty D^{d+2}_{L,2} q^L - \left( q^{d+1} +q^{-1}\right)\sum_{L=-1}^\infty D^{d+2}_{L,1} q^L  \right] 
\end{align}
is obtained by taking the $\Delta \to d$ and $\Delta \to d+1$ limits of \eqref{eq:spin2SH} and  \eqref{eq:spin1SH} respectively. Doing so would introduce IR divergent terms with non-positive powers of $q$; the terms
\begin{align}
	\log Z^0_{\rm det} =\int_0^\infty \frac{dt}{2t}\left[ -D^{d+2}_{-1,2} \left( q^{-1}-q\right) -D^{d+2}_{0,2} \left(q^{d}+1 \right)  + D^{d+2}_{-1,1}\left(q^{-2}-q^2 \right)  +D^{d+2}_{1,1} \left(q^{d+2}+1\right) \right]  
\end{align}
replace the negative powers $q^{-k}$ with $q^k$, and cancel against the $L=0$ terms and $L=1$ terms in the two sums in \eqref{eq:gravnaivedet}. What this does is to implement the absolute value signs and zero mode exclusion in \eqref{eq:gravPIheat}.

\paragraph{Bulk partition function}

Putting the $\Delta \to d$ and $\Delta \to d+1$ limits of \eqref{eq:spin2Zbulk} and \eqref{eq:vecZbulk} together yields the naive bulk partition function 
\begin{align}
	\log Z^{\rm naive}_{\rm bulk} = \log Z_{\rm bulk} + \log Z_{\rm bulk}^0 \; , 
\end{align}
where 
\begin{align}
	\log Z_{\rm bulk} =\int_0^\infty \frac{dt}{2t} \frac{1+q}{1-q}\chi(t) 
\end{align}
is the quasicanonical bulk partition function defined with the massless spin-2 character \eqref{eq:spin2char}, and 
\begin{align}\label{eq:Zbulk0}
	\log Z_{\rm bulk}^0= \int_0^\infty \frac{dt}{2t} \frac{1+q}{1-q}  \left[ D_2^d - D_1^d \left( q^{-1}+d+q\right) \right] 
\end{align}
corresponds to the ``flipping" terms in \eqref{eq:flipnotation} converting the naive character to the true character. 

\subsubsection{The edge partition function}

As in the Maxwell case, we combine all factors beside $Z_{\rm bulk}$ into an edge path integral, i.e. we define the latter by $Z_\text{PI} \equiv Z_{\rm bulk} Z_{\rm edge}$. Explicitly, 
\begin{align}\label{eq:zedgegravdef}
	 \log Z_{\rm edge} = \log \frac{i^{d+3}}{\text{Vol}(G)_{\rm PI}}  +\log Z^{\rm naive}_\text{edge} + \log Z_{\rm bulk}^0 + \log Z^0_{\rm det} \; . 
\end{align}
The naive edge partition function is defined as
\begin{align}
	\log Z^{\rm naive}_\text{det} = \log Z^{\rm naive}_\text{bulk}+\log Z^{\rm naive}_\text{edge} \;, 
\end{align}
and is obtained by taking $\Delta \to d+1$ in \eqref{eq:vecZedge} and $\Delta\to d$ in \eqref{eq:spin2Zedge}
\begin{align}\label{eq:Zedgenaive}
	&\log Z^{\rm naive}_\text{edge} \nn\\
	=&\, -\int_0^\infty \frac{dt}{2t}\Bigg[ \left( q^d +1\right) \left( \sum_{l=1}^\infty D_{l,1}^d q^{l-1} +\sum_{l=0}^\infty D_{l}^d q^{l}+\sum_{l=1}^\infty D_{l}^d q^{l-2}+2q\sum_{l=0}^\infty D_{l}^d q^{l-2}\right)  \nn\\
	& \qquad \qquad \qquad  -\left( q^{d+1} +q^{-1}\right) \sum_{l=0}^\infty D^{d}_{l} q^{l-1} \Bigg] \nn\\
	=&\,  -\int_0^\infty \frac{dt}{2t}\left[ \left( q^{d-1} +q^{-1}\right)  \sum_{l=-1}^\infty D_{l,1}^d q^l +\left( q^{d-2} +1\right) \sum_{l=0}^\infty D_{l}^d q^{l}+2\left( q^{d-1} +q^{-1}\right)  \sum_{l=0}^\infty D_{l}^d q^{l}\right]\;.
\end{align}
In the second equality we have used $D^{d}_{-1,1} =- D_0^d$ and $D_{0,1}^d=0$ to the first sum to $l=-1$. Notice how the ghost subtraction cancels against the would-be scalar contributions on the first line. 

Now, with the goal of arriving at a path integral interpretation, we would like to rewrite this in terms of determinants of Laplace operators that descend from some $S^{d-1}$ path integrals. Guided by this principle, we introduce flipping terms to convert negative powers into positive ones and subtracting off the $q^0$ the $l$-sums, and write
\begin{align}\label{eq:gravZedge0}
	\log Z^{\rm naive}_\text{edge}  = \log Z^\text{det}_\text{edge} +\log Z^0_\text{edge} \;. 
\end{align}
Here
\begin{align}\label{eq:Zedgepredet}
	\log Z^\text{det}_\text{edge} =&\,  \log Z_{1,0}+2 \log Z_{0,1}+\log Z_{0,0} \nn\\
	\log Z_{1,0} = &-\int_0^\infty \frac{dt}{2t}\left[ \left( q^{d-1} +q^{-1}\right)  \sum_{l=-1}^\infty D_{l,1}^d q^l - D_{-1,1}^d \left(q^{-2}-q^2 \right) - D^{d}_{1,1} \left(q^{d}+1 \right)  \right] \nn\\
	\log Z_{0,1} = &-\int_0^\infty \frac{dt}{2t} \left[ \left( q^{d-1} +q^{-1}\right)  \sum_{l=0}^\infty D_{l}^d q^{l} -  D_0^d  \left( q^{-1}-q\right)   -  D_{1}^d \left(q^{d}+1 \right) \right]  \nn\\
	\log Z_{0,0} = &-\int_0^\infty \frac{dt}{2t} \left[\left( q^{d-2} +1\right) \sum_{l=0}^\infty D_{l}^d q^{l}-\left( q^{d-2} +1\right)  \right]
\end{align}
are free of terms with non-positive exponents, and thus can be written in terms of determinants of absolute values of vector and scalar Laplace operators on $S^{d-1}$:
\begin{align}\label{eq:Zedgedet}
	Z_{1,0}  = \det\nolimits'_{-1} \left| -\nabla_1^2-\left( d-2\right)  \right|^{\frac12} \;, \quad Z_{0,1}  = \det\nolimits' \left| -\nabla_0^2-\left( d-1\right)  \right|^{\frac12} \; , \quad Z_{0,0}  = \det\nolimits' \left(-\nabla_0^2 \right) ^{\frac12} \;. 
\end{align}
The second term in \eqref{eq:gravZedge0}, i.e. 
\begin{align}
	\log Z^0_\text{edge} 
	=&\,  - \int_0^\infty \frac{dt}{2t}\bigg[ D_{-1,1}^d \left(q^{-2}-q^2 \right) + D^{d}_{1,1} \left(q^{d}+1 \right) \nn\\
	& \qquad  +  2 D_0^d  \left( q^{-1}-q\right)   +  2 D_{1}^d \left(q^{d}+1 \right) + \left( q^{d-2} +1\right)  \bigg]\;,
\end{align}
corresponds to the flipping terms we introduced in \eqref{eq:Zedgepredet}, which naturally combine with the last two terms in \eqref{eq:zedgegravdef}. Summarizing, we have
\begin{align}\label{eq:Zedgedetsplit}
	\log Z^{\rm naive}_\text{edge} + \log Z_{\rm bulk}^0+\log Z^0_{\rm det} 
	= \log Z^\text{det}_\text{edge}+ \log Z_\text{edge,G} \; .
\end{align}
The last term is given by $\log Z_\text{edge,G}=\log Z_{\rm bulk}^0+\log Z^0_{\rm det}+\log Z^0_\text{edge} $ and can be simplified to be  
\begin{align}\label{eq:ZedgeGPI}
	\log Z_\text{edge,G}
	= \int_0^\infty \frac{dt}{2t} \left[   D^{d+2}_{1,1} \left(q^{d+2}+1  -\frac{1+q}{1-q}\right) + D_0^d \left(  q^{d} -  q^{d-2} \right)   \right] \;. 
\end{align}
Regularizing this in the same way as \eqref{eq:Maxedgefactor}, we can compute
\begin{align}\label{eq:ZedgeG}
	\log Z_\text{edge,G}\big|_{\rm finite}  = \left( \frac{2\pi}{d+2}\right)^{\frac{D^{d+2}_{1,1}}{2}} \left( \frac{d-2}{d}\right)^{\frac{D^{d}_0}{2}} \;. 
\end{align}
To sum up, the edge partition function for linearized gravity on $S^{d+1}$ in any $d\geq 3$ is given by\footnote{The case $d=3$ is accommodated with the modification \eqref{eq:d=3modify}.}
\begin{empheq}[box=\fbox]{align}\label{eq:Zedgegravresult}
	&\qquad \qquad \qquad \qquad Z_{\rm edge }= Z^\text{det}_\text{edge} Z^\text{non-det}_\text{edge} \nn\\
	Z^\text{det}_\text{edge} & = \det\nolimits'_{-1} \left| -\nabla_1^2-\left( d-2\right)  \right|^{\frac12}  \det\nolimits' \left| -\nabla_0^2-\left( d-1\right)  \right| \det\nolimits' \left(-\nabla_0^2 \right)^{\frac12} \nn\\
	Z^\text{non-det}_\text{edge} &= \frac{i^{d+3}}{\text{Vol}(SO(d+2))_c}\left( \frac{2\pi \mathrm{g}^2} {\text{Vol}(S^{d-1})} \right)^{\frac{D^{d+2}_{1,1}}{2}} d^{\frac{D^{d}_{1,1}+2D^d_1}{2}}\left( d-2 \right)^{\frac{D^d_0}{2}}  \; . 
\end{empheq}
Here we have combined \eqref{eq:ZedgeG} with \eqref{eq:ZPIvolgrav} and defined
\begin{align}\label{eq:gGN}
	\mathrm{g}^2 \equiv  8 \pi G_N \; . 
\end{align}
We emphasize that \eqref{eq:Zedgegravresult} represents a significant refinement over the original formula \eqref{introeq:Zedgeoriginal}, where the $\mathfrak{so}(d)$ structure remains entirely opaque. As we will see next, this kinematic information already provides a glimpse into the underlying physics of $Z_{\rm edge}$ to a notable extent.

\subsection{Towards an interpretation of $Z_{\rm edge}$}\label{sec:gravZedgecan}

A quadratic action that could give rise to the determinants \eqref{eq:Zedgedet} takes the form
\begin{align}\label{eq:gravedgeaction}
	S^{(2)}\left[A ,{\bm\phi} , \chi \right] =&\,  S\left[A \right] + S\left[{\bm\phi}  \right] + S\left[ \chi \right] \nn\\
	S\left[A \right]  =&\,  \frac{1}{\mathrm{g}^2}\int_{S^{d-1}} \frac{1}{4} F_{\mu\nu}F^{\mu\nu} -\left( d-2\right) A_\mu A^\mu  \nn\\
	S\left[{\bm\phi} \right]  = &  \frac{1}{\mathrm{g}^2} \int_{S^{d-1}}  \frac12 \partial_\mu {\bm\phi} \cdot \partial^\mu {\bm\phi}-\frac{d-1}{2}  {\bm\phi}\cdot {\bm\phi} \nn\\
	S\left[\chi \right]  =&\,  \frac{1}{\mathrm{g}^2}\int_{S^{d-1}} \frac12 \partial_\mu \chi \partial^\mu \chi \; . 
\end{align}
We have grouped the two tachyonic scalars into a 2D vector ${\bm\phi} = \left( \phi^{d+1} , \phi^{d+2} \right)$ contracted with the metric $\delta_{ab}  , a, b = d+1,d+2$, i.e. ${\bm\phi}  \cdot {\bm\phi}= \delta_{ab}\phi^a \phi^b$ (the reason for this notation will become clear later). For any $d\geq 3$, the action \eqref{eq:gravedgeaction} is invariant under the abelian shift symmetries
\begin{align}\label{eq:shiftsym}
	A_\mu \to A_\mu + Y_{1,\mu} \; , \qquad \phi^a \to \phi^a + Y^{(a)}_1 \;, \qquad \chi \to\chi +Y_0 \; ,
\end{align}
with $Y_{1,\mu}$ an $l=1$ vector spherical harmonic, and $Y_l$ a $l$-th scalar spherical harmonic respectively. We put a superscript ${(a)}$ on $Y_1$ to emphasize the independence of the two shift symmetries for $\phi^{d+1}$ and $\phi^{d+2}$. The modes generating the shift symmetries are exactly the zero modes of the Laplace operators  \eqref{eq:Zedgedet}. In the terminology of \cite{Bonifacio:2018zex,Bonifacio:2019hrj}, $A_\mu$ has a level $k=0$ shift symmetry, while the scalars $\phi^a$ and $\chi$ have shift symmetries of level $k=1$ and $k=0$ respectively. The tachyonic vector can be obtained as the decoupled longitudinal modes of massive gravity in the massless limit \cite{Bonifacio:2019hrj}. The two $k=1$ scalars are the dS analogs of Galileons, and have been studied as a toy model for the conformal modes of gravity \cite{Folacci:1992}. 

Recalling the embedding space representations of the spherical harmonics \eqref{appeq:STSHpoly}, the abelian shift symmetries \eqref{eq:shiftsym} are parametrized by an antisymmetric tensor $ E_{ij}$, two vectors $E_{i ,a}$, and one $c$-number $E_{d+1, d+2}$ in $\mathbb{R}^d$ ($i,j = 1, \dots, d$). These can be grouped into an antisymmetric tensor $E_{AB}$ in $\mathbb{R}^{d+2}$ ($A,B=1,\dots , d+2$), which parametrizes a $SO(d+2)$ transformation, as reflected in the degeneracies: $D^{d+2}_{1,1}=D^{d}_{1,1}+2D^{d}_{1}+D^{d}_{0}$. 

As in the Yang-Mills case, we do not expect a clean bulk-edge split for the full interacting gravitational path integral, but it is conceivable that  the 1-loop edge partition function \eqref{eq:Zedgegravresult} descends from the 1-loop approximation of a sector of the full interacting theory. In the cases of Maxwell and Yang-Mills, we observed that their $Z_{\rm edge}$ equals the inverse of the 1-loop path integral for a sigma-type model on $S^{d-1}$ that nonlinearly realizes the global parts of their respective gauge groups. For linearized Einstein gravity, a reasonable hypothesis is that \eqref{eq:Zedgegravresult} descends from a theory on $S^{d-1}$ that nonlinearly realizes the isometry group $SO(d+2)$ of $S^{d+1}$, the global part of the diffeomorphism group, i.e.
	\begin{align}\label{eq:Zedgeguess}
		Z_{\rm edge}= \frac{1}{\mathcal{W}^\text{1-loop} _{\rm edge}} \;, \qquad \mathcal{W}_{\rm edge} =  \int \mathcal{D}\Psi \, e^{-\mathcal{S}\left[\Psi \right]}\;.
	\end{align}
	Here the action for $\Psi=\{{A_\mu, {\bm\phi},\chi }\}$
	\begin{align}\label{eq:Sedgeform}
		\mathcal{S}\left[\Psi \right] = S^{(2)}\left[\Psi \right] + S^{(3)}\left[\Psi \right] + \cdots 
	\end{align}
	has a quadratic part given by \eqref{eq:gravedgeaction}, while the interaction terms $S^{(n)}\left[\Psi \right] $ involves $n$ powers of $\Psi$. An $SO(d+2)$ transformation (parameterized by an antisymmetric tensor $E_{AB}$ in $\mathbb{R}^{d+2}$) acts nonlinearly on $\Psi$ while leaving the action invariant
		\begin{align}\label{eq:fullinvariance}
		\delta_E \mathcal{S}\left[\Psi \right]  =  0 \;, \qquad \delta_{E} \Psi = \delta^{(0)}_{E} \Psi+\delta^{(1)}_{E} \Psi +\cdots \;. 
	\end{align}
	Here $\delta^{(0)}_{E} \Psi$ coincides with the abelian symmetries \eqref{eq:shiftsym} and $\delta^{(n)}_{E} \Psi $ involves $n$ powers of $\Psi$. With a bracket defined by $\delta_{E_1} \delta_{E_2} -\delta_{E_2} \delta_{E_1} = \delta_{[[E_1,E_2]]} $, the symmetries $\delta_E$ generate the $\mathfrak{so}(d+2)$ algebra. 
	
	Because of the invariance \eqref{eq:fullinvariance}, the integrations over the modes generating the non-linear symmetries \eqref{eq:fullinvariance}, denoted as $\bar\Psi=\{{\bar A_\mu, \bar{\bm\phi},\bar\chi }\}$, are not weighted by the action in the full interacting path integral \eqref{eq:Zedgeguess}. This $D^{d+2}_{1,1}$-dimensional integral gives the volume of $SO(d+2)$, and 	
	remains intact in the 1-loop approximation, contributing as the non-determinant factors in \eqref{eq:Zedgegravresult} (except the phase $i^{-d-3}$, which we ignore for now). Our path integration measure induces a bilinear form on the Lie algebra $\mathfrak{so}(d+2)$:
	\begin{align}
		\braket{\bar \Psi| \bar\Psi'}_{\rm PI} =	 \braket{\bar A| \bar A'}_{\rm PI} + 	\braket{\bar{\bm\phi}| \bar{\bm\phi}'}_{\rm PI} +  \braket{\bar \chi| \bar \chi'}_{\rm PI} =  \frac{1}{2\pi \mathrm{g}^2} \int_{S^{d-1}} \bar A^\mu \bar A'_\mu + \bar{\bm\phi} \cdot \bar{\bm\phi}'+ \bar \chi\bar \chi' \;,
	\end{align}
	which determines the metric with respect to which the volume ${\rm Vol}\left( SO(d+2)\right) _{\rm PI}$ is measured.
%
	This is generally distinct from the canonical invariant bilinear, which takes the general form
	\begin{align}
		\braket{\bar \Psi| \bar\Psi'}_c =	\frac{2\pi \mathrm{g}^2}{{\rm Vol}(S^{d-1})} \left[B_{1,1}  \braket{\bar A| \bar A'}_{\rm PI} + B_{0,1} 	\braket{\bar{\bm\phi}| \bar{\bm\phi}'}_{\rm PI} + B_{0,0} \braket{\bar \chi| \bar \chi'}_{\rm PI} \right] \;. 
	\end{align}
	The relative coefficients can in principle be fixed by the $\mathfrak{so}(d+2)$ algebra, but it requires the explicit form of the deformations of the symmetries \eqref{eq:fullinvariance}. On the other hand, the non-determinant factors in \eqref{eq:Zedgegravresult} predict the coefficients to be
	\begin{align}
		B_{1,1} =d \;, \qquad B_{0,1}  =d \;, \qquad  B_{0,0}  =d-2\;. 
	\end{align}
	It would be interesting to confirm this by deriving a simple nonlinear edge theory $\mathcal{S}[\Psi]$ that fits into the considerations above. Unfortunately, this appears to be challenging and likely requires techniques beyond the scope of the present work. In spite of this, earlier investigations into shift-symmetric theories \cite{Bonifacio:2018zex,Bonifacio:2019hrj} have provided significant clues, allowing us to speculate on a plausible interpretation of the tentative edge theory. We turn to this next. 

\subsubsection{A brane interpretation?}

\begin{figure}[H]
	\centering
	\includegraphics[width=0.8\textwidth]{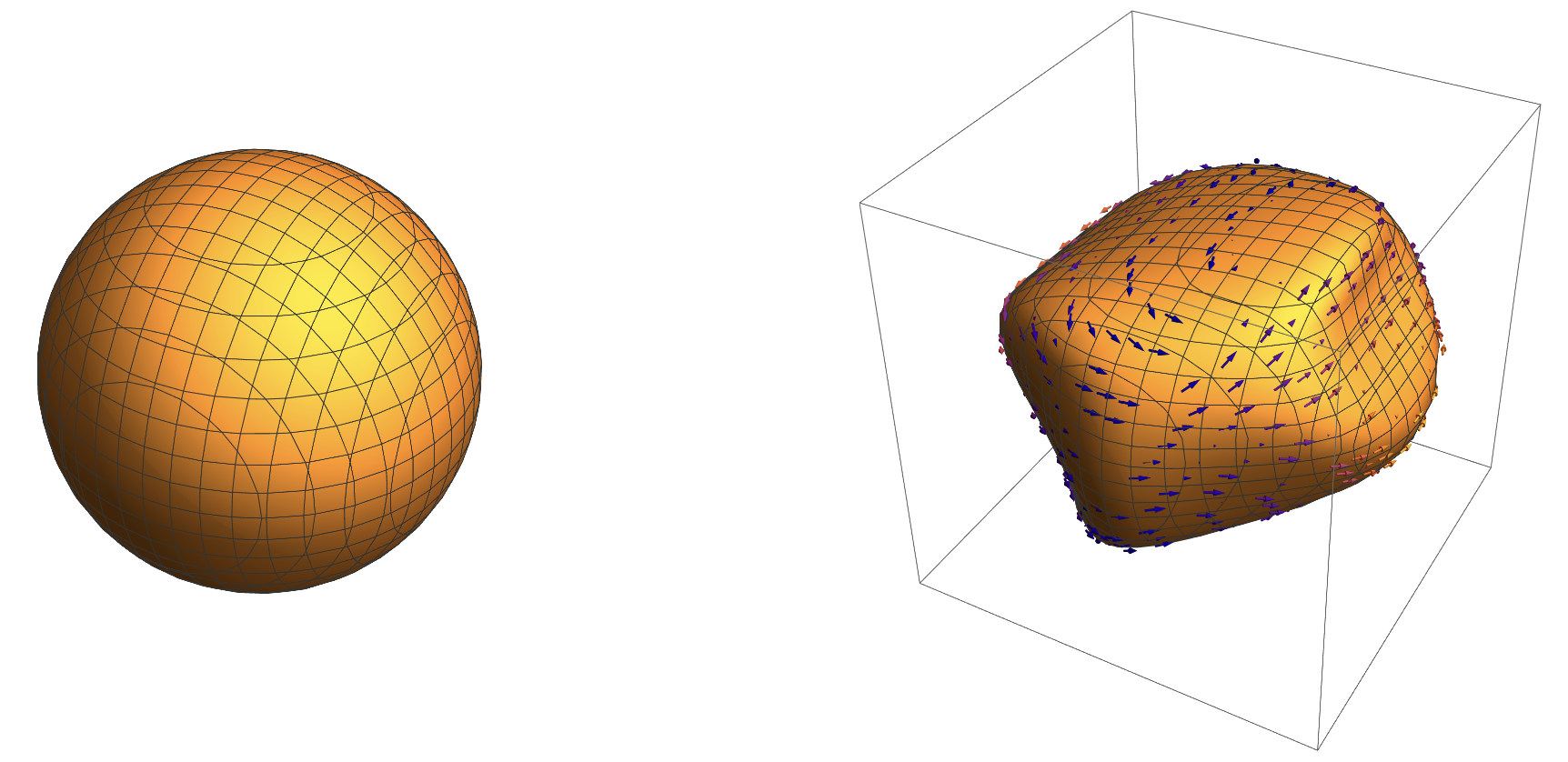}
	\caption{In the brane interpretation of $Z_{\rm edge}$, $\phi^a$ describes the deformations of the round $S^{d-1}$ (left) in the $X^a$-direction upon embedding in $S^{d+1}$. $A_\mu$, represented by a vector field in the right figure, describes diffeomorphisms on the sphere.}
	\label{pic:embed}
\end{figure}

The two conformal scalars $\phi^a$ are known to describe a $S^{d-1}$ or $dS_{d-1}$ brane embedded in some ambient space \cite{Goon:2011qf,Goon:2011uw,Burrage:2011bt}. Given that we started with a path integral on $S^{d+1}$, it seems reasonable to explore the possibility of embedding a $S^{d-1}$ brane in $S^{d+1}$, with $\phi^a$ describing its small oscillations in the two transverse directions. See figure \ref{pic:embed}. A simple realization of this idea is as follows. Recall that a round $S^{d+1}$ can be realized as a hypersurface 
\begin{align}\label{eq:sphereembed}
	\delta_{AB}X^AX^B=\left( X^1\right)^2 + \left( X^2\right)^2+\cdots +\left( X^{d+2}\right)^2 = 1 
\end{align}
in an ambient $\mathbb{R}^{d+2}$. We can describe a $S^{d-1}$ brane interior to this hypersurface by the (non-unique) parametrization \cite{Clark:2005ht,Clark:2007rn}
\begin{align}\label{eq:branepara}
	X^i (x) = \sqrt{1-{\bm \phi}(x)\cdot {\bm \phi}(x)}\, \Omega^i (x)\; , \quad 	X^a (x)=  \phi^a(x)  \;, \qquad i=1,\dots ,d, \quad a=d+1,d+2\;. 
\end{align}
Here $x^\mu$ are intrinsic coordinates parametrizing the rigid unit $S^{d-1}$: $\Omega^i \Omega_i =1$, on which the transverse coordinates $\phi^a(x) $ are dynamical fields. With \eqref{eq:branepara} one can compute the induced metric 
\begin{align}\label{eq:inducedmetric}
	G_{\mu\nu} [{\bm \phi}]\equiv \delta_{AB} \frac{\partial X^A}{\partial x^\mu} \frac{\partial X^B}{\partial x^\nu} =f^2 g_{\mu\nu} + B_{\mu\nu}\; ,
\end{align}
where $g_{\mu\nu} =\delta_{ij} \frac{\partial \Omega^i}{\partial x^\mu}\frac{\partial \Omega^j}{\partial x^\nu}$ is the standard round metric on the unit $S^{d-1}$, and 
\begin{align}\label{eq:fB}
	f^2 = 1-{\bm \phi} \cdot {\bm \phi}  \;, \qquad B_{\mu\nu}= \partial_\mu {\bm \phi} \cdot \partial_\nu {\bm \phi}+\frac{\left( {\bm \phi} \cdot \partial_\mu {\bm \phi} \right) \left( {\bm \phi} \cdot \partial_\nu {\bm \phi} \right) }{1-{\bm \phi} \cdot {\bm \phi} }\;. 
\end{align}
The simplest $SO(d+2)$ invariant worldvolume action is
\begin{align}\label{eq:brane action}
	S^{\rm brane} [{\bm \phi}] = \frac{1}{\mathrm{g}^2}\int_{S^{d-1}} d^{d-1} x \sqrt{G [{\bm \phi}]} =  \frac{1}{\mathrm{g}^2}\int_{S^{d-1}} d^{d-1} x  \sqrt{g}  \, f^{d-1}\sqrt{\det \left(I+\frac{1}{f^2}g^{-1}B \right) } \;,
\end{align}
with some brane tension $\frac{1}{\mathrm{g}^2}$ to be identified with the Newton's constant through \eqref{eq:gGN}. Expanding this in small ${\bm \phi}$, one recovers the quadratic action \eqref{eq:gravedgeaction} for ${\bm \phi}$. Note that the tachyonic mass squared $M^2=-(d-1)$ originates from the $d-1$ powers of $f$ and the negative sign in $f^2$ in \eqref{eq:fB}. Under the isometries generated by
\begin{align}
	J_{i a} = X_i \partial_a -X_a \partial_i \;, 
\end{align}
the conformal scalars transform  infinitesimally as
\begin{align}\label{eq:branetransform}
	\phi^a \to \phi^a +  E^{ia}X_i  = \phi^a +E^{ia}\Omega_i (x) + O(\phi^2) \;,
\end{align}
where $E^{ia}$ parametrizes the independent symmetries. Note that \eqref{eq:branetransform} coincides with the shift symmetries \eqref{eq:shiftsym} to the lowest order in $\phi$. The $SO(d)$ symmetries generated by 
\begin{align}
	J_{i j} = X_i \partial_j -X_j \partial_i 
\end{align}
act trivially on ${\bm \phi}$, while the $U(1)$ symmetry generated by
\begin{align}\label{eq:U1vec}
	J_{d+1,d+2} = X_{d+1}\partial_{d+2}- X_{d+2}\partial_{d+1} 
\end{align}
has become an internal $U(1)$ symmetry of \eqref{eq:brane action}: $\phi^a \to L\indices{^a_b}\phi^b$ with $L\indices{^a_b}$ a $2\times 2$ $SO(2)$ matrix.
While it is not clear to us at this point how to naturally incorporate $A_\mu$ and $\chi$ into the action \eqref{eq:brane action} so that the full $SO(d+2)$ symmetry is nonlinearly realized, they both seem to have natural geometric interpretations in this brane picture. Transforming under \eqref{eq:U1vec} like a compact scalar with a target $U(1)$, $\chi$ is clearly related to the angular direction in the $X^{d+1}$-$X^{d+2}$ plane. As for $A_\mu$, it is possible to rewrite its quadratic action \eqref{eq:gravedgeaction} in terms of the Lie derivative of the rigid round $S^{d-1}$ metric with respect to $A_\mu$:
\begin{align}\label{eq:vecactionrewrite}
	S[A] = \frac{1}{\mathrm{g}^2} \int_{S^{d-1}} \frac14 M^{\mu\nu}M_{\mu\nu} - \frac14 \left( M\indices{^\lambda_\lambda} \right) ^2 \;, \qquad M_{\mu\nu} \equiv \mathcal{L}_A g_{\mu\nu} = \nabla_\mu A_\nu +\nabla_\nu A_\mu \;.
\end{align}
It is thus natural to think of $A_\mu$ as (infinitesimal) diffeomorphisms on $S^{d-1}$ (see figure \ref{pic:embed}); the invariance by a shift by a Killing vector \eqref{eq:shiftsym} is then geometrically evident. One might want to incorporate $A_\mu$ in \eqref{eq:brane action} by transforming \eqref{eq:inducedmetric} with a diffeomorphism generated by $A_\mu$:
\begin{align}\label{eq:transformedinducedG}
	G_{\mu\nu}[A,{\bm\phi}] = G_{\mu\nu}[{\bm\phi}] + \mathcal{L}_A G_{\mu\nu}[{\bm\phi}]  + \cdots \;. 
\end{align}
This, however, cannot lead to the quadratic action \eqref{eq:vecactionrewrite}, because the physical transverse part of $A_\mu=A^T_\mu+A^L_\mu$ does not change the Jacobian: $G[A^T,{\bm\phi}] =G[{\bm\phi}]$. A non-trivial action for $A_\mu$ seems to necessarily involve the contractions of $G_{\mu\nu}[A,{\bm\phi}] $ and the reference background metric $g_{\mu\nu}$.

\paragraph{Polchinski's phase}

While the other non-determinant factors in \eqref{eq:Zedgegravresult} are naturally incorporated into $Z_{\rm edge}$ through considerations of locality and the nonlinear realization of $SO(d+2)$, the inclusion of the overall phase $i^{d+3}$ in $Z_{\rm edge}$ appears ad hoc. Let us nevertheless make a few comments.  First off, because of the tachyonic masses in \eqref{eq:gravedgeaction}, some modes will have a negative action--specifically, the constant modes $\phi_0^a$ of the two conformal scalars $\phi^a$ and all longitudinal modes $A_\mu^L=\partial_\mu \xi$ of the tachyonic vector $A_\mu$. To render their path integrals convergent, we follow the standard prescription \cite{Gibbons:1978ac} of Wick-rotating these modes in field space:
\begin{align}
	\mathcal{D}A^L &\to \mathcal{D}\left(\pm iA^L\right)  = \left( \prod_{l=1}^\infty i^{\pm D^d_l}\right) \mathcal{D}A^L =i^{\mp 1} \left( \prod_{l=0}^\infty i^{\pm D^d_l}\right) \mathcal{D}A^L\nn\\
	\mathcal{D}\phi_0^a & \to \mathcal{D}\left(\pm i \phi_0^a \right) = \pm i \, \mathcal{D}\phi_0^a 
\end{align}
Here we recall that $A^L_\mu$ are uniquely spanned by all the non-constant scalar spherical harmonics with $l\geq 1$; extending this to $l=0$ with a  compensating factor $i^{\mp 1}$, we can think of the infinite product as an overall rescaling of a local scalar Laplacian, which can be absorbed into a local counterterm. We have allowed $\pm$ to accommodate the different ways in which these modes could be Wick-rotated. All in all, we have a net overall factor of 
\begin{align}\label{eq:netphase}
	i^{\mp 1 \pm 1\pm 1}  \;.
\end{align}
This raises questions about whether it is appropriate to include the entire phase $i^{d+3}$ into \eqref{eq:Zedgegravresult}, because \eqref{eq:netphase} cannot match $i^{-d-3}$ in {\it all} $d\geq 3$ (recall that $Z_{\rm edge}$ is supposed to be the {\it inverse} of the $S^{d-1}$ partition function of the putative edge theory, i.e. \eqref{eq:Zedgeguess}), whatever signs we choose. 

There is one caveat, however: we have so far assumed the relative signs of the quadratic actions \eqref{eq:gravedgeaction}. To explore the implications of a sign change, consider replacing the action of one of the conformal scalars, say $\phi^{d+2}$, with one that includes an overall negative sign:
\begin{align}\label{eq:Wickaction}
	S\left[\phi^{0}\right]  =  - \frac{1}{\mathrm{g}^2} \int_{S^{d-1}}  \frac12\left(  \partial \phi^{0}\right)^2  -\frac{d-1}{2}  \left(   \phi^{0}\right)^2 \;. 
\end{align}
This modification causes all modes with $l \geq 2$ to acquire a negative action.  If we Wick-rotate them in field space to render the path integrals convergent, we have
\begin{align}
	\mathcal{D}\phi^0_{l\geq 2} \to \mathcal{D}\left(\pm i\phi^0_{l\geq 2}\right)  = \left( \prod_{l=2}^\infty i^{\pm D^d_l}\right) \mathcal{D}\phi^0_{l\geq 2} =i^{\mp (1+d)} \left( \prod_{l=0}^\infty i^{\pm D^d_l}\right) \mathcal{D}\phi^0_{l\geq 2} \;, 
\end{align}
where the infinite product, as before, can be absorbed into a local counterterm. While the Gaussian integrals still yield the conformal scalar determinant in \eqref{eq:Zedgedet}, the overall phase now changes to
\begin{align}
	i^{\mp 1\pm 1 \mp (1+d)}
\end{align}
which differ from \eqref{eq:netphase}. By adopting the following sign choices for the field-space Wick rotations:
\begin{align}\label{eq:signchoice}
	A_\mu^L \to + i A_\mu^L \; , \qquad \phi_0^{d+1} \to -i \phi^{d+1} \; , \qquad \phi_{l\geq 2}^0 \to +i \phi_{l\geq 2}^0 \;,
\end{align}
we obtain an overall phase of $i^{-d-3}$ for all $d\geq 3$. 

The action \eqref{eq:Wickaction} arises naturally by sending $\phi^{d+2} \to i \phi^0$ in the action of $\phi^{d+2}$. While speculative, this suggests a possible interpretation in our brane discussion: instead of $S^{d+1}$, the proposed $S^{d-1}$ brane theory might have Lorentzian dS spacetime, $dS_{d+1}$, as its target space. The specific sign choices in \eqref{eq:signchoice} might find justification through reasoning analogous to that in \cite{Maldacena:2024spf}.

\paragraph{Edge modes, observer and quantum reference frame}

The embedded brane interpretation for  $Z_{\rm edge}$, though primarily motivated by studies of shift-symmetric theories, resonates well with existing notions of “edge modes” in gravity. Edge modes are introduced as dynamical embeddings of co-dimension-2 corners of the causal diamond associated with a subregion of bulk spacetime \cite{Donnelly:2016auv,Freidel:2021dxw,Ciambelli:2021nmv}. These embedding fields address key challenges in defining the classical gravitational phase space for subregions, such as the non-integrability of charges associated with certain diffeomorphisms, and play a central role in the “corner proposal” \cite{Freidel:2020xyx,Freidel:2020svx,Freidel:2020ayo}, as reviewed in \cite{Ciambelli:2022vot}. Furthermore, edge modes have been proposed as natural realizations of quantum reference frames (QRFs) \cite{Carrozza:2021gju,Carrozza:2022xut,Goeller:2022rsx}, providing a framework for modeling observers in gravitational systems \cite{Susskind:2023rxm,DeVuyst:2024khu,AliAhmad:2024vdw}.

While gravity edge modes have been explored primarily through analyses of classical phase space and symmetries, our explicit results provide tentative support for these converging ideas when applied to the dS horizon, at the level of Euclidean partition functions.\footnote{The contribution from the observer or QRFs to gravitational free energies is explored with von Neumann algebraic techniques in \cite{Jensen:2023yxy,Kirklin:2024gyl,DeVuyst:2024fxc}.} Establishing a direct connection between our $Z_{\rm edge}$  and thermal partition functions associated with edge modes requires applying the edge mode analysis in the Lorentzian signature, and gaining a better understanding of their dynamics (e.g. their action or Hamiltonian).

Moreover, the appearance of shift-symmetric fields in our result suggests that edge modes may be understood as Goldstone modes associated with some spontaneously broken symmetries. In the static patch context, a natural symmetry-breaking pattern arises where the full group of diffeomorphisms (with  $SO(1,d+1)$  as its global part) is broken by the specification of a static patch to a subgroup (with  $SO(1,1) \times SO(d)$  as its global part) that preserves the static patch. In the corresponding Euclidean picture, specifying an origin similarly breaks the full group of diffeomorphisms (with  $SO(d+2)$  as its global part) to a subgroup (with  $U(1) \times SO(d)$  as its global part) that preserves its location. Clarifying these lines of thought could be an important step forward, potentially opening new possibilities for a bottom-up construction of effective theories for edge modes in more general gravitational settings using (or generalizing) existing EFT techniques.

Finally, it would be interesting to explore how the structures uncovered in this section might change for other proposals of modeling an observer, including inserting them as a particle worldline \cite{Anninos:2011af,Chandrasekaran:2022cip,Loganayagam:2023pfb,Kolchmeyer:2024fly}, or as an explicit boundary \cite{Banihashemi:2022jys,Anninos:2024wpy,Silverstein:2024xnr}. If edge modes should be thought of as Goldstones associated with spontaneous breaking of diffeomorphisms, a subset of the edge degrees of freedom might be killed in the latter scenario, where a part of diffeomorphisms are broken {\it explicitly} rather than spontaneously.  It has been argued that the inclusion of an observer as a particle worldline eliminates (the dimension dependence of) Polchinski's phase \cite{Maldacena:2024spf}.

\paragraph{When $d=3$} In this case, the putative edge theory resides on $S^2$. According to the formula \eqref{eq:massdim}, the tachyonic vector $A_\mu$ has $\Delta =2$, the two conformal scalars $\phi^a$ have  $\Delta =2$, and the massless scalar $\chi$ has  $\Delta =1$. These $\mathfrak{so}(1,1)$-weights correspond to discrete series representations of $SO(1,2)$ \cite{Sun:2021thf}. The Hilbert space realization of discrete series representations of $SO(1,2)$ is investigated in \cite{Anninos:2023lin}. Incidentally, $\mathcal{N}=2$ $dS_2$ supergravity includes one bosonic degree of freedom with $\Delta =1$ and another with  $\Delta =2$ \cite{Anninos:2023exn};\footnote{We thank Dio Anninos for pointing this out to us.} this curious observation may provide further inspiration for a non-linear completion of \eqref{eq:gravedgeaction}.

\paragraph{When $d=2$} While we have been focusing on $d \geq 3$, the topological nature of $dS_3$ gravity makes it a natural arena for an in-depth investigation of gravitational edge modes. In this case, the entire 1-loop Einstein gravity partition function on $S^3$ simplifies to its non-kinematic contributions \cite{Anninos:2020hfj}: 
\begin{align}\label{eq:3DZPI} 
		Z_{\rm PI} =\frac{ i^{5} }{\text{Vol}\left(SO(4) \right)_c } \left( \frac{16\pi^2 G_N} {\text{Vol}(S^{1}) \ell_\text{dS}}\right)^{3} (2\pi)^3 = (2\pi i)^5 \left( \frac{A}{4G_N}\right)^{-3} \;, 
\end{align} 
where we restored the dS length $\ell_\text{dS}$ and denoted the horizon area $A= 2\pi \ell_\text{dS}$. In \cite{Anninos:2020hfj}, an all-loop expression for the $S^3$ partition function of Einstein gravity formulated as a $SU(2) \times SU(2)$ Chern-Simons theory was obtained,\footnote{See \cite{Carlip:1992wg,Guadagnini:1994ahx,Castro:2011xb} for earlier work on applying the Chern-Simons formulation of gravity to calculate Euclidean $\Lambda>0$ partition functions.} with its first subleading term in the weak coupling expansion shown to agree with \eqref{eq:3DZPI}. From the perspective of Chern-Simons theory, it is natural to interpret \eqref{eq:3DZPI} (including Polchinski's phase $i^{5}$, as pointed out in \cite{Anninos:2020hfj,Castro:2023dxp}) as arising entirely from an edge mode contribution. Inspired by work on topological entanglement entropy \cite{Kitaev:2005dm,Levin:2006zz}, \cite{Anninos:2021ihe} investigated the hypothesis that (the all-loop version of) \eqref{eq:3DZPI} originates from edge modes by studying complexified Chern-Simons edge modes in both Euclidean and Lorentzian signatures. An important open question is to extend this to the case of gravity, whose putative edge-mode theory is a chiral $SL(2,\mathbb{C})$ WZW theory. 



\section{Discussions}\label{sec:discussion}

\subsection{$Z_{\rm edge}$ for (partially) massless higher-spin gauge fields}\label{ref:HS}

A massless rank-$s$ symmetric tensor gauge field $\phi_{\mu_1 \cdots \mu_s}$ has a quadratic action that is invariant under the abelian gauge symmetries \cite{Fronsdal:1978rb}
\begin{align}
	\phi_{\mu_1 \cdots \mu_s} \to \phi_{\mu_1 \cdots \mu_s} + \nabla_{(\mu_1} \xi_{\mu_2 \cdots \mu_s )}
\end{align}
with the spin-$(s-1)$ gauge parameters $\xi_{\mu_1\cdots \mu_{s-1}}$. The global part of the gauge symmetries corresponds to the spin-$(s-1)$ Killing tensors satisfying
\begin{align}\label{eq:globalHS}
	 \nabla_{(\mu_1} {\bar \xi}_{\mu_2 \cdots \mu_s )} =0 \; ,
\end{align}
which carries the representation $\rho^{d+2}_{s-1,s-1}$. For each $s\geq 1$, its $S^{d+1}$ partition function is \cite{Law:2020cpj} 
\begin{align}
	Z_{\rm PI}= \frac{i^{P_s} }{ \text{Vol}(G)_c } \left(\frac{8\pi G_N \left(d+2s-2 \right) \left(d+2s-4 \right) }{\text{Vol}(S^{d-1})}\right)^{\frac{N_{s-1}^{\rm KT}}{2}} \frac{\det\nolimits'_{-1}\left| -\nabla^2_{(s-1)}-\lambda_{s-1,s-1}\right|^{\frac12} }{\det\nolimits'_{-1}\left| -\nabla^2_{(s)}-\lambda_{s-2,s}\right|^{\frac12} }
\end{align}
In the non-determinant contributions, the overall phase $i^{P_s} $ is the higher-spin generalization of Polchinski's phase; explicitly, $P_s=N_{s-2}^{\rm CKT}+N_{s-1}^{\rm CKT}-N_{s-1}^{\rm KT}$ where $N^{\rm CKT}_{j}\equiv D^{d+3}_{j,j}$ and $N^{\rm KT}_j \equiv D^{d+2}_{j,j}$ are the numbers of spin-$j$ conformal Killing tensors (CKTs) and Killing tensors (KTs) on $S^{d+1}$ respectively. As in the YM and gravity, other factors arise from the global part of higher-spin gauge symmetries, and are intimately related to the interaction structure of the parent theory. Since higher-spin gravity generically involves infinite towers of gauge fields with different spins \cite{Vasiliev:1990en,Vasiliev:2003ev,Bekaert:2004qos},\footnote{One exception is in $d=2$, where higher-spin gravity involves a finite tower of spin $2,3,\dots , n$ and can be formulated as a $SU(n)\times SU(n)$ Chern-Simons theory in the Euclidean signature. This case has been explored in \cite{Anninos:2020hfj}.} a challenge arises in defining the volume of the higher-spin group (which involves a product of $\text{Vol}(G)_c $) itself, an issue we will not attempt to address in this work.

Once again we have $Z_{\rm PI} = Z_{\rm bulk}Z_{\rm edge} $, with \cite{Anninos:2020hfj}
\begin{align}\label{eq:HSZedge}
	\log Z_{\rm edge}= \log Z^\text{non-char}_{\rm edge}-\int_0^\infty \frac{dt}{2t}\frac{1+q}{1-q}\left[D_{s-1}^{d+2}\frac{q^{s+d-3}+q^{1-s}}{(1-q)^{d-2}}-D_{s-2}^{d+2}\frac{q^{s+d-2}+q^{-s}}{(1-q)^{d-2}}\right]_+ \;.
\end{align}
Here we recall $q\equiv e^{-t}$ and the flipping notation \eqref{eq:flipnotation}, and have denoted
\begin{align}\label{eq:HSZedgenonchar}
	Z^\text{non-char}_{\rm edge} = \frac{i^{P_s} }{ \text{Vol}(G)_c } \left(\frac{32\pi^3 G_N }{\text{Vol}(S^{d-1})}\right)^{\frac{N_{s-1}^{\rm KT}}{2}}\;. 
\end{align}
To determine the $\mathfrak{so}(d)$ content of \eqref{eq:HSZedge}, one approach is to use our branching rule method, generalized to the higher-spin case in section \ref{sec:HSPI}. However, inspired by both Yang-Mills theory and linearized gravity, we can take an alternative approach. Specifically, we infer the $\mathfrak{so}(d)$ content of \eqref{eq:HSZedge} by {\it requiring} the nonlinear realization of the abelian global spin-$(s-1)$ symmetries \eqref{eq:globalHS}.

Concretely, applying the $\mathfrak{so}(N+1)\to \mathfrak{so}(N)$ branching rule twice, one can deduce
\begin{align}\label{eq:Killingbranch}
	\rho^{d+2}_{s-1,s-1} =\bigoplus_{i=0}^{s-1} \bigoplus_{j=0}^i \left(i-j+1 \right) \rho^{d}_{i,j} \; . 
\end{align}
To achieve a nonlinear realization of the global spin-$(s-1)$ symmetries \eqref{eq:globalHS}, we include all spin-$j$ fields $\chi_{\mu_1 \cdots \mu_j}$ whose quadratic action is invariant under the shift-symmetry
\begin{align}\label{eq:HSshift}
	\chi_{\mu_1 \cdots \mu_j} \to \chi_{\mu_1 \cdots \mu_j} + f_{i,\mu_1 \cdots \mu_j}
\end{align}
 with multiplicities given in \eqref{eq:Killingbranch}. Here $ f_{i,\mu_1 \cdots \mu_j}$ is a spin-$j$ STT harmonic with angular momentum $i$ carrying the representation $\rho^{d}_{i,j}$. These shift-symmetric fields have a level-$(i-j)$ shift symmetry in the terminology of \cite{Bonifacio:2018zex}. They have the $\mathfrak{so}(1,1)$ weights
\begin{align}
	\Delta = d-2 + i  \; , \qquad \bar\Delta = -i\; .
\end{align}
We claim that \eqref{eq:HSZedge} comprises the reciprocal of $S^{d-1}$ partition functions of these fields:
\begin{gather}\label{eq:ZedgeHSmassless}
	Z_{\rm edge}\stackrel{?}{=} Z^{\rm det}_{\rm edge}Z^\text{non-det}_{\rm edge} \; , \qquad 	Z^{\rm det}_{\rm edge}=
	\prod_{i=0}^{s-1}\prod_{j=0}^i Z^{i-j+1 }_{ij}\;, \qquad Z_{ij}\equiv \det\nolimits'_{-1} \left| -\nabla_j^2 - \lambda_{i,j}\right|^{\frac{1}{2}} \; . 
\end{gather}
In the $S^{d-1}$ determinants, $\lambda_{i,j} = i(i+d-2)-j$ are the eigenvalues of the spin-$j$ STT Laplacians $-\nabla_j^2$. At these (effective) masses, the Laplace operators have zero modes, which are the signature of the shift symmetries \eqref{eq:HSshift}, and also a finite number of negative modes that have been Wick-rotated in field space. The non-determinant contribution $Z^\text{non-det}_{\rm edge}$ should be equal to \eqref{eq:HSZedgenonchar} up to the spin-$s$ analog of \eqref{eq:ZedgeG}. As a simple test for \eqref{eq:ZedgeHSmassless}, we take the appropriate limits of \eqref{eq:charPIstart} and write
\begin{align}
	\log Z_{ij} &= \log Z^{\rm naive}_{ij}+\log Z^{\rm flip}_{ij} \nn\\
	\log Z^{\rm naive}_{ij} &= -\int_0^\infty \frac{dt}{2t}  \left( q^{-i}+ q^{d-2+i}\right)  \sum_{l=-1}^\infty D_{l,j}^d q^l \nn\\
	\log Z^{\rm flip}_{ij} & =\int_0^\infty \frac{dt}{2t}\left[   1+q^{d-2+2i}+\sum_{l=-1}^{i-1} \left(q^{l-i}-q^{i-l} \right)  \right]  \;.
\end{align}
For $d\geq 4$, it can be verified for any $s\geq 1$ that the sum of $\log Z^{\rm naive}_{ij}$ matches precisely the integral in \eqref{eq:HSZedge} without $\left[ \cdots \right] _+$. For $d=3$, all contributions with $j\geq 2$ in \eqref{eq:ZedgeHSmassless} vanish. For $s=1,2$, the sum of $\log Z^{\rm naive}_{ij}$ exactly reproduces the integral  in \eqref{eq:HSZedge} without $\left[ \cdots \right] _+$, while for any $s\geq 3$, it agrees up to finite corrections. Thus, the kinematic parts on both sides of \eqref{eq:ZedgeHSmassless} must agree.

\paragraph{Partially massless fields}

In general, there exists partially massless (PM) fields \cite{Deser:1983tm,DESER1984396,Higuchi:1986py,Brink:2000ag,Deser:2001pe,Deser:2001us,Deser:2001wx,Deser:2001xr,Zinoviev:2001dt,Hinterbichler:2016fgl} in dS space. A spin-$s$ depth-$p$ PM field has a gauge symmetry parametrized by a spin-$p$ gauge parameter:
\begin{align}
	\phi_{\mu_1 \cdots \mu_s} \sim \phi_{\mu_1 \cdots \mu_s} + \nabla_{(\mu_{s-p}} \cdots \nabla_{\mu_s} \xi_{\mu_1 \cdots \mu_p)} + \cdots \; ,
\end{align}
where $\cdots$ stands for terms with fewer derivatives. The global part of the gauge transformations are parametrized by spin-$p$ tensors carrying the representation $\rho^{d+2}_{s-1,p}$. Their $S^{d+1}$ partition functions also exhibit a bulk-edge split, with their $Z_{\rm edge}$ found to be \cite{Anninos:2020hfj}
\begin{align}\label{eq:PMZedge}
	\log Z_{\rm edge}=\log Z^\text{non-char}_{\rm edge}- \int_0^\infty \frac{dt}{2t}\frac{1+q}{1-q}\left[D_{s-1}^{d+2}\frac{q^{d-2}+q^{-p}}{(1-q)^{d-2}}-D_{p-1}^{d+2}\frac{q^{s+d-2}+q^{-s}}{(1-q)^{d-2}}\right]_+ \; .
\end{align}
Here $Z^\text{non-char}_{\rm edge}$ is the PM analog of \eqref{eq:HSZedgenonchar} that we will not display. Curiously, for non-maximal depth $p<s-1$, a preliminary calculation shows that \eqref{eq:PMZedge} {\it cannot} be accounted for solely by shift-symmetric lower-spin fields non-linearly realizing the abelian global PM  symmetries. For example, the branching rule analysis yields the following edge field content on $S^{d-1}$ for a spin-2 depth-0 PM field on $S^{d+1}$:
\begin{center}
	\renewcommand{\arraystretch}{1.5} 
	\setlength{\tabcolsep}{10pt} 
	\begin{tabular}{|c|c|c|c|}
		\hline
		Spin & $\mathfrak{so}(1,1)$-weights & $M^2$ & Multiplicity \\ \hline
		\multicolumn{1}{|c|}{\makebox[2.5cm]{0}} & 
		\multicolumn{1}{c|}{\makebox[4.5cm]{$\Delta = d-2$, $\bar{\Delta} = 0$}} & 
		\multicolumn{1}{c|}{\makebox[3cm]{$0$}} & 
		\multicolumn{1}{c|}{\makebox[3cm]{2}} \\ \hline
		\multicolumn{1}{|c|}{\makebox[2.5cm]{0}} & 
		\multicolumn{1}{c|}{\makebox[4.5cm]{$\Delta = d-1$, $\bar{\Delta} = -1$}} & 
		\multicolumn{1}{c|}{\makebox[3cm]{$-(d-1)$}} & 
		\multicolumn{1}{c|}{\makebox[3cm]{1}} \\ \hline
		\multicolumn{1}{|c|}{\makebox[2.5cm]{0}} & 
		\multicolumn{1}{c|}{\makebox[4.5cm]{$\Delta = d-3$, $\bar{\Delta} = 1$}} & 
		\multicolumn{1}{c|}{\makebox[3cm]{$d-3$}} & 
		\multicolumn{1}{c|}{\makebox[3cm]{1}} \\ \hline
		\multicolumn{1}{|c|}{\makebox[2.5cm]{1}} & 
		\multicolumn{1}{c|}{\makebox[4.5cm]{$\Delta = d-2$, $\bar{\Delta} = 0$}} & 
		\multicolumn{1}{c|}{\makebox[3cm]{$-(d-3)$}} & 
		\multicolumn{1}{c|}{\makebox[3cm]{1}} \\ \hline
	\end{tabular}
\end{center}
The two massless scalars and the conformal scalar in the first two rows  nonlinearly realize 
the global part of the PM gauge symmetry, which carries the $\mathfrak{so}(d+2)$ irrep
\begin{align}
	\rho^{d+2}_{1,0} =2 \rho^{d}_{0,0}  \oplus \rho^{d}_{1,0} \; .
\end{align}
In addition, we have a tachyonic vector\footnote{The vector is tachyonic in the sense that the Proca mass squared $M^2$ is negative. If one takes into account the contribution from the curvature of $S^{d-1}$, the total effective mass $M^2+d-2=1$ remains positive.} and a massive scalar. 

We conclude this section by cataloging the edge theories identified in previous and this work:
\begin{table}[H]
	\centering
	\renewcommand{\arraystretch}{1.5} 
	\begin{tabular}{|c|c|c|c|}
		\hline
		\multicolumn{1}{|c|}{} & \multicolumn{1}{c|}{$p$-form} & \multicolumn{1}{c|}{Rank-$s$ totally symmetric tensors} & \multicolumn{1}{c|}{Mixed-symmetry} \\ 
		\hline
		\multicolumn{1}{|c|}{Massive} & \multicolumn{1}{c|}{Massive $(p-1)$-form} & \multicolumn{1}{c|}{Massive spin $\leq s-1$} & \multicolumn{1}{c|}{??} \\ 
		\hline
		\multicolumn{1}{|c|}{PM} & \multicolumn{1}{c|}{\cellcolor{gray!25}} & \multicolumn{1}{c|}{Shift-symmetric + massive spin $\leq s-1$} & \multicolumn{1}{c|}{??} \\ 
		\hline
		\multicolumn{1}{|c|}{Massless} & \multicolumn{1}{c|}{Massless $(p-1)$-form} & \multicolumn{1}{c|}{Shift-symmetric spin $\leq s-1$} & \multicolumn{1}{c|}{\cellcolor{gray!25}}  \\ 
		\hline
	\end{tabular}
\end{table}
This table encapsulates a wealth of structures and phenomena that invite further investigation. The first column summarizes the edge mode content for $p$-form theories previously obtained in \cite{Anninos:2020hfj,David:2021wrw,RiosFukelman:2023mgq,Mukherjee:2023ihb,Ball:2024hqe,Ball:2024xhf}. The second column summarizes the new detailed edge mode spectra for symmetric tensor fields obtained in this work.  In the massless case, it is particularly intriguing that totally symmetric and anti-symmetric tensor gauge fields give rise to two distinct classes of edge theories: shift-symmetric and gauge theories. A natural next step is to extend this table to the last column, namely the edge spectra for mixed-symmetry tensor fields that exist in $d\geq 4$, by applying the methods developed here. We have also not treated fermionic fields, which would require extending the $SO(d+2)\to U(1)\times SO(d)$ branching rule analysis.

\subsection{From $S^{d+1}$ to $S^{d+1}_\beta$}\label{sec:conical}

Throughout this work, we have restricted the periodicity of the thermal circle to $2\pi$. To discuss temperature dependence or Renyi entropy \cite{Klebanov:2011uf,Nishioka:2013haa}, one would like to conically deform the round sphere $S^{d+1}$ into $S^{d+1}_\beta$ with $\beta\neq 2\pi$,\footnote{One might worry that the background $S^{d+1}_\beta$ with $\beta \neq 2\pi$ deviates from being on-shell due to the presence of a conical singularity. Relatedly, because of the coupling to the Ricci tensor, the kinetic operators for spinning fields might contain $\delta$-function-like contributions supported at the origin. Although these concerns are valid, addressing them is beyond the scope of this work and is deferred to future studies.} which explicitly breaks the original $SO(d+2)$ isometry group to $U(1)\times SO(d)$. A common approach to compute 1-loop partition functions on such spaces is to directly obtain the spectra of Laplacians on $S^{d+1}_\beta$, as demonstrated for spin $s=0,1$ in \cite{Fursaev:1993hm,DeNardo:1996kp,DeNardo:1997gn}. However, this becomes cumbersome for higher-spin fields (see \cite{Beccaria:2017lcz} for a proposal for symmetric spin-$s$ totally symmetric tensors on $S_\beta^4$, generalizing the $s\leq 4$ results obtained there). $p$-forms on conically deformed spheres are considered in \cite{Dowker:2005eg,Dowker:2006bj,Dowker:2017flz,David:2021wrw,Dowker:2024cwy}.

The $SO(d+2)\to U(1)\times SO(d)$ branching rule analysis in section \ref{ref:branching} offers an elegant way to uplift results on the round $S^{d+1}$ to $S^{d+1}_\beta$, bypassing the process of directly solving the spectrum of Laplacians on $S^{d+1}_\beta$. As an example, consider a free massive scalar on $S^{d+1}_\beta$. In the generating function \eqref{scalar decom}, the recipe to uplift from the round $S^{d+1}$ to $S_\beta^{d+1}$ is to simply replace
\begin{align}\label{eq:deficitreplace}
	q^{|k|} \to q_\beta^{|k|}  
\end{align}
for the factors carrying the $\mathfrak{u}(1)$-weights $k$. At the level of the character integral, we then identify
\begin{align}\label{eq:betaiden}
	q_\beta = e^{-\frac{2\pi}{\beta}t} \; . 
\end{align}
Doing this, instead of \eqref{eq:massscalarPI}, we have
\begin{align}\label{eq:conPI}
	\log Z_{\rm PI} \left[ S^{d+1}_\beta\right]= \int_0^\infty \frac{dt}{2t}\frac{1+e^{-\frac{2\pi}{\beta}t}}{1-e^{-\frac{2\pi}{\beta}t}}\chi(t) \; . 
\end{align}
As a consistency check, one can start from the heat kernel representation
\begin{align}
	\log Z_{\rm PI} \left[ S^{d+1}_\beta\right]= \int_0^\infty \frac{d\tau}{2\tau} \sum_{k\in\mathbb{Z}}\sum_{l=0}^\infty D^d_l \, e^{-\left( \lambda^\beta_{kl} + M^2\right) \tau}
\end{align}
with the eigenvalues for the scalar Laplacian on $S^{d+1}_\beta$ found in \cite{DeNardo:1996kp}
\begin{align}
	\lambda^\beta_{kl} = \left(\frac{2\pi |k|}{\beta}+l\right)\left(\frac{2\pi |k|}{\beta}+l+d\right) \; , \qquad k\in \mathbb{Z} \; , \quad l=0,1,2,\dots \; ,
\end{align}
and follow the derivation outlined in appendix \ref{sec:heatkernel} leading to \eqref{eq:charPIstart}. Summing over $k\in \mathbb{Z}$ and $l\geq 0$ reproduces \eqref{eq:conPI} exactly. In \cite{Anninos:2020hfj}, the formula \eqref{eq:conPI} was obtained as a quasicanonical bulk partition function, independently defined in the Lorentzian signature, but without a Euclidean path integral interpretation for $\beta \neq 2\pi$. Our analysis here shows that this can indeed be thought of as a Euclidean path integral on $S^{d+1}_\beta$.

As opposed to solving for the spectrum of Laplacians on $S^{d+1}_\beta$, this uplifting approach is completely straightforward to generalize to higher-spin fields. Their 1-loop $S^{d+1}_\beta$ partition functions still have the bulk-edge split
\begin{gather}
	\log Z_\text{PI}\left[ S^{d+1}_\beta\right]=\log Z_\text{bulk} (\beta)+\log Z_\text{edge} (\beta)\;. 
\end{gather}
Since the bulk part of the $\mathfrak{so}(d+2)$-irreps generating functions is a product of a $\mathfrak{u}(1)$ and a $\mathfrak{so}(d)$ part, we always have 
\begin{gather}
	\log Z_\text{bulk} (\beta)=\int_0^\infty \frac{dt}{2t}\frac{1+e^{-\frac{2\pi}{\beta}t}}{1-e^{-\frac{2\pi}{\beta}t}}\chi(t) \; .
\end{gather}
The edge partition functions, in contrast, exhibit more intriguing features. By applying the lifting procedure \eqref{eq:deficitreplace} to \eqref{eq:vecedgesector} and \eqref{eq:2Vedge}, one can directly compute the edge partition functions for the massive spin-1 and spin-2 fields:
\begin{align}\label{eq:betaspin12}
	\log Z^{s=1}_\text{edge}(\beta)&=- \int_0^\infty \frac{dt}{2t} \frac{1+q}{1-q}  \frac{q^{\Delta-1} +q^{\bar{\Delta}-1}}{\left(1-q \right)^{d-2} } \nn\\
		\log Z^{s=2}_\text{edge}(\beta)
	&=- \int_0^\infty \frac{dt}{2t} \frac{1+q}{1-q} \left[ d \frac{q^{\Delta-1} +q^{\bar{\Delta}-1}}{\left(1-q \right)^{d-2} } +2 q_\beta \frac{q^{\Delta-2} +q^{\bar{\Delta}-2}}{\left(1-q \right)^{d-2} }  \right] \;. 
\end{align}
The spin-1 result is independent of $\beta$, as can be verified directly using the spectrum of the vector Laplacian on $S^{d+1}_\beta$ derived in \cite{DeNardo:1996kp}. For $d=3$, the spin-2 result can be obtained from the spectra provided in \cite{Beccaria:2017lcz}. In both cases, the eigenvalue sum requires an appropriate longitudinal mode extension, analogous to the $L=-1$ extension in the $\beta=2\pi$ case.

\paragraph{Photons}

The massless case requires additional care due to the presence of non-kinematic contributions. For instance, in the Maxwell partition function \eqref{eq:MaxPIheat}, the group volume factor differs from \eqref{eq:u1groupvolume} and is given by
\begin{align}\label{eq:groupbeta}
	\frac{1}{{\rm Vol}\left( U(1)\right)_{\rm PI}} \equiv \frac{ \mathrm{g}}{ \sqrt{2\pi\text{Vol}\left( S_\beta^{d+1}\right) }} =\frac{ \mathrm{g}}{ \sqrt{\beta \text{Vol}\left( S^{d+1}\right) }} \; , 
\end{align}
which modifies the original result \eqref{eq:Maxedgecomp} to
\begin{align}
	Z^{\rm U(1)}_\text{edge}\left( S_\beta^{d+1}\right) =\frac{ \mathrm{g}}{ \sqrt{\beta\text{Vol}\left( S^{d-1}\right) }}  \det\nolimits' \left( -\nabla_0^2\right)^{\frac{1}{2} } = \sqrt{\frac{2\pi}{\beta}}\frac{1}{Z_{\rm compact}(S^{d-1})}\;. 
\end{align}
 In \cite{David:2021wrw}, it was conjectured that $Z_{\rm edge}$ for massless $p$-form gauge theories is independent of  $\beta$. However, the group volume factor \eqref{eq:groupbeta}, which introduces $\beta$-dependence, was not taken into account in \cite{David:2021wrw}. Thus, the conjecture in \cite{David:2021wrw} should be understood as applying only to the kinematic contributions to $Z_{\rm edge}$, which can presumably be (dis)proved using our methods.

\paragraph{Gravitons}

The case for linearized gravity is even more intricate, as $\beta\neq2\pi$ breaks the isometry group from $SO(d+2)$ to $U(1)\times SO(d)$. This reduction in symmetry affects the dimension of the group volume factor \eqref{eq:ZPIvolgrav} and is also reflected in the determinant part of $Z_{\rm edge}$, which takes the form 
\begin{align}
	Z^\text{det}_\text{edge} (\beta) = \det\nolimits'_{-1} \left| -\nabla_1^2-\left( d-2\right)  \right|^{\frac12}  \det\nolimits\left| -\nabla_0^2+\left( d-2+\frac{2\pi}{\beta}\right) \left( \frac{2\pi}{\beta}-2\right) \right| \det\nolimits' \left( -\nabla_0^2 \right)^{\frac12} 
\end{align}
for general $\beta$. The second factor is the uplifted version of the conformal scalar determinant in \eqref{eq:Zedgegravresult}, where zero modes get lifted when $\beta\neq 2\pi$. This suggests that the edge theory would nonlinearly realize $U(1)\times SO(d)$ in this case. Moreover, the number of negative modes can vary with $\beta$. A more comprehensive analysis of the full graviton partition function $S^{d+1}_\beta$ is required to fully understand the $\beta$-dependence of these non-kinematic contributions.

\subsection{Outlook}\label{sec:outlook}

\paragraph{Generalizing Maxwell}

As a step towards understanding the {\it dynamical} origin of $Z_{\rm edge}$ in linearized gravity and higher-spin gauge theories,  it is natural to consider generalizing existing edge mode calculations for Maxwell theory. The results presented in this paper provide hints on how such calculations might be formulated, at least in the context of the dS static patch. For instance, drawing inspiration from lattice gauge theory, \cite{Donnelly:2014fua,Donnelly:2015hxa} proposed the following edge path integral:
\begin{align}\label{eq:DW}
	Z^{\rm DW}_{\rm edge} = \int \mathcal{D} E_\perp \, e^{-S^{\rm on-shell}[E_\perp]}
\end{align}
which successfully reproduces the partition function of a compact scalar field on the co-dimension-2 surface. One notable assumption in \eqref{eq:DW} is that only $U(1)$-invariant field strengths are included in the path integral measure. This aligns with our Maxwell theory results, which indicate that $Z_{\rm edge}$ does not carry any $U(1)$ weights. However, our analysis suggests that the analogous calculation for linearized gravity and higher-spin fields might require incorporating field strengths with non-trivial $U(1)$ weights. 

\paragraph{Other spacetimes}

Another interesting direction is to generalize our considerations to other spacetime backgrounds. For instance, the character formulas for 1-loop $S^{d+1}$ partition functions and their bulk-edge splits are closely mirrored in $EAdS_{d+1}$ \cite{Sun:2020ame}, relevant for thermal interpretations in Rindler-AdS$_{d+1}$. Given the structural similarities between the $S^{d+1}$ and $EAdS_{d+1}$ formulas, our results might immediately inform the field contents of $Z_{\rm edge}$ in the latter. 

For general black holes, the character formula for $Z_{\rm PI}$ is equivalent to the Denef-Hartnoll-Sachdev (DHS) formula \cite{Denef:2009kn,Law:2022zdq} applied to the dS static patch. The DHS derivation involves the Euclidean continuations of QNMs in Lorentzian black hole geometries, and $Z_{\rm edge}$ arises from QNMs that fail to Wick-rotate into smooth eigenfunctions of the Euclidean black hole Laplacian with low Matsubara frequencies \cite{Castro:2017mfj,Keeler:2018lza,Keeler:2019wsx,Grewal:2022hlo}. One might thus wonder if $Z_{\rm edge}$ can be obtained as a kind of dimension reduction of some would-be configurations in the original Euclidean black hole path integral. It is worth noting that in the rotating BTZ case, $Z_{\rm edge}$ plays a crucial role in capturing the non-local $T^\frac{3}{2}$ correction to the graviton partition function in the $T\to 0$ limit \cite{Kapec:2024zdj}. Finally, a key distinction between black hole and dS horizons is that the latter is observer-dependent. If part of $Z^{\rm dS}_{\rm edge}$ of gravity reflects observer degrees of freedom as discussed at the end of section \ref{sec:gravZedgecan}, there might be qualitative differences in the field content of $Z^{\rm BH}_{\rm edge}$ for a generic black hole. Evidence for such differences already appears in the static Nariai black hole \cite{Law:2025yec}.

\paragraph{Adding interactions}

Although shift-symmetric fields generally do not furnish UIRs of $SO(1,d+1)$,\footnote{From the relation \eqref{eq:massdim} between masses and $\mathfrak{so}(1,1)$ weights, shift-symmetric scalars appear to correspond to the exceptional series I representations \cite{Sun:2021thf} or discrete series in 2D. However, constructing such a Hilbert space realization is highly non-trivial, as discussed in \cite{Anninos:2023lin}.} their presence in $Z_{\rm edge}$ of gravity and higher-spin gauge fields underscores their physical relevance, and poses an interesting problem of constructing EFTs involving fields with different species that nonlinearly realize the $SO(d+2)$ or global higher-spin symmetries. Understanding the interaction structures of these theories might elucidate the physics of $Z_{\rm edge}$.

One way to proceed, in the gravity case for example, is to write down a general ansatz for the higher terms in \eqref{eq:Sedgeform} and the symmetry deformations \eqref{eq:fullinvariance}. The requirement \eqref{eq:fullinvariance} then leads to an infinite set of equations
\begin{gather}
	\delta^{(0)}_E S^{(2)}[\Psi] =  0 \nn\\
	\delta^{(1)}_E S^{(2)}[\Psi] +\delta^{(0)}_E S^{(3)}[\Psi]=  0 \nn\\
	\delta^{(2)}_E S^{(2)}[\Psi] +\delta^{(1)}_E S^{(3)}[\Psi]+\delta^{(0)}_E S^{(4)}[\Psi] =  0 \nn\\
	\vdots 
\end{gather}
The requirement that the deformed symmetries form the $\mathfrak{so}(d+2)$ algebra also leads to a similar set of equations. One can then solve these equations order by order.

Another possible approach is the coset construction \cite{Coleman:1969sm,Callan:1969sn,Volkov:1973vd} (see \cite{zumino1970effective,Ogievetsky1974} for a review), which requires us to identify the correct symmetry breaking pattern. This method has been applied to derive the EFT for a gravitating brane embedded in Minkowski space \cite{Clark:2006pd}, where the resulting brane theory includes vector degrees of freedom.

\section*{Acknowledgments} 

It is a great pleasure to thank Amr Ahmadain, Alejandra Castro, Bruno De Luca, Frederik Denef, Alexander Frenkel, Kurt Hinterbichler, Austin Joyce, Michael Landry, Juan Maldacena, Eva Silverstein, Gonzalo Torroba and Gabriel Wong for stimulating conversations, and especially Dionysios Anninos, Adam Ball, Luca Ciambelli, Klaas Parmentier and Zimo Sun for useful discussions and comments on the draft. AL was supported by NSF Grant PHY-2310429, the Stanford Science Fellowship, and a Simons Investigator Award. This work was performed in part at Aspen Center for Physics, which is supported by National Science Foundation grant PHY-2210452.

\begin{appendix}


\section{$SO(1,d+1)$ and its unitary irreducible representations}\label{sec:dSUIR}



$dS_{d+1}$ has a $SO(1,d+1)$ isometry whose algebra is generated by $L_{AB}$, where $A,B =0,1,\dots, d+1$, which satisfy the commutation relation
\begin{align}\label{eq:dS comm}
	[L_{AB},L_{CD}]=\eta_{BC} L_{AD}-\eta_{BD} L_{AC}+\eta_{AD} L_{BC}-\eta_{AC} L_{BD}\; .
\end{align}
Taking the linear combinations
\begin{align}\label{eq:congen}
	M_{ij} =L_{ij}\; , \qquad P_i =L_{d+1, i}+ L_{0, i} \; , \qquad K_i = L_{d+1, i}-L_{0, i}\; , \qquad D=L_{0,d+1}\;,
\end{align}
with $i,j=1,\dots ,d$, \eqref{eq:dS comm} can be recast into the Euclidean conformal algebra $\mathfrak{so}(1,d+1)$ on $\mathbb{R}^d$:
\begin{align}\label{eq:conformal algebra}
	&\qquad \qquad [M_{ij},M_{kl}]=\delta_{jk} M_{il}-\delta_{jl} M_{ik}+\delta_{il} M_{jk}-\delta_{ik} M_{jl}\nn\\
	&\qquad [M_{ij},P_k]=\delta_{jk}P_i-\delta_{ik}P_j \; , \qquad [M_{ij},K_k]=\delta_{jk}K_i-\delta_{ik}K_j\nn\\
	&[P_i,K_j]=-2D\delta_{ij}-2M_{ij} \; , \qquad 	[D,P_i]=P_i \; , \qquad  [D,K_i]=-K_i \;.
\end{align}
We refer the reader to \cite{Sun:2021thf} for a review on the representation theory of $SO(1,d+1)$. Here we present a quick survey for the dictionary between $SO(1, d+1)$ UIRs and quantum fields in $dS_{d+1}$ \cite{Basile:2016aen}. We focus on $d\geq 3$.

 A $SO(1,d+1)$ UIR $\mathcal{V}_{[\Delta, \mathbf{s}]}$ is labeled by an $\mathfrak{so}(d)$ highest weight $\mathbf{s}=[s_1,s_2,\dots,s_{\lfloor \frac{d}{2}\rfloor}]$ and a $\mathfrak{so}(1,1)$ weight $\Delta$. The former is nothing but the spin for the quantum field. We focus on scalars and symmetric tensors, i.e. those with $\mathbf{s}=[s,0,\dots,0]$ and $s\geq 0$ is an integer, whose mass is related to the $\mathfrak{so}(1,1)$ weight through 
\begin{align}\label{eq:massdim}
	M^2 \ell_\text{dS}^2=
	\begin{cases}
		\Delta (d-\Delta)\; , & s=0 \\
		(\Delta+s-2)(d+s-2-\Delta) \;, & s\geq 1 \\
	\end{cases}\;.
\end{align}

\paragraph{Principal series}

These UIRs describe massive fields in $dS_{d+1}$ that are heavy compared with the dS scale $\ell_\text{dS}$. For a spin-$s$ field, its conformal dimension and mass fall into the ranges
\begin{align}\label{eq:principaldelta}
	\Delta = \frac{d}{2}+ i\nu \; , \quad \nu \in \mathbb{R} \quad \Leftrightarrow \quad M \ell_\text{dS}\geq 
	\begin{cases}
		\frac{d}{2} &\text{ for } s=0 \\
		s+\frac{d}{2}-2 &\text{ for } s\geq 1
	\end{cases}\;.
\end{align}

\paragraph{Complementary series}

These UIRs describe massive fields in $dS_{d+1}$ that are light compared with the dS scale $\ell_\text{dS}$. For a light scalar,  it has
\begin{align}
	0 < \Delta <d \qquad \Leftrightarrow  \qquad 0<M\ell_\text{dS}<\frac{d}{2} \qquad  (s=0) \; .
\end{align}
For a light spin-$s$ field, the ranges are instead
\begin{align}\label{eq:compdelta}
	1<\Delta <d-1 \qquad \Leftrightarrow \qquad  \sqrt{(s-1)(s+d-3)}<M\ell_\text{dS}< s+\frac{d}{2}-2 \qquad (s\geq 1) \; .
\end{align}
The lower bound $\sqrt{(s-1)(s+d-3)}$ is known as the Higuchi bound \cite{Higuchi:1986py}.

\paragraph{Exceptional series II}

These UIRs only exist for $s\geq 1$ and describe the so-called partially massless (PM) particles \cite{Deser:1983tm,DESER1984396,Higuchi:1986py,Brink:2000ag,Deser:2001pe,Deser:2001us,Deser:2001wx,Deser:2001xr,Zinoviev:2001dt,Hinterbichler:2016fgl}.\footnote{When $d=3$, exceptional series II coincides with the discrete series \cite{Sun:2021thf}.
} These occur when $\mathfrak{so}(1,1)$ weight hits the exceptional points
\begin{align}
	\Delta = 1- p \qquad \Leftrightarrow \qquad m_{s,p}^2\ell_\text{dS}^2=(s-1-p)(d+s+p-3) \; , \qquad  p=0,1,\dots,s-1 \; ,
\end{align}
The local action describing these fields have the gauge symmetry that reads schematically
\begin{align}
	\phi_{\mu_1 \cdots \mu_s} \sim \phi_{\mu_1 \cdots \mu_s} + \nabla_{(\mu_{s-p}} \cdots \nabla_{\mu_s} \xi_{\mu_1 \cdots \mu_p)} + \cdots \; ,
\end{align}
where $\cdots$ stand for terms with fewer derivatives \cite{Hinterbichler:2016fgl}. We call the spin of the gauge parameter $p$ the ``depth" of the PM field. Note that the usual massless spin-$s$ field has the maximal depth $p=s-1$.



\section{$\mathfrak{so}(d)$ toolbox}\label{sec:sod}

In this appendix, we collect relevant facts about finite-dimensional irreducible representations (irreps) of $\mathfrak{so}(d)$. See for instance \cite{fulton_representation_2004} for a comprehensive review for the representation theory of $\mathfrak{so}(d)$. A finite-dimensional irrep $\rho^d_\mathbf{s}$ of $\mathfrak{so}(d)$ is labeled by a highest weight $\mathbf{s}=\left[ s_1, s_2 , \cdots , s_{r-1} ,  s_r \right] $, where $r=\lfloor \frac{d}{2}\rfloor$ and $s_i$ are integers satisfying
\begin{align}
	 s_1 \geq s_2 \geq \cdots \geq s_{r-1} \geq   |s_r| \geq 0 \; . 
\end{align}
When $d$ is odd, $s_r\geq 0$; when $d$ is even, $s_r$ can be positive or negative, distinguishing two chiral representations. Graphically $\rho^d_\mathbf{s}$ can be represented by a Young diagram:
\begin{center}
\begin{tikzpicture}
	\draw (0, 0) rectangle (6, 0.8);
	\node at (6.5, 0.4) {$s_1$};
	
	\draw (0, -0.8) rectangle (5, 0);
	\node at (5.5, -0.4) {$s_2$};
	
	\draw (0, -1.6) rectangle (2, -0.8);
	\node at (2.5, -1.2) {$\cdots$};
	
	\draw (0, -2.4) rectangle (2, -1.6);
	\node at (2.5, -2) {$s_{r-1}$};
	
	\draw (0, -3.2) rectangle (1, -2.4);
	\node at (1.5, -2.8) {$|s_r|$};
	
	\node at (3.5, -2) {$\vdots$};
	
\end{tikzpicture}
\end{center}
In this paper, we focus on 1- and 2-row represenations, denoting 1-row representations (which exist when $d\geq 2$) with \(l\) boxes by \(\rho^d_l\) and 2-row representations (which exist when $d\geq 4$) with \(l\) boxes in the first row and \(m\) in the second by \(\rho^d_{l,m}\). For generic $d$, their dimensions are
\begin{alignat}{2}\label{appeq:degen}
	D^d_l &\equiv {\rm dim} \, \rho^d_l =\frac{2l+d-2}{d-2}  \binom{l+d-3}{d-3}  && \quad (d\geq 3)\nn\\	
	D^d_{l,m} &\equiv {\rm dim} \, \rho^d_{l,m}= D^{d-2}_{m}\frac{(l-m+1)(l+m+d-3)(2l+d-2)(l+d-4)!}{(d-2)!(l+1)!} && \quad (d\geq 5)
	\; . 
\end{alignat}
We note that $\rho^d_{l,m=0}=\rho^d_l$ and $D^d_{l,m=0} = D^d_l $. 

%

\paragraph{When $d\leq 4$}

The degenerate cases of lower dimensions $d=2,3,4$ require a separate discussion. First, when $d=2$, irreducible $\mathfrak{so}(2)\cong \mathfrak{u}(1)$ representations are all 1-dimensional and labeled by $l\in \mathbb{Z}$, and therefore
\begin{align}
	D^2_l =1 \quad (l\in\mathbb{Z}) \; .
\end{align}
When $d=4$, because of the possible chiralities of the 2-row representations, the second formula \eqref{appeq:degen} needs to be modified:
\begin{align}\label{eq:so24dim}
	 D^4_{l,m} =  (l+m+1)(l-m+1)  \quad (m= 0 , \pm 1 , \dots , \pm l )
	\; . 
\end{align}
To treat the case of $d=3$ (relevant for $dS_4$ or $S^4$) in our analysis, it is useful to formally define
\begin{align}\label{eq:so3reps}
	\rho^3_{l,s} = 
	\begin{cases}
		\rho^3_{l}  &\; , \quad s = 1 \; , \quad l\geq 1 \\
		\emptyset &\; , \quad s \geq 2 
	\end{cases}\; . 
\end{align}
The $s=1$ case can be understood as the would-be $\rho^3_{l,1}$ representation as being ``dualized" to become $\rho^3_{l}$; for $s\geq 2$, \eqref{eq:so3reps} is equivalent to the non-existence of such representations when $d=3$. Note that \eqref{eq:so3reps} is consistent with taking the $d\to3$ limit of \eqref{appeq:degen}:
\begin{align}\label{eq:so3degen}
	\lim_{d\to 3} D^d_{l,s} = 
	\begin{cases}
		2l+1 &\; , \quad s = 0 \; , \quad l\geq 0 \\
		2l+1 &\; , \quad s = 1 \; , \quad l\geq 1 \\
		0 &\; , \quad s \geq 2 
	\end{cases}\; . 
\end{align}

\subsection{Symmetric transverse traceless spherical harmonics on $S^{d-1}$}\label{sec:STSH}



Spin-$s$ symmetric transverse traceless (STT) spherical harmonics $f_{l,\mu_1 \cdots \mu_s}$ on a unit round $S^{d-1}$ with orbital angular momentum $l\geq s$ are eigenfunctions of STT Laplacians \cite{rubin1984eigenvalues,Higuchi:1986wu} 
\begin{gather}\label{STSH eq}
	-\nabla_s^2f_{l,\mu_1 \cdots \mu_s}=\lambda_{l,s} f_{l,\mu_1 \cdots \mu_s} \; , \qquad \lambda_{l,s}=l(l+d-2)-s \; ,  \nn\\
	\nabla^\lambda f_{l,\lambda \mu_1 \cdots \mu_{s-1}}=0 \; , \qquad f\indices{_{l,}_{\mu_1 \cdots \mu_{s-2}}_\lambda^\lambda}=0 \; .
\end{gather}
For $s=0$, both the transversality and tracelessness condition are trivial, and $f_{l}$ are the ordinary scalar spherical harmonics; for $s=1$, the tracelessness condition is trivial. When $d\geq 5$, the vector space of $f_{l,\mu_1 \cdots \mu_s}$ furnishes the irreducible $\mathfrak{so}(d)$-module $\rho^d_{l,s}$, and therefore $f_{l,\mu_1 \cdots \mu_s}$ has degeneracy $D^d_{l,s}$ defined in \eqref{appeq:degen}. When $d=4$, the vector space is instead a direct sum of the chiral pair $\rho^4_{l,s}\oplus \rho^4_{l,-s}$ and the degeneracy is $2D^4_{l,s}$ instead. When $d=3$, transverse vector spherical harmonics are constructed from the scalar spherical harmonics through $f_{l,\mu} \propto \epsilon_{\mu\nu} \partial^\nu f_l$, while $f_{l,\mu_1 \cdots \mu_s}$ do not exist for $s\geq 2$ \cite{rubin1984eigenvalues}. This is also reflected in \eqref{eq:so3reps} and \eqref{eq:so3degen}. 


Regarding the unit round $S^{d-1}$ as a hypersurface $X^2=1$ in an ambient $\mathbb{R}^d$, we can constuct the STT harmonics as 
degree-$l$ homogeneous ploynomials 
\begin{align}\label{appeq:STSHpoly}
	 Y_s^l (X, U ) =  E_{i_1 \cdots i_l, j_1 \cdots j_s}X^{i_1}\cdots X^{i_l} U^{j_1}\cdots U^{j_s}\; ,\qquad l\geq s \;, 
\end{align}
where 
$U^j$ is an auxiliary vector encoding the tensor indices. The tensor $E_{i_1 \cdots i_l, j_1 \cdots j_s}$ has a structure of the traceless Young diagram: 
\begin{center}
	\begin{tikzpicture}
		\draw (0,0) rectangle (5,0.8); 
		\node at (2.5, 0.4) {$l$}; 
		
		\draw (0,-0.8) rectangle (3,0); 
		\node at (1.5, -0.4) {$s$}; 
	\end{tikzpicture}
\end{center}
From the polynomial \eqref{appeq:STSHpoly}, one obtains the STT harmonics as the projection
\begin{align}
	f_{l,\mu_1 \cdots \mu_s} = \frac{\partial X^{i_1}}{\partial x^{\mu_1}} \cdots \frac{\partial X^{i_s}}{\partial x^{\mu_s}} Y_{s, i_1 \cdots i_s}^l (X) \Big|_{X^2=1}\;. 
\end{align}





\paragraph{Killing tensors}

A spin-$s$ Killing tensor (KT) is a totally symmetric traceless tensor satisfying the Killing equation
\begin{align}
	 \nabla_{(\mu_1}\epsilon_{\mu_2\cdots \mu_{s+1})}=0.
\end{align}
Taking the trace of this equation shows that they are divergenceless, while taking the divergence we recover \eqref{STSH eq} with $l=s$. In terms of $\mathbb{R}^d$ homogeneous polynomials, they are represented by
\begin{align}
	Y_s^s (X, U ) =  E_{i_1 \cdots i_s, j_1 \cdots j_s}X^{i_1}\cdots X^{i_l} U^{j_1}\cdots U^{j_s}\;. 
\end{align}

%

\subsection{Branching laws of $\mathfrak{so}(d+2)$- into $\mathfrak{u}(1)\oplus \mathfrak{so}(d)$-modules ($d\geq 3$)}\label{branch}

The branching laws of $\mathfrak{so}(d+2)$- into $\mathfrak{u}(1)\oplus \mathfrak{so}(d)$-modules have been derived in \cite{branch}. We focus on 2-row $\mathfrak{so}(d+2)$-representations $\rho^{d+2}_{n,s}$. For $d\geq 4$, the branching law says:
\begin{enumerate}
	\item For a given $\rho^{d+2}_{n,s}$,  $\rho^2_k \otimes \rho^d_{l,m}$ (recall that $p\in \mathbb{Z}$ and $m=0,\pm 1, \dots , \pm l$ when $d=4$) is contained in the former if and only if
	\begin{align}
		n\geq l \geq 0 \quad  \text{ and } \quad s\geq |m| \geq0\;. 
	\end{align}
	
	\item Defining the integers
	\begin{align}
		L_0 = n- \text{max}(s,l),\quad L_1= \min(s,l)-|m| \; ,
	\end{align}
	the multiplicity of $\rho^2_k \otimes \rho^d_{l,m}$ in the direct sum decomposition is the coefficient of $x^k$ in the power series expansion of in $x$ of
	\begin{align}
		\frac{x^{L_0+1}-x^{-L_0-1}}{x-x^{-1}}\frac{x^{L_1+1}-x^{-L_1-1}}{x-x^{-1}}\;. 
	\end{align}
	
\end{enumerate}
The rule above for $d=3$ is modified with the formal definitions \eqref{eq:so3reps}. The case of $d=2$ is special and the $\mathfrak{so}(4)\to \mathfrak{u}(1)\oplus \mathfrak{u}(1)$ branching rule will be derived in appendix \ref{sec:so4}.


\subsection{$\mathfrak{so}(4)$}\label{sec:so4}


The Lie algebra \(\mathfrak{so}(4)\) is special due to its isomorphism:
\begin{equation}\label{eq:so4sio}
	\mathfrak{so}(4) \cong \mathfrak{su}(2)_+ \oplus \mathfrak{su}(2)_- \; .
\end{equation}
Its irreducible representations can therefore be alternatively labeled by pairs of spins \((j_+, j_-)\), where \(j_\pm \) are non-negative integers or half-integers, related to the $\mathfrak{so}(4)$-weights \((l, m)\) by
\begin{equation}
	l = j_+ + j_-, \quad m = j_+ - j_- \; , \qquad j_\pm \in \frac12 \mathbb{Z}_{\geq 0}\;. 
\end{equation}
The dimension of the representation is $	(2j_++1) (2j_- +1)= (l + m + 1)(l - m + 1)$ as given in \eqref{eq:so24dim}. When $m\neq 0$, the two chiralities with $m>0$ or $m<0$ are distinguished by $ j_+ > j_- $ or $ j_+ <  j_- $.

%

Because of the isomorphism \eqref{eq:so4sio}, the characters of the \(\mathfrak{so}(4)\) representations can be written as products of \(\mathfrak{su}(2)\) characters:
\begin{equation}
	\chi^{\mathfrak{so}(4)}_{l,m}(\theta_+, \theta_-) = \chi^{\mathfrak{su}(2)}_{j_+}(\theta_+) \, \chi^{\mathfrak{su}(2)}_{j_-}(\theta_-) \; .
\end{equation}
To derive the branching rule into \(\mathfrak{u}(1) \oplus \mathfrak{u}(1)\) modules, we expand the characters:
\begin{equation}
	\chi^{\mathfrak{su}(2)}_{j}(\theta) = \frac{\sin\left((2j + 1)\frac{\theta}{2}\right)}{\sin\left(\frac{\theta}{2}\right)} = \sum_{m = -j}^{j} e^{i m \theta} \; .
\end{equation}
Thus,
\begin{equation}
	\chi^{\mathfrak{so}(4)}_{l,m}(\theta_+, \theta_-) = \sum_{m_+ = -j_+}^{j_+} \sum_{m_- = -j_-}^{j_-} e^{i m_+ \theta_+} e^{i m_- \theta_-} \; .
\end{equation}
Each term corresponds to a weight \((m_+, m_-)\) of a \(\mathfrak{u}(1) \oplus \mathfrak{u}(1)\)-module. We thus obtain the decomposition rule:
\begin{equation}
	\rho^4_{l,m} \to \bigoplus_{m_+ = -j_+}^{j_+} \bigoplus_{m_- = -j_-}^{j_-} \rho^2_{m_+} \otimes  \rho^2_{m_-}\; .
\end{equation}
This shows that each \(\mathfrak{u}(1) \oplus \mathfrak{u}(1)\)-weight \((m_+, m_-)\) appears once, and the total number of modules equals the dimension \(D^4_{l,m}\).


\section{Path integral measure and character integral manipulations}\label{sec:charintegral}

\subsection{Path integral measure}

1-loop $S^{d+1}$ partition functions for a bosonic field $\phi$ (which can be spinning or contain other labels associated with global/gauge symemtries) involves quadratic path integrals of the general form 
\begin{align}\label{appeq:PIstart}
	Z =\int \mathcal{D}\phi\,  e^{-\frac{1}{2\mathrm{g}^2}\int_{S^{d+1}}\phi (-\mathcal{Q})\phi}\; , 
\end{align}
where $\mathcal{Q}$ is a Laplace type operator, and $\mathrm{g}$ is a coupling constant in theories with gauge fields or nonlinearly realized symmetries. Suppose $\mathcal{Q}$  has mass dimension $p$ and an expansion in terms of orthonormal modes $f_{n}$, i.e.
\begin{align}
	\phi = \sum_n a_{n} f_{n}\; , \qquad \int_{S^{d+1}} f_{n}f_{m} =\delta_{nm} \; .
\end{align}
We define the path integral measure to be 
\begin{align}\label{EPImeasure}
	\mathcal{D}\phi \equiv \prod_n \frac{M^p}{\sqrt{2\pi}\mathrm{g}}d a_{n} \; .
\end{align}
Here $M$ is a parameter with dimension of mass, inserted so that \eqref{appeq:PIstart} remains dimensionless; throughout this paper we will set $M=1$ and one can restore it by dimension analysis. We can think of the measure \eqref{EPImeasure} as putting the following metric on the field space
\begin{align}\label{PI metric}
	ds^2=\frac{M^{2p}}{2\pi \mathrm{g}^2}\int_{S^{d+1}} (\delta\phi)^2=\frac{M^{2p}}{2\pi \mathrm{g}^2}\sum_n d a_n^2\;. 
\end{align}
We can see that the field $\phi$ satisfies the normalization condition
\begin{align}
	\int \mathcal{D}\phi \, e^{-\frac{1}{2\mathrm{g}^2}\int_{S^{d+1}}\phi \cdot \phi}=1\; , 
\end{align}
and that the Gaussian integration \eqref{appeq:PIstart} results in a determinant without any extra factor other than the dimensionful parameter $M$:
\begin{align}
	Z  = \int \mathcal{D}\phi\, e^{-S[\phi]} = \det(-\frac{\mathcal{Q}}{M^{2p}})^{-1/2} \; .
\end{align}

\subsection{From heat kernal to character integral}\label{sec:heatkernel}

The heat kernel representation of the 1-loop partition function for a massive spin-$s$ field on $S^{d+1}$ takes the form \cite{Anninos:2020hfj,Law:2020cpj}
\begin{align}
	\log Z_{\rm PI} = \int_0^\infty \frac{d\tau}{2\tau}\, e^{-\frac{\epsilon^2}{4\tau}} \, \Tr_{-1} e^{-\left( -\nabla_s^2+ M^2+M_s^2\right) \tau} \; ,
\end{align}
where $M_s^2 = s-(s-2)(s+d-2)$ and $M^2=\left(\Delta +s-2 \right) \left(\bar\Delta +s-2 \right) $.  A factor $e^{-\frac{\epsilon^2}{4\tau}} $ is inserted to regulate the UV divergence at $\tau=0$. We focus on $d\geq 3$. With the eigenvalues \eqref{STSH eq} and degeneracies \eqref{appeq:degen}, we write
\begin{align}\label{eq:PInext}
	\log Z_{\rm PI} = \int_0^\infty \frac{d\tau}{2\tau}\, e^{-\frac{\epsilon^2}{4\tau}-\nu^2 \tau}  \sum_{L=-1}^\infty D^{d+2}_{L,s} e^{-\left(\frac{d}{2}+L\right)^2 \tau}
\end{align}
where we have written $\Delta = \frac{d}{2}+i \nu$ with $\nu\in \mathbb{R}$, corresponding to the principal series \eqref{eq:principaldelta}. The case of complementary series \eqref{eq:compdelta} can be obtained by analytic continuation. Next, we use the Hubbard-Stratonovich trick to write
\begin{equation}
	e^{-\left(\frac{d}{2}+L\right)^2 \tau}= \int_{ \mathbb{R} + i\delta} du \ \frac{e^{-u^2/4\tau}}{\sqrt{4\pi \tau}}  e^{i\left(\frac{d}{2}+L\right)u} \; ,
\end{equation}
with $\epsilon>\delta >0$. Subtituting this into \eqref{eq:PInext}, we can perform the $\tau$-integral, yielding
\begin{align}
	\log Z_{\rm PI} = \int_{ \mathbb{R} + i\delta}  \frac{du}{2\sqrt{u^2+\epsilon^2}} e^{-\nu\sqrt{u^2+\epsilon^2}}   \sum_{L=-1}^\infty D^{d+2}_{L,s} e^{i\left(\frac{d}{2}+L\right)u} \;. 
\end{align}
Deforming the contour by folding it up along the two sides of the branch cut from $u = i \epsilon$ to $u=i \infty$, changing variables $u=it$ and using that the square root takes opposite signs on both sides of the cut, we transform this to an integral
\begin{align}
	\log Z_{\rm PI} = &\int_\epsilon^\infty \frac{dt}{2\sqrt{t^2-\epsilon^2}} \left( e^{-i\nu\sqrt{t^2-\epsilon^2}}  +e^{i\nu\sqrt{t^2-\epsilon^2}}  \right)  \sum_{L=-1}^\infty D^{d+2}_{L,s} e^{-\left(\frac{d}{2}+L\right)t} \;. 
\end{align}
Formally putting $\epsilon =0$, we arrive at 
\begin{align}\label{eq:charPIstart}
	\log Z_{\rm PI}	= & \int_0^\infty \frac{dt}{2t} \left( e^{-\Delta t}  +e^{-\bar\Delta t}  \right)  \sum_{L=-1}^\infty D^{d+2}_{L,s} e^{-L t } \;.
\end{align}

\end{appendix}


\bibliographystyle{utphys}
\bibliography{ref}

\end{document}